%% file: bb.tex
\documentclass[smallcondensed]{svjour2}

\smartqed  
\usepackage{color}
\usepackage{graphicx}
\usepackage{mathptmx}      
\usepackage{amsmath}
\usepackage{amssymb}
\usepackage{times}
\usepackage{epsfig}
\usepackage{psfrag}
  
\journalname{J Stat Phys}
\begin{document}
\title{
The depinning transition in presence of disorder: a toy model
}

\author{Bernard Derrida \and Martin Retaux}

\institute{   B. Derrida \and M. Retaux \\
              Laboratoire de Physique Statistique,\\
              \'Ecole Normale Sup\'erieure,
 Universit\'e Pierre et Marie Curie, Universit\'e Denis Diderot, CNRS \\
              24, rue Lhomond,
              75231 Paris Cedex 05 - France \\
              \email{derrida@lps.ens.fr}\and\email{martin.retaux@ens.fr}
}

\date{Received: date / Accepted: date}

\maketitle

\begin{abstract}
{We introduce a toy model, which represents a  simplified version of the problem of the depinning transition in the limit of strong disorder. This toy  model can be formulated as a    simple renormalization transformation for the probability distribution of a single real variable. For this toy model, {the critical line is known exactly in one particular case} and it can be    calculated  perturbatively in the general case. One can also show that, at the transition,  there is no  strong disorder fixed distribution   accessible by renormalization.
Instead,  both our  numerical and analytic approaches indicate   a transition of infinite order (of  the  Berezinskii-Kosterlitz-Thouless type).  We give numerical evidence  that this infinite order transition persists
 for the problem of the depinning transition with disorder on the hierarchical lattice.}
\end{abstract}
\ \\ \ \\
\today \\ \ \\

The depinning transition in presence of impurities \cite{Al,DHV,FORG1,MG3,GT2,KL,MG4,MG2,TC} (in the version of the Poland  Scheraga (PS) model  with uncorrelated disorder) is one of the simplest  problems for which the effect of disorder, at a phase transition,  is non trivial.

Though the problem is very simple to   formulate and  despite all the  progress done over the last 30 years \cite{G1}, many basic questions on the precise location of the critical surface or on the nature of the depinning transition when disorder is relevant are still debated.

Several authors have  studied a simplified version of the problem,  by considering the depinning problem on a hierarchical lattice  \cite{DHV,L2,MG2,TC}, but in this case too, the same basic questions remain hard to  answer.

Here we try to look at an even simpler problem, a toy model, which resembles the depinning problem on the hierarchical lattice in the limit of strong disorder. Our toy model can be formulated as a very simple renormalization transformation for a probability distribution of a single variable. Our main result is that, in contrast to usual critical phenomena,   the transition is not characterized by a critical fixed distribution. Instead,  the transition is of infinite order (of the  Berezinskii-Kosterlitz-Thouless type \cite{KT}).

This article is organized as follows. First, in order to show the connection between  our toy model and the depinning problem, we make in section 1 a short review on past results on the Poland-Scherga model  with disorder and on its simplified version on the hierarchical lattice.
Then, in section 2,  we present several numerical results on the location of the transition and on the nature of the  critical behaviour of our toy model.
In section 3, we explain how the infinite order transition can be understood analytically. This is confirmed by the absence of  fixed critical distributions. We also show how to characterize the critical manifold perturbatively.
Lastly in section 4, we present some numerical results  which indicate that the hierachical model and our toy model have similar critical behaviors, namely an infinite order transition.

\section{From the Poland  Scheraga model with disorder to our toy model}
\begin{enumerate}
\item{ \it The Poland Scheraga model in presence of disorder} \\
The Poland Scheraga model \cite{PS}  is a model for the denaturation of the DNA molecule  (the transition from a double strand molecule into two single strands) or for the depinning transition of a line from a substrate. 
In the Poland Scheraga model, one represents the two strands of DNA as in figure 1. There is a binding  energy  $\epsilon_i$ when the two strands are in contact at position $i$. In addition there is an entropy factor  $\omega_l$ for each loop of length $l$ between  two consecutive contacts 
(a loop of length $l$ corresponds to $l-1$ consecutive unpaired bases).
\begin{figure}[h]
\begin{center}
\includegraphics[width=8.5cm]{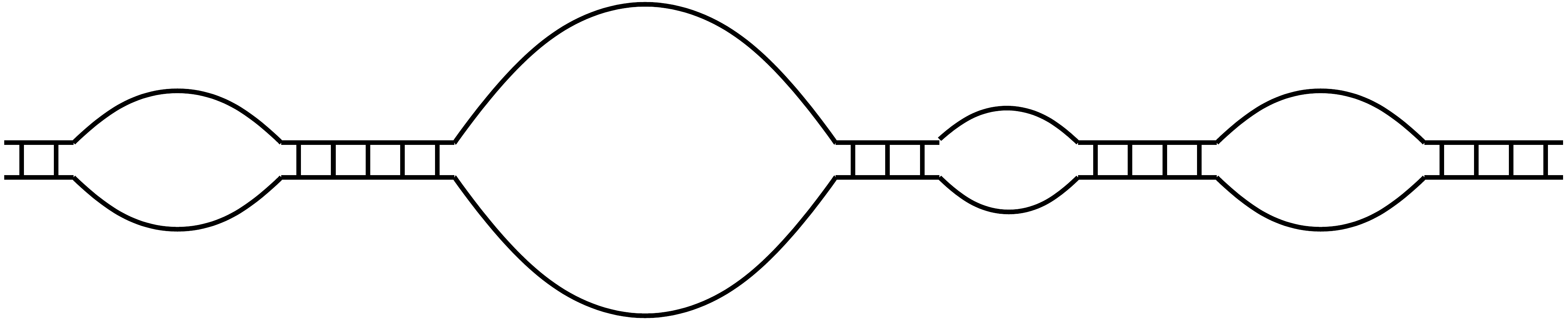}
\end{center}
\caption{ In the Poland Scheraga model there is an energy for each pair of bases in contact and an entropy  factor (\ref{entropy-factor}) for each loop of unpaired bases.
}
\label{fig1}
\end{figure}
The partition function $Z_L$  of a molecule of length $L$ is then given by
$$Z_L= \sum_{k \ge 2} \  \ \sum _{1<i_2 <i_3 \cdots < i_{k-1} < L} \omega(i_2-i_1) \cdots \omega(i_k-i_{k-1}) \exp\left[- {\epsilon_{i_1}  + \epsilon _{i_2} +   \cdots \epsilon_{i_k} \over T} \right]$$
where,  in the sum, $k$ is the number of contacts, $i_1, \cdots i_k$ are the positions of the contacts  (we have chosen here  to impose contacts at positions $1$ and $L$ so that   $i_1=1$ and $i_k=L$). In the disordered version of the model, the $\epsilon_i$'s are quenched i.i.d. random variables.

In the Poland Scheraga model
it is well known \cite{Fisher,KMP,PS,Guttman} that the nature of the transition depends on the large $l$ bevavior of $\omega(l)$. Usually one chooses a large $l$ dependence of the form 
\begin{equation}
\omega(l)  \sim {s^l \over l^c} \ .
\label{entropy-factor}
\end{equation}
where $s$ and $c$ are two parameters
( $ l\log s  $ is the extensive part of the entropy of  a large loop of size $l$
while the critical behavior  at the transition depends \cite{Fisher,KMP,PS,Guttman} on  the parameter $c$ ).

 Depending on the large $L$ behavior of $\log Z_L$, the system is either  in the unpinned   or in the pinned  phase
$$
\lim_{L \to \infty} {\langle \log Z_L \rangle \over L} 
 =  \log s    \ \ \ \  \ \   \text{in the unpinned phase} \ \ \ \  $$ $$ 
\lim_{L \to \infty} {\langle \log Z_L \rangle \over L} 
 > \log s   \ \ \ \   \ \   \text{in the pinned  phase}  $$
where $\langle. \rangle$ denotes an average over the disorder (i.e. over the random energies $\epsilon_i$)
and the simplest questions one may ask about  of the denaturation transition  are:
\begin{itemize}
\item Where is  the precise  location of the transition temperature $T_c$  which separates these two phases ?
\item
How does the difference ${\log Z_L / L} - \log s$  vanish as $T \to T_c$?
\end{itemize} 
\ \\ \ \\ 

{\it In the  pure case}, i.e. when 
 all the $\epsilon_i$ are equal,
these questions have  well known  answers  \cite{Fisher,KMP,PS,Guttman}
and 
it is known that there is a phase transition (for attractive energies, i.e. for negative $\epsilon$) whenever $c>1$.   For $1 < c < 2$  the transition is 
 second order  with an exponent $\nu  $ which varies continuously  with $c$  while for  $c> 2$ , it  becomes first order
$$ \lim_{L \to \infty} {\log Z_L \over L}  - \log s \sim (T_c^\text{pure}-T)^\nu     \ \ \ \text{ with} \ \ \    \left\{\begin{array}{lll}
 \nu    =1/(c-1)  &  \ \ \ \text{for} \ \ \ & 1<c <2  \\ 
 \nu    =1 &  \ \ \ \text{for} \ \ \ & c >2  \  . 
  \end{array} \right.$$
 \\
In this  pure case, let us define  $u_c ^\text{pure}$ by
$$u_c^\text{pure}= \exp \left[-{\epsilon \over T_c^\text{pure}}  \right] $$

which will be useful below.
\\

{\it In the  random case}, that is when the $\epsilon_i$ are i.i.d. random variables, there is still a transition   for $c>1$ but, for a general distribution of the energies $\epsilon_i$, the transition temperature $T_c^\text{quench}$ is not known and the nature of the transition is still debated. In this random case one can however   calculate the annealed  partition function $\langle Z_L \rangle$ (where as above  $\langle. \rangle$  
is an average over the $\epsilon_i$'s)  and show that 
that it undergoes a transition at a temperature  $T_c^\text{annealed}$ given by
$$\left\langle \exp \left[-{\epsilon_i \over T_c^\text{annealed }}  \right] \right\rangle  = u_c ^\text{pure} .$$
Using then Jensen's inequality one has \cite{CG}  that $\langle \log Z_L \rangle \leq \log \langle Z_L \rangle$ and therefore
\begin{equation}
T_c^\text{quenched} \leq T_c^\text{annealed} \ . 
\label{TT}
\end{equation}

Since the mid seventies, one knows after the work of Harris \cite{harris}  
under what condition  the  critical behavior of a pure system is modified by a weak amount of disorder.  For the depinning transition the Harris criterion predicts that  disorder is irrevelant when $c<3/2$, (meaning that a weak enough disorder should not change the critical behavior at the transition) while it is relevant for $c>3/2$.
\\ \ \\ 
 For $1<c<3/2$ these predictions have been confirmed rigorously: it has even been shown that for a weak enough disorder (i.e. if the distribution of the $\epsilon_i$'s is narrow enough)  the quenched and  the annealed models  have the same transition temperature  \cite{Al,L1,T1} (and (\ref{TT}) becomes an equality) and  the same  critical behavior  near the transition \cite{GT1,L1}.
When the distribution of the $\epsilon_i$'s is broader, one expects
(\ref{TT}) to become a strict inequality \cite{B1,T1,T2} and the nature of the transition remains debated \cite{Az1,Az2,KM,KL,TC}.
\\ \ \\
For $3/2<c$ , the relevance of disorder has also been confirmed
\cite{Rel1,Rel2,Rel3}.  Then, even for a narrow distribution of  the
$\epsilon_i$'s,  the inequality (\ref{TT}) is strict and the nature of the
transition is still debated (as for  a  broad enough distribution when
$c<3/2$). One however 
has bounds for the difference
$ T_c^\text{annealed}
-T_c^\text{quenched}$
 when disorder is small \cite{AlS,AZ,T2}.
One also
knows that for all values of $c>1$  the transition is
smooth \cite{GT2,GT3,AW}.  
\\ \ \\
The case $c=3/2$  has been the most difficult to analyse, as it is the case where, according to the Harris criterion,  a weak disorder is marginal and it has been debated for  years whether disorder was marginally relevant with $ T_c^\text{annealed}
 \neq T_c^\text{quenched}$ or irrelevant   with $ T_c^\text{annealed} =
  T_c^\text{quenched}$
\cite{MB1,MB2,DHV,FORG1,FORG2,GN,MG1}. The problem has finally been  settled
\cite{Rel2,Rel3} and the weak disorder behavior of the difference 
$ T_c^\text{annealed}- T_c^\text{quenched}$ has been estimated  \cite{DHV,Rel2,Rel3}.

\ \\

In conclusion for the Poland Scheraga model,  
with  strong enough disorder when $c < 3/2$ or with arbitrary disorder when $c \ge 3/2$,  
  the precise location of $  T_c^\text{quenched} $
and  the critical behavior as $T \to  T_c^\text{quenched} $ remain debated
questions. In particular one does not know how the critical behavior
depends on $c$ if it does at all.

\ \\
\item { \it The Hierarchical lattice} \\
In order to gain some insight on the previous problem, several authors have studied a simpler version of the problem: the depinning transition on a
 hierarchical lattice. On such a lattice the problem can be formulated as follows:
the partition  $Z_{n}$ of an interface of length $L_n=2^n$ can be
calculated  (up to a trivial normalization factor)   by
the following recursion relation \cite{Berger,DHV,L2,L4,L3,MG2,TC}
\begin{equation}
Z_{n}= {Z_{n-1}^{(1)} \ Z_{n-1}^{(2)}  + b-1 \over b} \ .
\label{hie}
\end{equation}
In (\ref{hie})  $ Z_{n-1}^{(1)}$ and $Z_{n-1}^{(2)} $ are two independent realizations of the partition function of an interface  of length $2^{n-1}$ and
$b$ is a parameter which characterizes the lattice.

As for the Poland Scheraga model, the pinned and the unpinned phases are defined by
$$
\lim_{n \to \infty} {\langle \log Z_n  \rangle \over 2^n} 
 =  0    \ \ \ \  \ \   \text{in the unpinned phase} \ \ \ \  $$ $$ 
\lim_{n \to \infty} {\langle \log Z_n \rangle  \over  2^n} 
 > 0   \ \ \ \   \ \   \text{in the pinned  phase}  \ . $$

To make the connection  with the Poland Scheraga model, each partition function $Z_0$ (which corresponds to a strand of length $2^0=1$) is randomly distributed according to a given distribution $P_0(Z)$ or equivalently one can write
$$Z_0 = \exp \left(-{\epsilon \over T}  \right)$$
where each energy $\epsilon$   is chosen according to a given distribution $\rho(\epsilon)$.
\ \\  \ \\
{\it In the pure case}, i.e. when $P_0(Z)$ is delta distributed, the critical
value of $Z_0$ is given by the unstable fixed point of the map $Z \to (Z^2 + b-1)/b$
$$Z_0^\text{critical} =  \left\{ \begin{array}{lll} 1 &  \text{for}  \ \ \  &1 < b < 2
 \\ b-1  & & 2 < b  \end{array}\right. $$
and  the critical behavior is given by

\begin{equation}
\label{hie5}
 \lim_{n \to \infty} {\log Z_n  \over 2^n}   \sim (Z_0 - Z_0^\text{critical})^\nu     \ \ \ \text{ with} \  \left\{\begin{array}{lr}
 \nu    ={\log 2 / \log(2/b)} & \text{for ~ } 1<b <2  \\ 
 \nu    ={\log 2 / \log(\frac{2(b-1)}{b})} & \text{for ~ } b >2 .
  \end{array} \right.
\end{equation}
So in the pure case the transition is always second order, but the
exponent varies with $b$. Thus  $b$ plays a role similar to the parameter $c$ in the Poland Scheraga model.
\ \\ \ \\
{\it In the random  case}, the Harris criterion  tells us \cite{DHV} that disorder is irrelevant for  $ \sqrt{2}  < b < 2 + \sqrt{2}$ while it is relevant for $1<b < \sqrt{2}$ and for $b > 2 + \sqrt{2}$;

When disorder is irrelevant, very much like in the PS model, the quenched and the annealed  models have the same transition point and the same critical behavior when $P_0(Z)$ is narrow enough.

When  disorder is relevant, or when disorder is irrelevant but  strong enough, the main results  established so
far are similar to those of the Poland Scheraga model: the annealed and the quenched models
have different transition temperatures, the transition is smooth  \cite{LT}  and
for $b= \sqrt{2}$ and $b= 2 + \sqrt{2}$, disorder is marginally relevant
\cite{L2,L4,L3}.
The precise position of the transition in the quenched case is not known
(only bounds are known \cite{LT})
and the nature of the transition is still debated \cite{DHV,MG2,TC}.


 One can remark that  the recursion (\ref{hie}) is invariant under the transformation
$\{ b, Z \} \to \{b'=b/(b-1) ,Z'=Z/(b-1) \}$.  It is  therefore sufficient to consider the range
$$ 1 < b < 2 \ .$$
It is easy to check that if     $Z_{n-1}^{(1)} $ and $Z_{n-1}^{(2)} $  are  both larger than $b-1$, then the recursion (\ref{hie}) gives $Z_n> b-1$. 

In terms of the free energy $X= \log Z$, the recursion (\ref{hie}) becomes
\begin{equation}
X_{n} = X_{n-1}^{(1)}+X_{n-1}^{(2)}  + \log \left( {1 + (b-1) e^{-X_{n-1}^{(1)}-X_{n-1}^{(2)}}  \over b} \right) 
\label{hie1}
\end{equation}
and the range  $X_n > \log(b-1)$ is stable under the recursion.
 For $b$ close to $1$ we see that the third term in the r.h.s. of (\ref{hie1}) is essentially 0 except when the sum  $X_{n-1}^{(1)}+X_{n-1}^{(2)} $ is close to
or less than  $\log(b-1)$.
\\

\item{ \it The toy model studied in the present paper}

Our toy model is a simplified  version   of  recursion (\ref{hie1}):
\begin{equation}
X_{n} = \max \left[X_{n-1}^{(1)}+X_{n-1}^{(2)} , -a\right] 
\label{toy}
\end{equation}
where $a$ is a fixed positive  number (which plays the role of $-\log(b-1)$ in (\ref{hie1})). As in (\ref{hie1}), the range $X > -a$ is stable and  at each step one  essentially adds two independent variables $X_{n-1}^{(1)}$ and $X_{n-1}^{(2)} $ except when the sum is close to or less than the boundary value  $-a$ (see figure \ref{rc}). 
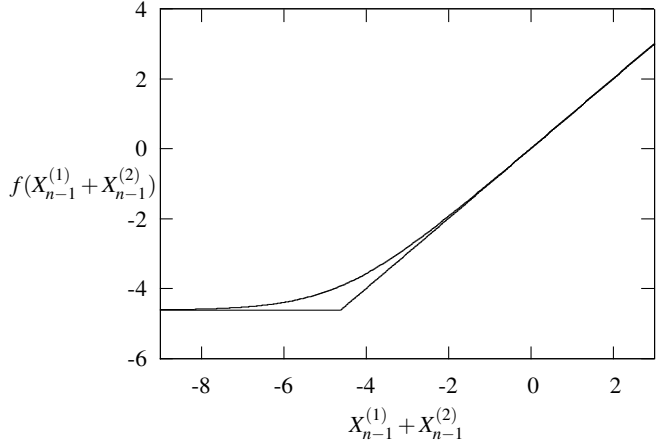
\begin{figure}[h]
\begin{center}
\input{./rc.tex}
\end{center}
\caption{ The right hand side of (\ref{hie1}) and of (\ref{toy}) are plotted versus $X_{n-1}^{(1)}+X_{n-1}^{(2)}$ in the case $b=1.01$ when $a=-\log(b-1)$.
}
\label{rc}
\end{figure}

The question is as before:
 given an initial distribution $P_0(X)$ of $X_0$, what is the large $n$ limit of the free energy
$$F_\infty = \lim_{n\to \infty} {\langle X_n \rangle  \over 2^n}  \ \ \  .$$

In this toy model the two phases can be identified by
\begin{eqnarray}
F_\infty
\nonumber
 =  0     &&  \ \ \ \ \ \ \    \text{in the unpinned phase}     \\
F_\infty   
\label{2-phases}
 > 0   &&   \ \ \ \ \ \ \   \text{in the pinned  phase} \end{eqnarray} 
 and by varying the initial distribution $P_0(X)$ one can observe a transition between these two phases.\\

{\it In the pure case}, that is when the initial distribution is a delta function
$$P_0(X) = \delta(X-\mu) ~ , $$
it is easy to see that the transition  is  first order:
$$
\begin{array}{llll}
 F_\infty & =0   \ \ \ \ & \ \ \ \ \text{for} \ \ \ \ & \mu \le 0 \\ 
    & =\mu   &   & \mu \ge 0 ~ . \\ 
 \end{array}  $$
(the transition is first order at $\mu=0$ because  $dF_\infty / d \mu$ is discontinuous).
\ \\ \ \\ \ 
{\it In the random case}
imagine that (for $a>1$) the initial distribution depends on a parameter $\lambda$ as in the following  example
\begin{equation}
P_0(X)= (1- \lambda) \delta(X+1) + \lambda \delta(X-1) \ .
\label{initial-distribution}
\end{equation}
By varying  the parameter $\lambda$  one can observe a transition from the pinned phase to the unpinned phase.

In the example (\ref{initial-distribution}), 
for $\lambda=1$, 
it is obvious that  $\langle X_n \rangle = 2^n $
 and  so $\lambda=1$ belongs to the pinned phase (\ref{2-phases}). For $\lambda=0$ it is also obvious that
  $-a \leq \langle X_n \rangle <0$  and so $\lambda=0$ belongs to the unpinned phase (\ref{2-phases}). As  $\langle X_n \rangle$ increases with $\lambda$, the phase transition should occur at  some  critical value $\lambda_c$.

One can obtain a sequence of upper bounds $\lambda_n$  for $\lambda_c$  by looking at the value $\lambda_n$ such that 
\begin{equation}
\langle X_n \rangle_{\lambda_n} =0 \ .
\label{bound}
\end{equation}
To see that $\lambda_n$ defined by (\ref{bound}) is an upper bound of $\lambda_c$ one can use the fact that
$$ \langle X_{n} \rangle_\lambda = \left\langle  \max \left[X_{n-1}^{(1)}+X_{n-1}^{(2)} , -a\right] 
 \right\rangle_\lambda \geq 2 \langle X_{n-1} \rangle_\lambda \ . $$
Since $ \langle X_{n} \rangle_\lambda $ is a continuous function of $\lambda$, and  as soon as $ \langle X_{n} \rangle_\lambda >0$, one has
$$F_\infty \ge {\langle X_{n} \rangle_\lambda \over 2^n}  > 0 \  . $$

We are going to  see that one  signature of the infinite order transition is that the upper bounds $\lambda_n$ defined in (\ref{bound}) satisfy  for large $n$
\begin{equation}
\lambda_n - \lambda_c = O\left({1 \over n^2} \right)  \ . 
\label{n2}
\end{equation}
To relate (\ref{n2}) to the infinite order transition, one can use the following argument:
from (\ref{toy}) one can easily show that
$$2 \langle X_{n-1} \rangle_\lambda \leq \langle X_{n} \rangle_\lambda  \leq 2 \langle X_{n-1} \rangle_\lambda + a \   $$ 
(we have seen that the range $X \ge -a$ is stable and the second inequality follows from the fact that $X_{n-1} \ge -a$). Therefore
if $ \langle X_{n} \rangle_\lambda=y$, for some positive $y$, 
one has
$$2^m y   \leq \langle X_{n+m} \rangle_\lambda \leq 2^m y + (2^m -1) a $$
and  one can be sure  that
\begin{equation}
 \langle X_{n} \rangle_\lambda = y \ \ \ \ \ \   \ \Rightarrow
\ \ \ \ \ \ 
{y \over 2^n} \leq F_\infty \leq {y + a  \over 2^n} \ .
\label{large-n}
\end{equation}
If one defines 
$\mu_n(y)$ 
as the value of $\lambda$ such that 
\begin{equation}
\langle X_{n} \rangle_{\mu_n(y)} = y
\label{mu-eq}
\end{equation}
and if, as in (\ref{n2}), one has
\begin{equation}
\label{mun}
\mu_n(y)- \lambda_c \sim {A \over n^2}  \ ,
\end{equation}
and  from (\ref{large-n}) one gets
$$F_\infty(\mu_n(y)) \sim   \exp \left( - {\sqrt{A} \log 2 \over \sqrt{\mu_n(y) - \lambda_c} }\right) \ . $$

In principle the  amplitude $A$ in (\ref{mun}) could depend on $y$. One can however argue that it does not:  for example,   for large $y$, changing $y$ by a factor $2$ has the effect of changing $n$ into $n+1$ and this does not change the amplitude  $A$. So for large $n$ one expects 
(\ref{mun}) to hold with a constant $A$ independent of $y$ (for $y \ge 0$).

This is confirmed in figure  \ref{rv3} where we  plot $\mu_n(y)$  defined by (\ref{mu-eq}) versus  $1/n^2$ for $y=0,1,10$  in the case $a=1$ and we see that for large $n$ the data are consistent with (\ref{mun}) and an amplitude $A$ independent of $y$.
\begin{figure}[h]
\begin{center}
\input{./rv3.tex}
\end{center}
\caption{ Values of $\mu_n(y)$ solutions of (\ref{mu-eq}) for $y=0,1$ and $10$. The convergence to $\lambda_c$ is as predicted in (\ref{mun}) with an amplitude $A$ which seems to be independent of $y$ (in the figure, $A$ is the slope at the origin).
}
\label{rv3}
\end{figure}
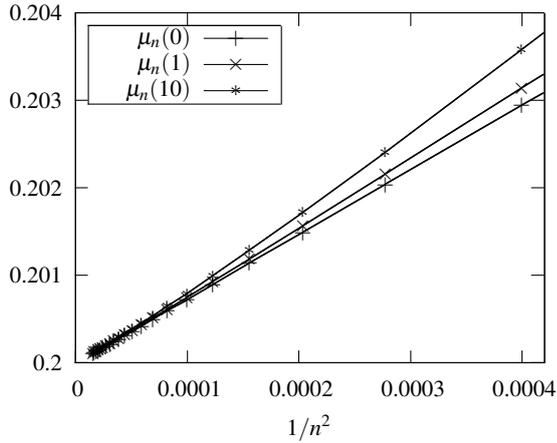

Therefore we expect that as $\lambda \to \lambda_c$
\begin{equation}
\label{infinite-order}
F_\infty(\lambda) \sim   \exp \left( - {\sqrt{A} \log 2 \over \sqrt{\lambda - \lambda_c} }\right) \ . 
\end{equation}

{\it Remark:}
One can also find lower bounds  for $\lambda_c$ by noticing that a consequence of (\ref{toy}) is that for any $\alpha$
$$
\langle e^{\alpha X_{n+1} }\rangle  \leq \langle e^{\alpha X_{n}} \rangle^2 + e^{-\alpha a}  \ .
$$ 
Therefore  $\lambda \leq \lambda_c$
whenever one can find some $\alpha >0 $ for which $  \langle e^{\alpha X_{n}} \rangle \leq {1 + \sqrt{1 - 4 e^{-\alpha a}} \over 2}$.
This kind of lower bound is in the spirit of those obtained from estimates of non-integer moments of the partition function in disordered systems \cite{Rel1}.
\end{enumerate}

\section{Numerical evidence of the infinite order transition}
\begin{figure}[h]
\input{./rv2.tex}
\caption{ The upper bounds $\lambda_n$ obtained by solving (\ref{bound}) for $10 \leq n \leq 200$. One sees clearly the $ 1/n^2$ convergence  which is expected for an infinite order transition of the form (\ref{infinite-order}).
}
\label{rv2}
\end{figure}
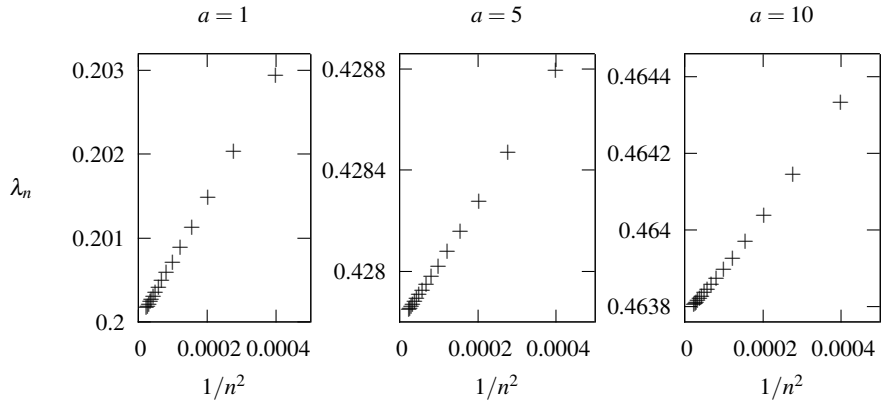

We saw in the previous section that one  signature (\ref{n2})  of the infinite 
order transition 
(\ref{infinite-order})
is  that the upper bounds $\lambda_n$ (solutions of (\ref{bound})) converge to $\lambda_c$ as $1/n^2$. In figure \ref{rv2}  we see  clearly this  $1/n^2$ convergence when we plot $\lambda_n$ versus $1/n^2$ for the initial distribution (\ref{initial-distribution}). 

In table \ref{tab} we give estimates of the critical values $\lambda_c$ and of the amplitude $A$ in (\ref{n2}) for several choices of $a$.

 \begin{table}[h!]
\begin{center}
 \begin{tabular}{|c|c|c|c|c|c|c|}
  \hline
 & & & & & & \\
$  a$ = &  1 & 2 & 5 & 10 & 20 & 50 \\
  \hline
 & & & & & & \\
  $\lambda_c  \simeq $& \  .2000 \  
   &\  .3333 \  
   &\  .4278\  
   &\  .4638 \     
   & \ .48182 \    
   &\  .4927   \   \\
$A\simeq $& 7 & 5.2 & 2.3 & 1.1 & .7 &   \\
$a(\lambda_c - 1/2)\simeq $& .3 & .333 & .361 & .362 & .364 & .364  \\
  \hline
\end{tabular}
\end{center}
\caption{The values $\lambda_c$  are estimated by extrapolating the bounds $\lambda_n$ obtained by solving (\ref{bound}). The amplitude $A$ is estimated as the slope  at the origin of the data of figure  \ref{rv2}. }
\label{tab}
\end{table}

One  can notice from  the last line of table \ref{tab} that, for large $a$,
\begin{equation}
\label{large-a}
\langle X_0 \rangle_{\lambda_c} = 2 \left(\lambda_c- {1 \over2} \right) \sim {.36 \over a} \ .
\end{equation}
In the appendix A we give a general  argument for  this $1/a$ dependence of $\langle X_0 \rangle_{\lambda_c}$.

\begin{figure}[h]
\begin{center}
\input{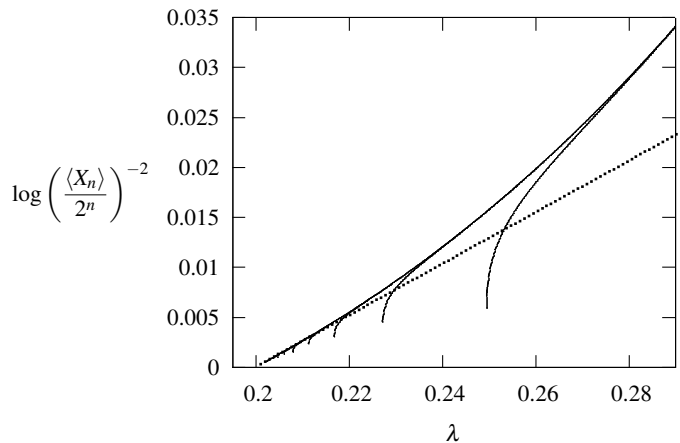}
\end{center}
\caption{ $[\log(\langle X_n \rangle/2^n)]^{-2} $ versus $\lambda$. The 
envelope seems to vanish linearly  as $\lambda \to \lambda_c$. This is consistent with the infinite order transition (\ref{infinite-order}). The thin dashed line is the linear behaviour expected from (\ref{infinite-order})  with the value $A$ given by the numerical estimates of table \ref{tab}.
}
\label{rv}
\end{figure}

Another way of visualizing  the infinite order transition (\ref{infinite-order})
is to try to plot $1/\log(F_\infty)^2$  as a function of $\lambda$. If (\ref{infinite-order}) is valid one should   observe a linear crossing with the real axis. In figure \ref{rv} we plot $1/\log[\langle X_n \rangle / 2^n]^2$ versus $\lambda$ for $n=10,15, \cdots 60$. The envelope appears to cross the positive real axis with a non-zero slope, at values of $\lambda_c$ consistent with the estimates of table \ref{tab} and the slope 
$(1/A/\log(2)^2)$ 
(estimated with the value $A$ of table \ref{tab})
 shown as a thin line  
 seems to be tangent to the envelope as expected from (\ref{infinite-order}).

\begin{figure}[h]
\begin{center}
\input{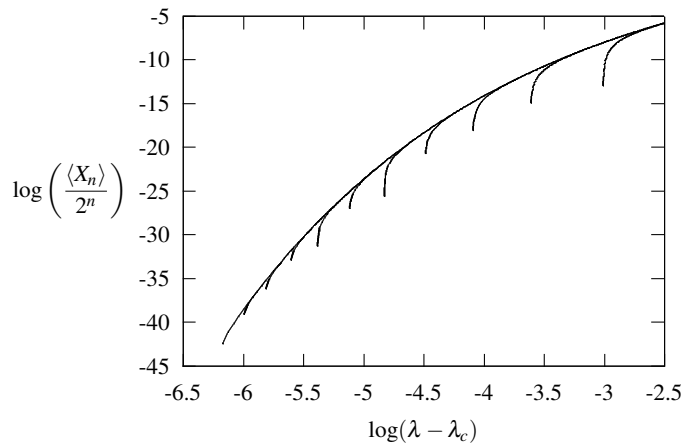}
\end{center}
\caption{
$\log( \langle X_n \rangle /2^n)$  versus $\log(\lambda- \lambda_c)$ in the case $a=1$. The envelope is expected to give $F_\infty$. There is nowhere evidence of a power law as $\lambda \to \lambda_c$.
}
\label{contrario}
\end{figure}

A contrario, a  power singularity  $F_\infty \sim (\lambda-\lambda_c)^\gamma$
does not  seem compatible with our direct calculation of $F_\infty$:  in figure \ref{contrario}   $\log F_\infty$  seems to be nowhere  a linear  function of $\log(\lambda-\lambda_c)$.

\section{Analytic arguments in favour of the infinite order transition}
If $a$ is an integer and  the initial distribution  $P_0(X)$ of $X$ is concentrated on integer values, then
the distribution $P_n(X)$  obtained
 under  the renormalization (\ref{toy})
remains concentrated on integers, and is of the form
\begin{equation}
P_n(X) = \sum_{k=-a}^\infty p_n^{(k)}  \  \delta(X-k)  \ ,
\label{Pn}
\end{equation}
 and     the renormalisation transformation (\ref{toy}) reads
\begin{eqnarray}
p_{n+1}^{(k)}=  \sum_{q=-a}^{k+a} p_n^{(q)} \  p_n^{(k-q)}  \ \ \ \ \ \ \    \text{for} \ \ \ \  \ \ \  k > -a 
\label{renorm}
 \\   p_{n+1}^{(-a)} = \sum_{q_1=-a}^{0}  \ \   \sum_{q_2=-a}^{-a-q_1} \   \ p_n^{(q_1)} \ p_n^{(q_2)}   \ .
\label{renorm1}
\end{eqnarray}
In terms of the generating function
\begin{equation}
H_n(x) = \int P_n(X)  \ z^{X+a} \ dX = \sum_{k=-a}^\infty p_n^{(k)}  \ z^{k+a} 
\label{generating}
\end{equation}
the transformation  (\ref{renorm},\ref{renorm1})  becomes
\begin{equation}H_{n+1}(z) = {H_n^2(z) - Q_n(z) \over z^a} + Q_n(1) \label{map} \end{equation}
 where $Q_n(z)$ is the polynomial of degree $a-1$  obtained by keeping the first $a$ coefficients of the expansion of $H_n^2(z)$ around $z=0$ :
\begin{equation}
  \ \ \ H_n^2(z) = \sum_{k=0}^\infty \beta^{(k)}  \ z^k \ \ \ \ 
 \ \ \ \ \Rightarrow \ \ \ \ Q_n(z) = \sum_{k=0}^{a-1}  \beta^{(k)}   \ z^k  \ .
\label{Q-def} \end{equation}

The unpinned phase corresponds to the fixed point $H(z)=1$, the pinned phase  to the fixed point $H(z)=0$  and the critical point of the pure case to the fixed point $H(z)=z^a$ of the transformation (\ref{map}).
One can then imagine two possible scenarios for the critical behavior in the strong disorder case:
\begin{enumerate}
\item  The existence of a new    unstable fixed point $H_c(z)$ corresponding to the transition in the strong disorder case: this would imply  a critical behavior given by a power law, with an exponent related, as usual in critical phenomena, to the repulsive eigenvalue of the linearised map around the fixed point $H_c(z)$.  We will see below that there is no accessible fixed point $H_c(z)$ (here accessible means a fixed distribution  with non negative weights on the integers). 
\item A transition of infinite order with no critical fixed point, as for example in the renormalization equations (\ref{evol-BD}) below.
\end{enumerate}

\subsection{The fixed points of the map (\ref{map})}
\underline{In the case $a=1$,} the map (\ref{map})
can be written as
\begin{equation}
H_{n+1}(z)=  {H_n(z)^2 - H_n(0)^2 \over z}+ H_n(0)^2 \ .
\label{map1}
\end{equation}
This  map  has been studied by 
 Collet, Eckmann, Glaser and Martin \cite{Eckmann}. 
 They have shown in particular that there is no other fixed points 
(accessible when all the $p_n^{(k)} \ge 0$)
than $H(z)=0$, $H(z)=1$, $H(z)=z $ 
and that there are no  periodic orbits.
They also identified the critical manifold given by the condition  
\begin{equation}
H_n(2) - 2 H'_n(2)=0
\label{critical-manifold}
\end{equation}  (condition which should be supplemented by the fact that $H_n(z)$ is analytic for $|z| <2$) which separates the basins of attraction of the two fixed points $H(z)=0$ (pinned phase) and $H(z)=1$ (unpinned phase). 

For the initial distribution (\ref{initial-distribution}), one has $H_0(z)= 1- \lambda+ \lambda z^2$ and the condition (\ref{critical-manifold}) gives  $\lambda_c=1/5$ in full agreement with what we saw numerically in section 2.

Another result of \cite{Eckmann} is that along the critical manifold (i.e. if $H_0(2) -2 H_0'(2)=0$ and $H(z)$ is analytic in the disc $|z | < 2$) one has for large $n$
\begin{equation}
1- H_n(0)   \sim{4 \over n^2} \ .
\label{convergence}
\end{equation}
We will recover this large $n$ dependence in our perturbative approach below.
\\ \ \\
{\it Remark:}
For $a=2$ and an initial distribution of the form (\ref{initial-distribution}), one has $H_0(z)=(1-\lambda) z + \lambda z^3$ and
$H_1(z)=(1-\lambda)^2 + 2 \lambda (1- \lambda) z^2 + \lambda^2 z^4$.
It is clear that for  $n \ge 1$, the distribution $P_n(X)$ is concentrated on even values of $X$. Therefore the case $a=2$ with $H_1(z)=(1-\lambda)^2 + 2 \lambda (1- \lambda) z^2 + \lambda^2 z^4$ is equivalent to the case $a=1$ with $H_1(z)=(1-\lambda)^2 + 2 \lambda (1- \lambda) z  + \lambda^2 z^2$  for which the transition  according to (\ref{critical-manifold}) is located at  $\lambda_c=1/3$ in  agreement with the estimate of table \ref{tab}.
\ \\ \ \\
\underline{For $a \ge 2$}  one can show (see appendix B) as in the case $a=1$  that the only accessible fixed points of the map (\ref{map}) are $H(z)=0$ (the pinned phase), $H(z)=1$ (the unpinned phase) and $H(z)=z^a$ (the critical fixed point of the pure case).
 
As for the case $a=1$ the critical fixed point of (\ref{map}) in  the pure case is unstable  in presence of disorder:  for small $\epsilon$,   one gets from (\ref{map}) 
$$ \ \ \  H_n(z)=z^a + \epsilon h(z) \ \ \ \ \Rightarrow 
 \ \ \ \ \ H_{n+1} \simeq  z^a + 
2 \epsilon h(z)$$
so that disorder is relevant.

In contrast to the case $a=1$, the critical manifold, which would generalize (\ref{critical-manifold}), is not known.

\ \\ \ \\ {\it Remark:} When one tries to study the iteration (\ref{toy}) numerically, one has to limit the number of possible values of $X_n$. One way of doing it is to replace (\ref{toy}) by
\begin{equation}
X_{n+1} = f\left(X_{n-1}^{(1)} + X_{n-1}^{(2)} \right) \ \ \  \text{with} 
 \ \ \ \ f(X)= \left\{\begin{array}{lll} -a  & \text{ \ \ \ if \ \ \ } & X   \le -a \\ X & & -a \le X \le a' \\
a' & & X > a' \end{array} \right.
\label{cut-of}
\end{equation} where $a'$ is a large positive number. Thus $a'$ plays the role of a cut-off for large values of $X$.
One can show (see Appendix C) that as long as  $a'$ is finite there exists an
 (accessible) unstable  fixed distribution. This fixed point disappears in the limit $a' \to \infty$, so the artefact of the cut-off $a'$ is to give rise to an unstable fixed distribution. It is possible that the fixed point distribution found in \cite{MG2} on the hierarchical lattice is due to a similar artefact.

\subsection{The perturbative approach}
It is too difficult to  calculate $F_\infty$ analytically by  iterating the renormalization transformation  (\ref{renorm}) or (\ref{map})  for a general  $P_0(X)$ or $H_0(z)$. Here our analytic approach is limited to distributions $P_n(x)$ 
of the form (\ref{Pn}) with
\begin{equation}
p_n^{(k)}  =  v  \, \phi^k \, R_n( u\, k) \ \ \ \ \text{for} \ \ \  k >  -a
\label{initial-dist}
\end{equation}
with 
\begin{equation}
p_n^{(-a)}   =  1- \sum_{k > -a} p_n^{(k)} 
\label{norm}
\end{equation}
where $u$ is a small parameter, $ v$ is also small, the function $R_n(w)$ is  smooth  
and   $ \phi $ is a positive constant which  satisfies
\begin{equation}
 2 \phi^a=1 \ . 
\label{c-def}
\end{equation}
The normalization  (\ref{norm}) gives to leading order in $u$  
\begin{equation}
p_n^{(-a)} = 1 - {2 \phi \over 1-\phi} \,  v \,  R_n(0)  + O(u^3)  \ .
\label{pma}
\end{equation}
The  main reason for the choice (\ref{initial-dist}) is
that we observed  numerically that for generic initial distributions close to the transition, the renormalization after a few steps leads to distributions of the   form
(\ref{initial-dist}), and then after these few steps, the distribution changes slowly for many iterations
of (\ref{renorm}).
In fact one can show, using the Euler MacLaurin formula,  that  when 
$$
v= O(u^2) \  , 
$$
the distribution keeps the form (\ref{initial-dist}) 
   under the transformation (\ref{renorm}),  with
$$
R_{n+1}(x)=R_n(x) +  u \  a {d R_n(x)  \over d x} + {v \over u} \int_0^x R_n(x_1) \ R_n(x -x_1) \ d x_1 +  O(u^2) \ . $$
As $u$ is small, $R_n(x)$ takes the scaling form
$$R_n(x) = r (x, n \, u)$$ where the scaling function $r$  satisfies
\begin{equation}
\label{renor1}
{\partial r(x,\tau) \over \partial \tau} = a {\partial  r(x,\tau)  \over \partial x}  + w \int_0^x r(x_1,\tau) \ r(x -x_1,\tau) \ d x_1
\end{equation}
where $w$ is defined by
\begin{equation}
v = w  \, u^2 \  .
\label{w-def}
\end{equation}
Therefore to understand the problem when $u$ is small, one needs to predict the large $\tau$ behaviour of $r(x,\tau)$ solution of  (\ref{renor1}) as the initial distribution  $r(x,0)$ varies. This
is still  a  problem   difficult to solve for an arbitrary $r(x,0)$. For a particular choice however, when  $r(x,0)$ is of the form
\begin{equation}
r(x,0) = B(0) e^{-D(0) x} \ ,
\label{expo}
\end{equation}
\\ the problem can be solved.
It is easy to check that $r(x,\tau)$  solution of (\ref{renor1}) remains of the same form
\begin{equation}
r(x,\tau) = B(\tau) e^{-D(\tau) x}
\label{rwtau}
\end{equation}
with the parameters $B(\tau)$ and $D(\tau)$ evolving according to
\begin{equation}
{d B(\tau) \over d \tau} =-a D(\tau) B(\tau)\ \ \ \ ; \ \ \ \ 
{d D(\tau) \over d \tau} =- w         B(\tau) \  . 
\label{evol-BD}
\end{equation}
This type of renormalization equation is characteristic of an infinite order transition \cite{KT,TC}.
They can be integrated and   three different  behaviors emerge:
 
\begin{itemize}
\item  For $ 2 B(0)    w    - a D(0)^2 <0$ the solution is 
$$ D(\tau)=  D(0) - {2 B(0) \,     w   \,   \sinh( \kappa \tau) \over 
2 \,  \kappa\,  \cosh(\kappa \tau) + a \, D(0)\,  \sinh(\kappa \tau) } $$
$$B(\tau)=  {a \,  D(\tau)^2    \over 2    w   } - {2 \kappa^2 \over a \,    w   }$$
where  $\kappa$ is the positive  root of $$\kappa^2 =
- {a B(0)    w    \over 2}
+{a^2 D(0)^2 \over 4} 
 $$
(one could  choose  as well  the negative root as $D(\tau)$ and $B(\tau)$ are even functions of $\kappa$).
One can easily check that in the limit $\tau \to \infty$
$$D(\tau ) \to {2 \kappa \over a} \ \ \ \ ; \ \ \ \ 
B(\tau) \to 0 $$
and this corresponds to the unpinned phase.
\\ 
\item  For $ 2 B(0)    w    - a D(0)^2 =0$ the solution is 
\begin{equation}
 D(\tau)= { 2   D(0)  \over 
2    + a D(0) \tau } \label{D-critical} \end{equation}
and
\begin{equation}
B(\tau) = { a D(\tau)^2 \over 2    w   } \ .
\label{B-critical} \end{equation}
We see using (\ref{pma}) that
\begin{equation}
1-H_n(0) = 1-p_n^{(-a)} =  {2 \phi \over 1-\phi}  v  \, B(\tau) \simeq {4  \phi \over a(1-\phi) n^2} \ .
\label{convergence1}
\end{equation}
In the case $a=1$, one has $\phi=1/2$ and one recovers    the result (\ref{convergence}) of 
\cite{Eckmann}.
We expect, for other values of $a$, the amplitude  $4 \phi/((1-\phi) a)$ of the $1/n^2$ decay in     (\ref{convergence1}) to be generic for all initial distributions at criticality
(there are exceptions however such as (\ref{initial-distribution}) when $a$ is even: after one step of renormalization, the distribution is concentrated on even values  of $X$ and  by iterating one can never reach a distribution of the form (\ref{initial-dist}) with a smooth function $R_n(x)$).
\\ 
\item  For $ 2 B(0)    w    - a D(0)^2 >0$ the solution is 
$$ D(\tau)=  D(0) - {2 B(0)     w     \sin( \kappa \tau) \over 
2  \kappa \cos(\kappa \tau) + a D(0) \sin(\kappa \tau) } $$
$$B(\tau)=   {a \,  D(\tau)^2   \over 2    w   } + {2 \kappa^2 \over a    w   }$$
where $$\kappa^2 =
 {a B(0)    w    \over 2}
-{a^2 D(0)^2 \over 4} \ .
$$
This solution predicts that $D(\tau)$ and $B(\tau) \sim D(\tau)^2$ diverge as $\tau \to \tau_c$ (where $\tau_c$ is solution of $\tan  \kappa \tau_c= - a D(0)/2/\kappa$).
For $\tau \sim \tau_c$ the distribution 
cannot be described by the form (\ref{initial-dist}) and the renormalization (\ref{renor1}) is no longer   a valid approximation of the true renomalization.

This solution indicates however that $p_n^{(-a)}$ decreases as $\tau \to \tau_c$ and this corresponds to the pinned phase.
Close to the transition (i.e. for $\kappa$ small) the number $n=\tau/u$ to reach the regime $\tau  \sim \tau_c $ is given by $\tau \sim  \pi /\kappa$. Therefore for  small $\kappa$ and $u$   one has
$$F_\infty \sim {1 \over 2^n}= 2^{-\pi  / (u \, \kappa)} = \exp\left[ - {\pi \log 2 \over u \sqrt{{a     w    B(0)  / 2} - a^2 D(0)^2/4}}\right] \ .$$
If $D(0)$ is fixed and $B(0)$ varies, the depinning transition occurs at  $B_c(0)$ 
\begin{equation}
B_c(0)={ a \, D(0)^2 \over 2 \,    w   }
\label{Bc(0)}
\end{equation}
 and this gives for $F_\infty$
$$F_\infty \sim \left[ - { \sqrt{A} \log 2 \over  \sqrt{  B(0) -B_c(0)}}\right]$$
as in (\ref{infinite-order})
with  the amplitude $A$ given by $$A={ 2 \pi^2 \over u^2  \, a  \,    w    } \ .$$
\\ \ \\ {\it Remark:} In the case $a=1$ a distribution of the form (\ref{initial-dist},\ref{expo})  gives for the generating function $H_0(z)$ defined in (\ref{generating})
$$H_0(z)= 1 + 4 v  B(0)  { (z-1) \over (2 - \exp(-D(0) u))(2 - z \exp(-D(0) u))}
$$
and the critical manifold  (\ref{critical-manifold}) gives  to leading order, for $v=    w    u^2$,
\begin{equation}
B_c(0) =  D(0)^2 /    w    /2
\label{crit1}
\end{equation}
which agrees with (\ref{Bc(0)}) in this particular case $a=1$.
\end{itemize}
\subsection{The critical manifold}
We have seen that  (\ref{crit1}) gives the critical manifold to leading order in $u$ for an initial condition of the form (\ref{initial-dist})
\begin{eqnarray}
p_0^{(k)}  =  v \, \phi ^k \, R_0( u\, k) \ \ \ \ \text{for} \ \ \  k >  -a
\end{eqnarray}
with (see (\ref{expo}))
$$R_0(x) = B(0) e^{-D(0) x} \ .$$

One can try to develop a pertubation theory to determine this critical manifold to higher order in $u$. We are going now to explain how this can be done to generate the first correction to (\ref{crit1}).

To obtain the next order in $u$ one  considers, for $k > -a$,  distributions of the form
\begin{equation}
p_n^{(k)} =  \phi ^{k}    w     u^2  e^{- D(\tau ) u \,  k} \left[   B( \tau) + u F(  u \,  k, \tau)  + O(u^2) \right] 
\label{correction}
\end{equation}
 where $ n=\tau/ u $ with $\tau$ of order 1. 
The reason  for this choice is that  a  function $F$ of the form written in the r.h.s. of (\ref{correction}) is generated by the iteration even if one starts without it.
For such a choice, using (\ref{c-def}) one can show that for $n u= \tau$ 
\begin{equation}
\label{correction-norm}
p_n^{(-a)}= 1 - 2 u^2    w    {\phi  \over 1-\phi }  \left[ B( \tau  ) + u \left( F( 0, \tau)+  a D(\tau) B(\tau) -  {D(\tau) B(\tau) \over 1-\phi }  \right)+ O(u^2)  \right] \ .
\end{equation} 
Inserting (\ref{correction},\ref{correction-norm}) into (\ref{renorm}) one gets (\ref{evol-BD}) to leading order in $u$ and to the next order
\begin{eqnarray}
{\partial F \over \partial \tau}  =  
 (D'  x - a D) F  +   a {\partial F \over \partial  x} +  2    w    B \int_0^x F(x_1,\tau) d x_1
 \ \ \ \ \ \ \ \ \ \ \ \ \ \ 
 \ \ \ \ \ \ \ \ \ \ \ \ \ \ 
\ \ \ \ \
\label{F-evol}
\\    - { D'^2 B  \over 2} x^2  +{ D'' B  +  2 D' B' \over 2}  x+     w    \left( 2 a - {1 +\phi  \over 1-\phi }\right)B^2 +{ a^2 B D^2  - B'' \over 2} \ .
\nonumber
\end{eqnarray}
This evolution equation for  $F$ is linear. It is nevertheless  not easy to  solve. 
One can  write the evolution equations of the moments  $I_n(\tau)$ defined by 
$$I_n(\tau ) = \int_0^\infty x^n \, F(x,\tau )  \, e^{-D(\tau ) x}  \, d x $$
but they are all coupled and we were not able to solve them.

For example from (\ref{F-evol}) one has
\begin{eqnarray*}
{ d I_1(\tau)   \over d \tau}  =  
 \left({ 2    w    B
 \over D^2} - a \right)   I_{0}(\tau) + {2    w    B \over D} I_1(\tau)
 \ \ \ \ \ \ \ \ \ \ \ \ \ \ 
 \ \ \ \ \ \ \ \ \ \ \ \ \ \ 
\ \ \ \ \
\\    - { 3  D'^2 B  \over  D^4 }   +{ D'' B  +  2 D' B' \over  D^3 }  +     w    \left( 2 a - {1 +\phi  \over 1-\phi }\right){B^2 \over D^2}  +{ a^2 B D^2  - B'' \over 2 D^2} 
\end{eqnarray*}
which becomes using  the evolution equations (\ref{evol-BD}) of $B(\tau)$ and $D(\tau)$ 
$${d I_1(\tau ) \over d \tau}=
 \left({ 2    w    B \over D^2} - a \right)   I_{0}(\tau) + {2    w    B \over D} I_1(\tau) 
 - 3    w   ^2 {B^3 \over D^4} +  w \left( {9        a \over 2} -   {1+\phi  \over 1-\phi }\right) {B^2 \over D^2} 
\ . $$
Along  the critical manifold (\ref{crit1}) however, where $2    w    B(\tau)= a D(\tau)^2$,  the evolution of $I_1(\tau)$ becomes autonomous
and one gets
$${d I_1(\tau ) \over d \tau}=  \left( {3 a^3 \over 4    w   }-{a^2(1+\phi ) \over 4 (1-\phi )    w   }  \right) D(\tau)^2
 + { a D(\tau)} I_1(\tau)  $$
with $D(\tau)$ given by (\ref{D-critical}). This can be integrated
$$I_1(\tau) = - {2  \over 3  a } \times  \left( {3 a^3 \over 4    w   }-{a^2(1+\phi ) \over 4 (1-\phi )    w   }  \right)   D(\tau) + K (2 + a D(0) \tau)^{2 }$$
where $K$ is an integration constant. As $F$ is a correction, the condition to remain on the critical manifold is that $F$ does not grow with $\tau$. Therefore
the integration constant  $K$ should vanish, and the critical manifold should be given by
$$I_1(0) = - {2  \over 3  a } \times  \left( {3 a^3 \over 4    w   }-{a^2(1+\phi ) \over 4 (1-\phi )    w   }  \right)   D(  0) \ .$$
In the case $a=1$,  one has $\phi =1/2$ and $I_1(0)=0$  and one can check that this is in  agreement with (\ref{critical-manifold}).


\section{ Numerical evidence of 
an infinite order transition
for  the hierarchical lattice 
}

In this  section, we will give numerical evidence that the hierarchical lattice model in
the limit of strong disorder and our toy model have very similar critical
behaviors. Our numerical analysis is entirely based on the recursion (\ref{hie1})  in terms of the free energy $X=\log Z$:
\begin{equation}
\label{recX}
X_n=X_{n-1}^{(1)}+X_{n-1}^{(2)}+\log
\left(\dfrac{1+(b-1)e^{-X_{n-1}^{(1)}-X_{n-1}^{(2)}}}{b}\right) \ .
\end{equation}
To make a simpler connection with the toy model, we call 
$$a=-\log(b-1) .$$ 
It is then easy to check in (\ref{recX})  that  the range $X_n
>-a$ is stable.

The Harris criterion in terms of $a$ tells us that for  $ -\log(\sqrt{2}-1) \simeq 0.88137  < a < \infty $, an arbitrary amount of disorder  is relevant while in the range $0 < a < -\log(\sqrt{2}-1) $, a weak enough
disorder stays irrelevant.

We are going to see that there is numerical evidence of an infinite order transition for the hierarchical lattice case when the disorder is relevant. On  the other hand, for a case where disorder is irrelevant our data will show a transition very similar to the pure case, with a power singularity.

\subsection{ A case where disorder is relevant}

\begin{figure}[h]
\centering
\input{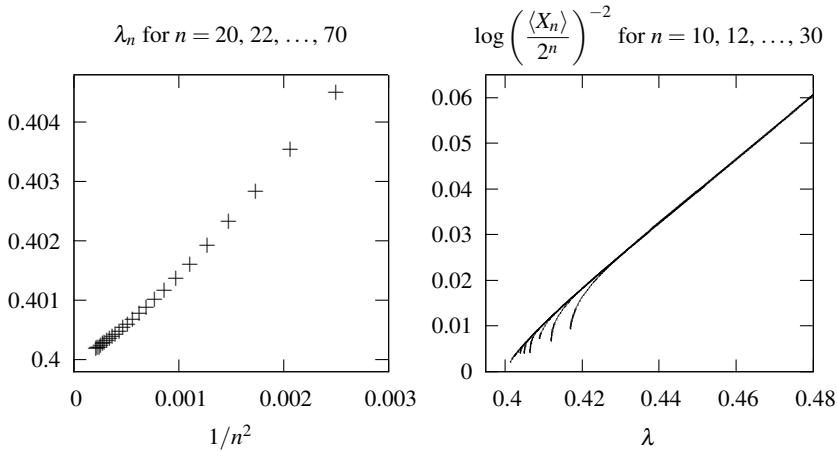}
\caption{These two plots correspond to the case $a=2$, for which disorder is relevant. The initial distribution is a double peak $P_0(X)=(1-\lambda) \delta (X+1/2)+\lambda \delta (X-1/2)$.  As the recursion spreads the weights on the real numbers,  we used a linear interpolation on a grid to make the iterations.}
\label{desordrerel}
\end{figure}

When disorder is relevant, the quenched critical point $\lambda _c$ is not
known. One can  nevertheless calculate a sequence of upper bounds $\lambda _n$
very much like in  the toy model (\ref{bound}):
from  a convexity argument in (\ref{recX}) and the fact that the range $X_n \ge \log(b-1)$ is stable one can show that
$$    2 ( \langle X_n \rangle - \log b)  \le  \langle X_{n+1}  \rangle - \log b   \le 2  \langle X_n \rangle - \log (b) -\log(b-1) \ .
$$
Therefore   the values  $\lambda_n$ such that $\left\langle X_n \right\rangle _{\lambda _n} = \log b $ are upper bounds for $\lambda_c$. In figure \ref{desordrerel}, we plot $\lambda_n$ versus $1/n^2$ for the particular case $a=2$. Our data are consistent with a $1/n^2$ convergence very similar to what we saw in figure \ref{rv2} for the toy model.

\bigskip

 One can also test the infinite order behavior (\ref{infinite-order})
by plotting  in figure \ref{desordrerel}  the function $\log (\left\langle X_n \right\rangle/2^n) ^{-2}$ versus $\lambda$.  Very much like in figure \ref{rv}, we observe  a linear  crossing of the x-axis at the transition point.

\bigskip

Figures \ref{desordrerel} and figures \ref{rv2} and \ref{rv} show very similar behaviors and therefore give a good evidence that there is an infinite order transition in the hierarchical lattice when disorder is relevant. This confirms the prediction of Tang and Chat\'e \cite{TC}.

\subsection{A case where disorder is irrelevant}
We now give numerical evidence that disorder is irrelevant for $a = .7$
as expected from the Harris criterion.
If disorder is irrelevant, one expects that the critical point is given by $\langle e^{X_0 } \rangle=1$ as in the annealed problem.
For the example of  initial distribution $P_0(X)=(1-\lambda) \delta (X+1/2)+\lambda \delta (X-1/2)$, this gives : $\lambda _c = (1-e^{-1/2})/(e^{1/2}-e^{-1/2}) \simeq 0.37754$.  
One also expects the same  power law singularity (\ref{hie5}) as for the pure case.

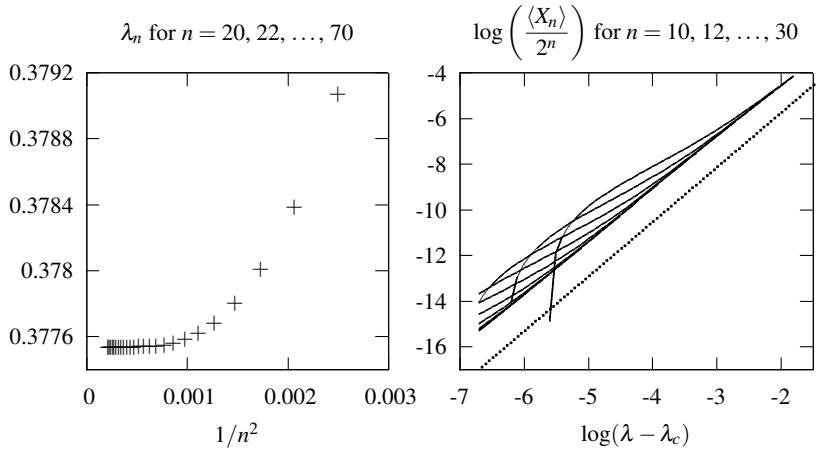
\begin{figure}[h]
\centering
\input{./desordrenonrel.tex}
\caption{These two plots correspond to the case  $a=0.7$, for which a small amount of disorder is expected to be irrelevant. The initial distribution is still a double peak $P_0(X)=(1-\lambda) \delta (X+1/2)+\lambda \delta (X-1/2)$. 
The bounds $\lambda_n$ do not converge like $1/n^2$.
The dotted line is a linear function with the  slope  $\log (2) / \log (2/b)$. This shows that the quenched and the annealed free energies have the same critical exponent.}
\label{desordrenonrel}
\end{figure}

Figure \ref{desordrenonrel}  shows that,   for $a=.7$  and for the distribution $P_0(X)=(1-\lambda) \delta (X+1/2)+\lambda \delta (X-1/2)$, the bounds $\lambda_n$ do not converge anymore like $1/n^2$. A log-log plot of $F_\infty$ versus $\log(\lambda-\lambda_c)$ exhibits  a power law singularity  with the same exponent as the pure model.

\bigskip

\bigskip

\section{Conclusion}
In this work we have studied a toy model (\ref{toy}) inspired by the problem of the depinning transition in presence  of disorder.
Both our numerical results and our analytical approaches support the existence of an infinite order transition  (\ref{infinite-order}) in agreement with what had been predicted by Tang and Chat\'e \cite{TC}. 
Our numerical results indicate also a very similar behavior for the depinning transition  on the hierarchical lattice.

 For our toy model, our analytical approach was limited to a simple class of initial distributions (\ref{initial-dist},\ref{rwtau},\ref{correction}). An interesting question would be to see whether our results applies to a broader class of initial distributions (as predicted by our numerical study of section 2). In particular it would be nice to be able to calculate the critical manifold  of (\ref{renor1}) for arbitrary distributions $r(x,\tau)$ and to develop a perturbation method able  to extend our results on the critical manifold to higher orders in $u$.

Determining the precise location of the critical manifold in the hierarchical model (for strong disorder) is also a possible interesting extension of the present work. Of course, more work is needed to confirm the infinite order transition for the hierarchical model and to see whether it appears in more realistic models of depinning with strong disorder \cite{CH,CY,GMO,KMP,Lubensky-Nelson}.

It could also be interesting to see whether the simple renormalization (\ref{toy}) studied here could be used to investigate other systems with strong disorder \cite{Igloi,Monthus}.

\section{Appendix A: the large $a$ limit of the toy model  (\ref{toy})}
In this appendix we try to explain the $1/a$   dependence of  $\lambda_c$ that we observed  for large $a$ in table  \ref{tab}. 

Consider a distribution $P_0(X)$ characterized by its first two moments
$$\int P_0(X) X d X = \mu  \ \ \ \ \ \ \ ; \ \ \ \ \ \ \
\int P_0(X) X^2 d X = \mu^2 + \sigma  \ . $$
The parameters $\mu$ and $\sigma$ change if one varies the initial distribution $P_0(X)$.
 Let us call $\mu_c(a, \sigma)$ the critical value of $\mu $ at the depinning transition (\ref{2-phases}).

By   the simple change of scale  $ a\to a', \mu \to a' \mu / a, \sigma \to a'^2 \sigma / a^2$ it is clear that
$\mu_c(a,\sigma)$ should  be of the form
$$\mu_c(a, \sigma) =  a \  F\left({\sigma \over a^2} \right)  \ . $$
On the other hand, if $a$ is very large compared to $\mu$ and $\sqrt{\sigma}$, for many iterations of the renormalization,  the distribution is renormalized as if $a$ was infinite: after $n$ steps, one has $\mu_n \simeq 2^n \mu$ and
$\sigma_n \simeq 2^n \sigma$.  Therefore for $\mu \ll a$ and $\sigma \ll a^2$ one has
$$\mu_c(a,  \sigma) \simeq \sigma G(a) \ .$$
Combining these two equations leads to the result that
$$\mu_c(a,\sigma) \sim {\sigma \over a} \ .  $$
For the particular case of the binary distribution (\ref{initial-distribution})
one has $\mu = 2 \lambda-1$ and therefore $\lambda_c -1/2 \sim 1/a$ as observed in table \ref{tab}.
\section{Appendix B: absence of critical fixed distributions of the map (\ref{map})}
 In this appendix, we show that the only fixed distributions accessible by the renormalization group (\ref{renorm}) are  $H(z)=0$ (which corresponds to the pinned phase), $H(z)=1$ (which corresponds to the unpinned phase) and $H(z)=z^a$ (which corresponds to the critical point of the pure system).

We have seen that in terms of the generating function $H_n(z)$, the renormalization transformation (\ref{toy}) can be written as
\begin{equation}H_{n+1}(z) = {H_n^2(z) - Q_n(z) \over z^a} + Q_n(1)   \label{map2}
\end{equation}
 where $Q_n(z)$ is the polynomial of degree $a-1$  obtained by keeping the first $a$ coefficients of the expansion of $H_n^2(z)$ around $z=0$ :
\begin{equation}
 \text{If } \ \ \ \ \ H_n^2(z) = \sum_{k=0}^\infty \beta^{(k)}  \ z^k \ \ \ \ 
 \ \ \ \ \text{then } \ \ \ \ Q_n(z) = \sum_{k=0}^{a-1}  \beta^{(k)}   \ z^k \ .
\label{ren-H}
\end{equation}

It is clear that if the $p_n^{(k)}$'s in (\ref{renorm},\ref{generating}) are non-negative  then the $\beta^{(k)}$'s are also non-negative.
If one looks for a fixed distribution (i.e. for a  fixed point $H_*(z)$ of the map (\ref{map})) 
 one gets
\begin{equation}
H_*(z) = {z^a \pm \sqrt{\Delta(z)} \over 2} 
\label{fixed-point}
\end{equation}
with 
\begin{equation}
\label{Delta-def}
\Delta(z)= z^{2 a} + 4 Q_n(z) - 4 z^a Q_n(1) \ . 
\end{equation}

Let us first observe that
 $\Delta(z)$ cannot vanish on the positive real axis:
\begin{itemize}
\item \underline{for $z \le 1$}, 
by writing 
$$\Delta(z) = z^{2 a} + 4  \sum_{k=0}^{a-1} \beta^{(k)} (z^k - z^a)$$
it is obvious that $\Delta(z) >0$ .

\item \underline{for $ z \ge 1$}, using the fact that $$\sum_{k=-a}^\infty p_n^{(k]}=1= \sum_{k=0}^\infty \beta^{(k)}$$
one can rewrite $\Delta(z)$ as
$$\Delta(z)=  z^{2 a} \sum_{k=a}^\infty \beta^{(k)} +  \sum_{k=0}^{a-1} \beta^{(k)}(z^a-2)^2 + 4 \sum_{k=0}^{a-1}  \beta^{(k)}( z^k - 1)$$
\end{itemize}
and in this case too, $\Delta(z)$ is strictly positive on the whole positive 
axis.

If $\sqrt{\Delta(z)}$ is not a polynomial, then, because $\Delta(z)$ does not vanish on the positive real axis, the closest singularity of $H(z)$ is not on the positive real axis, and so the coefficients of $H(z)$ cannot be all positive. It is  known (Pringsheim's theorem \cite{Flajolet}) that a series with positive coefficients has a singularity at the intersection of the positive real axis and its  circle of convergence of the series.
One then gets from (\ref{Delta-def})
$\Delta(z) = z^{2 a} $ or $\Delta(z) = (z^{ a}-2)^2$ which gives $H_*(z)=0$, $z^a$ or $1$.

If $\sqrt{\Delta(z)} $ is  a polynomial, then from (\ref{Delta-def}) and the fact that $Q_n(z)$ is of degree $a-1$, the only possibilities are $Q_n(z)=Q_n(1)^2$ which implies
either $Q_n(z)=0$ or $Q_n(z)=1$.
\section{Appendix C: the extra critical fixed point of the  truncated transformation
(\ref{cut-of}) } 
In this appendix we show that if one truncates the transformation  as in (\ref{toy}) by (\ref{cut-of}), there appears a new critical fixed distribution. This new fixed point  disappears in the  $a' \to \infty$ limit.

For the sake of simplicity, let us limit the discussion to the case $a=1$ and to a large  integer value of $a'$.
For integer  $a $ and $a'$ the transformation  (\ref{renorm}) remains the same for $k < a'$. The only change is for $k=a'$ where  it becomes

\begin{equation}
 p_{n+1}^{(a')} = \sum_{q_1=0}^{a'}  \ \   \sum_{q_2=a'-q_1}^{a'} \   \ p_n^{(q_1)} \ p_n^{(q_2)} \ .
\label{cut1}
\end{equation}
Because the transformation   for $ k < a'$ does not depend on $a'$,  the recursion relation of  the  generating function $H_n(z)$ 
\begin{equation}
H_n(z) =  \sum_{k=-a}^{a'} p_n^{(k)}  \ z^{k+a} 
\label{generating1}
\end{equation}
is the same as (\ref{map1})  	up to terms of order $z^{a+a'-1}$
\begin{equation}
H_{n+1}(z)=  {H_n(z)^2 - H_n(0)^2 \over z}+ H_n(0)^2 + O\left( z^{a+a'-1} \right) \ .
\label{map2}
\end{equation}
Therefore a fixed point of (\ref{map2})  should be of the form
$$H_*(z) = {z \over 2}  +   H(0)\sqrt{1 - z + {z^2 \over 4 H(0)^2}} + O\left( z^{a+a'} \right) \ .$$
For  $k < a+a'-1$ (here $a=1$) one gets for the weights of the fixed distribution
$$p_*^{(k)} = {1 \over 2 \pi i} \oint {dz \over z^{k+2}} \left[{z \over 2} + H(0) \sqrt{1 - z + {z^2 \over 4 H(0)^2}}\right]$$
where the integration contour is a small circle around the origin.

For $H(0)$ close to $1$   this gives 

$$p_*^{(k)} \simeq {1 \over 2^{k-1}} {1- H(0)  \over \pi} \int_0^1 dy \sqrt{1 - y^2} \ \cos \left(k \sqrt{2 (1 - H(0))} y \right) .$$
If $q$ is the first zero of
$ \int_0^1 dy \sqrt{1 - y^2} \ \cos ( q  y )$
all the $p_*^{(k)}$ are positive as long as $k < q/\sqrt{2 (1- H(0)}$.
Adjusting the boundary condition at $k=a'$ selects one particular value $H(0)$ which should  be such that
$$1- H(0) \simeq {q^2 \over 2 a'^2} \ . $$
In the limit $a' \to \infty$, obviously  $H(0) \to 1$ and this fixed merges with the fixed point $H_*(z)=1$.

\begin{acknowledgements}
{We would like to thank J.P. Eckmann, G. Giacomin, L.H. Tang and F. Werner for useful and stimulating discussions.}
\end{acknowledgements}

\end{document}

%% file: rc.tex
\setlength{\unitlength}{0.240900pt}
\ifx\plotpoint\undefined\newsavebox{\plotpoint}\fi
\sbox{\plotpoint}{\rule[-0.200pt]{0.400pt}{0.400pt}}%
\begin{picture}(960,720)(0,0)
\sbox{\plotpoint}{\rule[-0.200pt]{0.400pt}{0.400pt}}%
\put(131.0,131.0){\rule[-0.200pt]{4.818pt}{0.400pt}}
\put(111,131){\makebox(0,0)[r]{-6}}
\put(879.0,131.0){\rule[-0.200pt]{4.818pt}{0.400pt}}
\put(131.0,241.0){\rule[-0.200pt]{4.818pt}{0.400pt}}
\put(111,241){\makebox(0,0)[r]{-4}}
\put(879.0,241.0){\rule[-0.200pt]{4.818pt}{0.400pt}}
\put(131.0,350.0){\rule[-0.200pt]{4.818pt}{0.400pt}}
\put(111,350){\makebox(0,0)[r]{-2}}
\put(879.0,350.0){\rule[-0.200pt]{4.818pt}{0.400pt}}
\put(131.0,460.0){\rule[-0.200pt]{4.818pt}{0.400pt}}
\put(111,460){\makebox(0,0)[r]{ 0}}
\put(879.0,460.0){\rule[-0.200pt]{4.818pt}{0.400pt}}
\put(131.0,569.0){\rule[-0.200pt]{4.818pt}{0.400pt}}
\put(111,569){\makebox(0,0)[r]{ 2}}
\put(879.0,569.0){\rule[-0.200pt]{4.818pt}{0.400pt}}
\put(131.0,679.0){\rule[-0.200pt]{4.818pt}{0.400pt}}
\put(111,679){\makebox(0,0)[r]{ 4}}
\put(879.0,679.0){\rule[-0.200pt]{4.818pt}{0.400pt}}
\put(195.0,131.0){\rule[-0.200pt]{0.400pt}{4.818pt}}
\put(195,90){\makebox(0,0){-8}}
\put(195.0,659.0){\rule[-0.200pt]{0.400pt}{4.818pt}}
\put(323.0,131.0){\rule[-0.200pt]{0.400pt}{4.818pt}}
\put(323,90){\makebox(0,0){-6}}
\put(323.0,659.0){\rule[-0.200pt]{0.400pt}{4.818pt}}
\put(451.0,131.0){\rule[-0.200pt]{0.400pt}{4.818pt}}
\put(451,90){\makebox(0,0){-4}}
\put(451.0,659.0){\rule[-0.200pt]{0.400pt}{4.818pt}}
\put(579.0,131.0){\rule[-0.200pt]{0.400pt}{4.818pt}}
\put(579,90){\makebox(0,0){-2}}
\put(579.0,659.0){\rule[-0.200pt]{0.400pt}{4.818pt}}
\put(707.0,131.0){\rule[-0.200pt]{0.400pt}{4.818pt}}
\put(707,90){\makebox(0,0){ 0}}
\put(707.0,659.0){\rule[-0.200pt]{0.400pt}{4.818pt}}
\put(835.0,131.0){\rule[-0.200pt]{0.400pt}{4.818pt}}
\put(835,90){\makebox(0,0){ 2}}
\put(835.0,659.0){\rule[-0.200pt]{0.400pt}{4.818pt}}
\put(131.0,131.0){\rule[-0.200pt]{0.400pt}{132.013pt}}
\put(131.0,131.0){\rule[-0.200pt]{185.011pt}{0.400pt}}
\put(899.0,131.0){\rule[-0.200pt]{0.400pt}{132.013pt}}
\put(131.0,679.0){\rule[-0.200pt]{185.011pt}{0.400pt}}
\put(30,405){\makebox(0,0){$ f(X_{n-1}^{(1)}+X_{n-1}^{(2)} )$  ~ ~}}
\put(515,29){\makebox(0,0){$ X_{n-1}^{(1)}+X_{n-1}^{(2)} $}}
\put(185,207.67){\rule{0.723pt}{0.400pt}}
\multiput(185.00,207.17)(1.500,1.000){2}{\rule{0.361pt}{0.400pt}}
\put(131.0,208.0){\rule[-0.200pt]{13.009pt}{0.400pt}}
\put(218,208.67){\rule{0.723pt}{0.400pt}}
\multiput(218.00,208.17)(1.500,1.000){2}{\rule{0.361pt}{0.400pt}}
\put(188.0,209.0){\rule[-0.200pt]{7.227pt}{0.400pt}}
\put(238,209.67){\rule{0.723pt}{0.400pt}}
\multiput(238.00,209.17)(1.500,1.000){2}{\rule{0.361pt}{0.400pt}}
\put(221.0,210.0){\rule[-0.200pt]{4.095pt}{0.400pt}}
\put(256,210.67){\rule{0.723pt}{0.400pt}}
\multiput(256.00,210.17)(1.500,1.000){2}{\rule{0.361pt}{0.400pt}}
\put(241.0,211.0){\rule[-0.200pt]{3.613pt}{0.400pt}}
\put(268,211.67){\rule{0.723pt}{0.400pt}}
\multiput(268.00,211.17)(1.500,1.000){2}{\rule{0.361pt}{0.400pt}}
\put(259.0,212.0){\rule[-0.200pt]{2.168pt}{0.400pt}}
\put(280,212.67){\rule{0.723pt}{0.400pt}}
\multiput(280.00,212.17)(1.500,1.000){2}{\rule{0.361pt}{0.400pt}}
\put(271.0,213.0){\rule[-0.200pt]{2.168pt}{0.400pt}}
\put(288,213.67){\rule{0.723pt}{0.400pt}}
\multiput(288.00,213.17)(1.500,1.000){2}{\rule{0.361pt}{0.400pt}}
\put(283.0,214.0){\rule[-0.200pt]{1.204pt}{0.400pt}}
\put(297,214.67){\rule{0.723pt}{0.400pt}}
\multiput(297.00,214.17)(1.500,1.000){2}{\rule{0.361pt}{0.400pt}}
\put(291.0,215.0){\rule[-0.200pt]{1.445pt}{0.400pt}}
\put(306,215.67){\rule{0.723pt}{0.400pt}}
\multiput(306.00,215.17)(1.500,1.000){2}{\rule{0.361pt}{0.400pt}}
\put(300.0,216.0){\rule[-0.200pt]{1.445pt}{0.400pt}}
\put(312,216.67){\rule{0.723pt}{0.400pt}}
\multiput(312.00,216.17)(1.500,1.000){2}{\rule{0.361pt}{0.400pt}}
\put(309.0,217.0){\rule[-0.200pt]{0.723pt}{0.400pt}}
\put(318,217.67){\rule{0.723pt}{0.400pt}}
\multiput(318.00,217.17)(1.500,1.000){2}{\rule{0.361pt}{0.400pt}}
\put(315.0,218.0){\rule[-0.200pt]{0.723pt}{0.400pt}}
\put(324,218.67){\rule{0.723pt}{0.400pt}}
\multiput(324.00,218.17)(1.500,1.000){2}{\rule{0.361pt}{0.400pt}}
\put(321.0,219.0){\rule[-0.200pt]{0.723pt}{0.400pt}}
\put(330,219.67){\rule{0.723pt}{0.400pt}}
\multiput(330.00,219.17)(1.500,1.000){2}{\rule{0.361pt}{0.400pt}}
\put(327.0,220.0){\rule[-0.200pt]{0.723pt}{0.400pt}}
\put(336,220.67){\rule{0.723pt}{0.400pt}}
\multiput(336.00,220.17)(1.500,1.000){2}{\rule{0.361pt}{0.400pt}}
\put(339,221.67){\rule{0.723pt}{0.400pt}}
\multiput(339.00,221.17)(1.500,1.000){2}{\rule{0.361pt}{0.400pt}}
\put(333.0,221.0){\rule[-0.200pt]{0.723pt}{0.400pt}}
\put(344,222.67){\rule{0.723pt}{0.400pt}}
\multiput(344.00,222.17)(1.500,1.000){2}{\rule{0.361pt}{0.400pt}}
\put(347,223.67){\rule{0.723pt}{0.400pt}}
\multiput(347.00,223.17)(1.500,1.000){2}{\rule{0.361pt}{0.400pt}}
\put(342.0,223.0){\rule[-0.200pt]{0.482pt}{0.400pt}}
\put(353,224.67){\rule{0.723pt}{0.400pt}}
\multiput(353.00,224.17)(1.500,1.000){2}{\rule{0.361pt}{0.400pt}}
\put(356,225.67){\rule{0.723pt}{0.400pt}}
\multiput(356.00,225.17)(1.500,1.000){2}{\rule{0.361pt}{0.400pt}}
\put(359,226.67){\rule{0.723pt}{0.400pt}}
\multiput(359.00,226.17)(1.500,1.000){2}{\rule{0.361pt}{0.400pt}}
\put(350.0,225.0){\rule[-0.200pt]{0.723pt}{0.400pt}}
\put(365,227.67){\rule{0.723pt}{0.400pt}}
\multiput(365.00,227.17)(1.500,1.000){2}{\rule{0.361pt}{0.400pt}}
\put(368,228.67){\rule{0.723pt}{0.400pt}}
\multiput(368.00,228.17)(1.500,1.000){2}{\rule{0.361pt}{0.400pt}}
\put(371,229.67){\rule{0.723pt}{0.400pt}}
\multiput(371.00,229.17)(1.500,1.000){2}{\rule{0.361pt}{0.400pt}}
\put(374,230.67){\rule{0.723pt}{0.400pt}}
\multiput(374.00,230.17)(1.500,1.000){2}{\rule{0.361pt}{0.400pt}}
\put(377,231.67){\rule{0.723pt}{0.400pt}}
\multiput(377.00,231.17)(1.500,1.000){2}{\rule{0.361pt}{0.400pt}}
\put(380,232.67){\rule{0.723pt}{0.400pt}}
\multiput(380.00,232.17)(1.500,1.000){2}{\rule{0.361pt}{0.400pt}}
\put(383,233.67){\rule{0.723pt}{0.400pt}}
\multiput(383.00,233.17)(1.500,1.000){2}{\rule{0.361pt}{0.400pt}}
\put(386,234.67){\rule{0.723pt}{0.400pt}}
\multiput(386.00,234.17)(1.500,1.000){2}{\rule{0.361pt}{0.400pt}}
\put(389,235.67){\rule{0.723pt}{0.400pt}}
\multiput(389.00,235.17)(1.500,1.000){2}{\rule{0.361pt}{0.400pt}}
\put(392,236.67){\rule{0.723pt}{0.400pt}}
\multiput(392.00,236.17)(1.500,1.000){2}{\rule{0.361pt}{0.400pt}}
\put(395,237.67){\rule{0.723pt}{0.400pt}}
\multiput(395.00,237.17)(1.500,1.000){2}{\rule{0.361pt}{0.400pt}}
\put(398,238.67){\rule{0.482pt}{0.400pt}}
\multiput(398.00,238.17)(1.000,1.000){2}{\rule{0.241pt}{0.400pt}}
\put(400,239.67){\rule{0.723pt}{0.400pt}}
\multiput(400.00,239.17)(1.500,1.000){2}{\rule{0.361pt}{0.400pt}}
\put(403,240.67){\rule{0.723pt}{0.400pt}}
\multiput(403.00,240.17)(1.500,1.000){2}{\rule{0.361pt}{0.400pt}}
\put(406,242.17){\rule{0.700pt}{0.400pt}}
\multiput(406.00,241.17)(1.547,2.000){2}{\rule{0.350pt}{0.400pt}}
\put(409,243.67){\rule{0.723pt}{0.400pt}}
\multiput(409.00,243.17)(1.500,1.000){2}{\rule{0.361pt}{0.400pt}}
\put(412,244.67){\rule{0.723pt}{0.400pt}}
\multiput(412.00,244.17)(1.500,1.000){2}{\rule{0.361pt}{0.400pt}}
\put(415,245.67){\rule{0.723pt}{0.400pt}}
\multiput(415.00,245.17)(1.500,1.000){2}{\rule{0.361pt}{0.400pt}}
\put(418,247.17){\rule{0.700pt}{0.400pt}}
\multiput(418.00,246.17)(1.547,2.000){2}{\rule{0.350pt}{0.400pt}}
\put(421,248.67){\rule{0.723pt}{0.400pt}}
\multiput(421.00,248.17)(1.500,1.000){2}{\rule{0.361pt}{0.400pt}}
\put(424,250.17){\rule{0.700pt}{0.400pt}}
\multiput(424.00,249.17)(1.547,2.000){2}{\rule{0.350pt}{0.400pt}}
\put(427,251.67){\rule{0.723pt}{0.400pt}}
\multiput(427.00,251.17)(1.500,1.000){2}{\rule{0.361pt}{0.400pt}}
\put(430,252.67){\rule{0.723pt}{0.400pt}}
\multiput(430.00,252.17)(1.500,1.000){2}{\rule{0.361pt}{0.400pt}}
\put(433,254.17){\rule{0.700pt}{0.400pt}}
\multiput(433.00,253.17)(1.547,2.000){2}{\rule{0.350pt}{0.400pt}}
\put(436,255.67){\rule{0.723pt}{0.400pt}}
\multiput(436.00,255.17)(1.500,1.000){2}{\rule{0.361pt}{0.400pt}}
\put(439,257.17){\rule{0.700pt}{0.400pt}}
\multiput(439.00,256.17)(1.547,2.000){2}{\rule{0.350pt}{0.400pt}}
\put(442,258.67){\rule{0.723pt}{0.400pt}}
\multiput(442.00,258.17)(1.500,1.000){2}{\rule{0.361pt}{0.400pt}}
\put(445,260.17){\rule{0.700pt}{0.400pt}}
\multiput(445.00,259.17)(1.547,2.000){2}{\rule{0.350pt}{0.400pt}}
\put(448,262.17){\rule{0.700pt}{0.400pt}}
\multiput(448.00,261.17)(1.547,2.000){2}{\rule{0.350pt}{0.400pt}}
\put(451,263.67){\rule{0.723pt}{0.400pt}}
\multiput(451.00,263.17)(1.500,1.000){2}{\rule{0.361pt}{0.400pt}}
\put(454,265.17){\rule{0.482pt}{0.400pt}}
\multiput(454.00,264.17)(1.000,2.000){2}{\rule{0.241pt}{0.400pt}}
\put(456,267.17){\rule{0.700pt}{0.400pt}}
\multiput(456.00,266.17)(1.547,2.000){2}{\rule{0.350pt}{0.400pt}}
\put(459,268.67){\rule{0.723pt}{0.400pt}}
\multiput(459.00,268.17)(1.500,1.000){2}{\rule{0.361pt}{0.400pt}}
\put(462,270.17){\rule{0.700pt}{0.400pt}}
\multiput(462.00,269.17)(1.547,2.000){2}{\rule{0.350pt}{0.400pt}}
\put(465,272.17){\rule{0.700pt}{0.400pt}}
\multiput(465.00,271.17)(1.547,2.000){2}{\rule{0.350pt}{0.400pt}}
\put(468,274.17){\rule{0.700pt}{0.400pt}}
\multiput(468.00,273.17)(1.547,2.000){2}{\rule{0.350pt}{0.400pt}}
\put(471,276.17){\rule{0.700pt}{0.400pt}}
\multiput(471.00,275.17)(1.547,2.000){2}{\rule{0.350pt}{0.400pt}}
\put(474,277.67){\rule{0.723pt}{0.400pt}}
\multiput(474.00,277.17)(1.500,1.000){2}{\rule{0.361pt}{0.400pt}}
\put(477,279.17){\rule{0.700pt}{0.400pt}}
\multiput(477.00,278.17)(1.547,2.000){2}{\rule{0.350pt}{0.400pt}}
\put(480,281.17){\rule{0.700pt}{0.400pt}}
\multiput(480.00,280.17)(1.547,2.000){2}{\rule{0.350pt}{0.400pt}}
\put(483,283.17){\rule{0.700pt}{0.400pt}}
\multiput(483.00,282.17)(1.547,2.000){2}{\rule{0.350pt}{0.400pt}}
\put(486,285.17){\rule{0.700pt}{0.400pt}}
\multiput(486.00,284.17)(1.547,2.000){2}{\rule{0.350pt}{0.400pt}}
\put(489,287.17){\rule{0.700pt}{0.400pt}}
\multiput(489.00,286.17)(1.547,2.000){2}{\rule{0.350pt}{0.400pt}}
\put(492,289.17){\rule{0.700pt}{0.400pt}}
\multiput(492.00,288.17)(1.547,2.000){2}{\rule{0.350pt}{0.400pt}}
\put(495,291.17){\rule{0.700pt}{0.400pt}}
\multiput(495.00,290.17)(1.547,2.000){2}{\rule{0.350pt}{0.400pt}}
\put(498,293.17){\rule{0.700pt}{0.400pt}}
\multiput(498.00,292.17)(1.547,2.000){2}{\rule{0.350pt}{0.400pt}}
\put(501,295.17){\rule{0.700pt}{0.400pt}}
\multiput(501.00,294.17)(1.547,2.000){2}{\rule{0.350pt}{0.400pt}}
\put(504,297.17){\rule{0.700pt}{0.400pt}}
\multiput(504.00,296.17)(1.547,2.000){2}{\rule{0.350pt}{0.400pt}}
\put(507,299.17){\rule{0.700pt}{0.400pt}}
\multiput(507.00,298.17)(1.547,2.000){2}{\rule{0.350pt}{0.400pt}}
\put(510,301.17){\rule{0.482pt}{0.400pt}}
\multiput(510.00,300.17)(1.000,2.000){2}{\rule{0.241pt}{0.400pt}}
\put(512,303.17){\rule{0.700pt}{0.400pt}}
\multiput(512.00,302.17)(1.547,2.000){2}{\rule{0.350pt}{0.400pt}}
\put(515,305.17){\rule{0.700pt}{0.400pt}}
\multiput(515.00,304.17)(1.547,2.000){2}{\rule{0.350pt}{0.400pt}}
\put(518,307.17){\rule{0.700pt}{0.400pt}}
\multiput(518.00,306.17)(1.547,2.000){2}{\rule{0.350pt}{0.400pt}}
\multiput(521.00,309.61)(0.462,0.447){3}{\rule{0.500pt}{0.108pt}}
\multiput(521.00,308.17)(1.962,3.000){2}{\rule{0.250pt}{0.400pt}}
\put(524,312.17){\rule{0.700pt}{0.400pt}}
\multiput(524.00,311.17)(1.547,2.000){2}{\rule{0.350pt}{0.400pt}}
\put(527,314.17){\rule{0.700pt}{0.400pt}}
\multiput(527.00,313.17)(1.547,2.000){2}{\rule{0.350pt}{0.400pt}}
\put(530,316.17){\rule{0.700pt}{0.400pt}}
\multiput(530.00,315.17)(1.547,2.000){2}{\rule{0.350pt}{0.400pt}}
\put(533,318.17){\rule{0.700pt}{0.400pt}}
\multiput(533.00,317.17)(1.547,2.000){2}{\rule{0.350pt}{0.400pt}}
\put(536,320.17){\rule{0.700pt}{0.400pt}}
\multiput(536.00,319.17)(1.547,2.000){2}{\rule{0.350pt}{0.400pt}}
\multiput(539.00,322.61)(0.462,0.447){3}{\rule{0.500pt}{0.108pt}}
\multiput(539.00,321.17)(1.962,3.000){2}{\rule{0.250pt}{0.400pt}}
\put(542,325.17){\rule{0.700pt}{0.400pt}}
\multiput(542.00,324.17)(1.547,2.000){2}{\rule{0.350pt}{0.400pt}}
\put(545,327.17){\rule{0.700pt}{0.400pt}}
\multiput(545.00,326.17)(1.547,2.000){2}{\rule{0.350pt}{0.400pt}}
\put(548,329.17){\rule{0.700pt}{0.400pt}}
\multiput(548.00,328.17)(1.547,2.000){2}{\rule{0.350pt}{0.400pt}}
\multiput(551.00,331.61)(0.462,0.447){3}{\rule{0.500pt}{0.108pt}}
\multiput(551.00,330.17)(1.962,3.000){2}{\rule{0.250pt}{0.400pt}}
\put(554,334.17){\rule{0.700pt}{0.400pt}}
\multiput(554.00,333.17)(1.547,2.000){2}{\rule{0.350pt}{0.400pt}}
\put(557,336.17){\rule{0.700pt}{0.400pt}}
\multiput(557.00,335.17)(1.547,2.000){2}{\rule{0.350pt}{0.400pt}}
\multiput(560.00,338.61)(0.462,0.447){3}{\rule{0.500pt}{0.108pt}}
\multiput(560.00,337.17)(1.962,3.000){2}{\rule{0.250pt}{0.400pt}}
\put(563,341.17){\rule{0.700pt}{0.400pt}}
\multiput(563.00,340.17)(1.547,2.000){2}{\rule{0.350pt}{0.400pt}}
\put(566,343.17){\rule{0.482pt}{0.400pt}}
\multiput(566.00,342.17)(1.000,2.000){2}{\rule{0.241pt}{0.400pt}}
\multiput(568.00,345.61)(0.462,0.447){3}{\rule{0.500pt}{0.108pt}}
\multiput(568.00,344.17)(1.962,3.000){2}{\rule{0.250pt}{0.400pt}}
\put(571,348.17){\rule{0.700pt}{0.400pt}}
\multiput(571.00,347.17)(1.547,2.000){2}{\rule{0.350pt}{0.400pt}}
\put(574,350.17){\rule{0.700pt}{0.400pt}}
\multiput(574.00,349.17)(1.547,2.000){2}{\rule{0.350pt}{0.400pt}}
\multiput(577.00,352.61)(0.462,0.447){3}{\rule{0.500pt}{0.108pt}}
\multiput(577.00,351.17)(1.962,3.000){2}{\rule{0.250pt}{0.400pt}}
\put(580,355.17){\rule{0.700pt}{0.400pt}}
\multiput(580.00,354.17)(1.547,2.000){2}{\rule{0.350pt}{0.400pt}}
\put(583,357.17){\rule{0.700pt}{0.400pt}}
\multiput(583.00,356.17)(1.547,2.000){2}{\rule{0.350pt}{0.400pt}}
\multiput(586.00,359.61)(0.462,0.447){3}{\rule{0.500pt}{0.108pt}}
\multiput(586.00,358.17)(1.962,3.000){2}{\rule{0.250pt}{0.400pt}}
\put(589,362.17){\rule{0.700pt}{0.400pt}}
\multiput(589.00,361.17)(1.547,2.000){2}{\rule{0.350pt}{0.400pt}}
\put(592,364.17){\rule{0.700pt}{0.400pt}}
\multiput(592.00,363.17)(1.547,2.000){2}{\rule{0.350pt}{0.400pt}}
\multiput(595.00,366.61)(0.462,0.447){3}{\rule{0.500pt}{0.108pt}}
\multiput(595.00,365.17)(1.962,3.000){2}{\rule{0.250pt}{0.400pt}}
\put(598,369.17){\rule{0.700pt}{0.400pt}}
\multiput(598.00,368.17)(1.547,2.000){2}{\rule{0.350pt}{0.400pt}}
\multiput(601.00,371.61)(0.462,0.447){3}{\rule{0.500pt}{0.108pt}}
\multiput(601.00,370.17)(1.962,3.000){2}{\rule{0.250pt}{0.400pt}}
\put(604,374.17){\rule{0.700pt}{0.400pt}}
\multiput(604.00,373.17)(1.547,2.000){2}{\rule{0.350pt}{0.400pt}}
\put(607,376.17){\rule{0.700pt}{0.400pt}}
\multiput(607.00,375.17)(1.547,2.000){2}{\rule{0.350pt}{0.400pt}}
\multiput(610.00,378.61)(0.462,0.447){3}{\rule{0.500pt}{0.108pt}}
\multiput(610.00,377.17)(1.962,3.000){2}{\rule{0.250pt}{0.400pt}}
\put(613,381.17){\rule{0.700pt}{0.400pt}}
\multiput(613.00,380.17)(1.547,2.000){2}{\rule{0.350pt}{0.400pt}}
\multiput(616.00,383.61)(0.462,0.447){3}{\rule{0.500pt}{0.108pt}}
\multiput(616.00,382.17)(1.962,3.000){2}{\rule{0.250pt}{0.400pt}}
\put(619,386.17){\rule{0.700pt}{0.400pt}}
\multiput(619.00,385.17)(1.547,2.000){2}{\rule{0.350pt}{0.400pt}}
\put(622.17,388){\rule{0.400pt}{0.700pt}}
\multiput(621.17,388.00)(2.000,1.547){2}{\rule{0.400pt}{0.350pt}}
\put(624,391.17){\rule{0.700pt}{0.400pt}}
\multiput(624.00,390.17)(1.547,2.000){2}{\rule{0.350pt}{0.400pt}}
\put(627,393.17){\rule{0.700pt}{0.400pt}}
\multiput(627.00,392.17)(1.547,2.000){2}{\rule{0.350pt}{0.400pt}}
\multiput(630.00,395.61)(0.462,0.447){3}{\rule{0.500pt}{0.108pt}}
\multiput(630.00,394.17)(1.962,3.000){2}{\rule{0.250pt}{0.400pt}}
\put(633,398.17){\rule{0.700pt}{0.400pt}}
\multiput(633.00,397.17)(1.547,2.000){2}{\rule{0.350pt}{0.400pt}}
\multiput(636.00,400.61)(0.462,0.447){3}{\rule{0.500pt}{0.108pt}}
\multiput(636.00,399.17)(1.962,3.000){2}{\rule{0.250pt}{0.400pt}}
\put(639,403.17){\rule{0.700pt}{0.400pt}}
\multiput(639.00,402.17)(1.547,2.000){2}{\rule{0.350pt}{0.400pt}}
\multiput(642.00,405.61)(0.462,0.447){3}{\rule{0.500pt}{0.108pt}}
\multiput(642.00,404.17)(1.962,3.000){2}{\rule{0.250pt}{0.400pt}}
\put(645,408.17){\rule{0.700pt}{0.400pt}}
\multiput(645.00,407.17)(1.547,2.000){2}{\rule{0.350pt}{0.400pt}}
\multiput(648.00,410.61)(0.462,0.447){3}{\rule{0.500pt}{0.108pt}}
\multiput(648.00,409.17)(1.962,3.000){2}{\rule{0.250pt}{0.400pt}}
\put(651,413.17){\rule{0.700pt}{0.400pt}}
\multiput(651.00,412.17)(1.547,2.000){2}{\rule{0.350pt}{0.400pt}}
\multiput(654.00,415.61)(0.462,0.447){3}{\rule{0.500pt}{0.108pt}}
\multiput(654.00,414.17)(1.962,3.000){2}{\rule{0.250pt}{0.400pt}}
\put(657,418.17){\rule{0.700pt}{0.400pt}}
\multiput(657.00,417.17)(1.547,2.000){2}{\rule{0.350pt}{0.400pt}}
\put(660,420.17){\rule{0.700pt}{0.400pt}}
\multiput(660.00,419.17)(1.547,2.000){2}{\rule{0.350pt}{0.400pt}}
\multiput(663.00,422.61)(0.462,0.447){3}{\rule{0.500pt}{0.108pt}}
\multiput(663.00,421.17)(1.962,3.000){2}{\rule{0.250pt}{0.400pt}}
\put(666,425.17){\rule{0.700pt}{0.400pt}}
\multiput(666.00,424.17)(1.547,2.000){2}{\rule{0.350pt}{0.400pt}}
\multiput(669.00,427.61)(0.462,0.447){3}{\rule{0.500pt}{0.108pt}}
\multiput(669.00,426.17)(1.962,3.000){2}{\rule{0.250pt}{0.400pt}}
\put(672,430.17){\rule{0.700pt}{0.400pt}}
\multiput(672.00,429.17)(1.547,2.000){2}{\rule{0.350pt}{0.400pt}}
\multiput(675.00,432.61)(0.462,0.447){3}{\rule{0.500pt}{0.108pt}}
\multiput(675.00,431.17)(1.962,3.000){2}{\rule{0.250pt}{0.400pt}}
\put(678,435.17){\rule{0.482pt}{0.400pt}}
\multiput(678.00,434.17)(1.000,2.000){2}{\rule{0.241pt}{0.400pt}}
\multiput(680.00,437.61)(0.462,0.447){3}{\rule{0.500pt}{0.108pt}}
\multiput(680.00,436.17)(1.962,3.000){2}{\rule{0.250pt}{0.400pt}}
\put(683,440.17){\rule{0.700pt}{0.400pt}}
\multiput(683.00,439.17)(1.547,2.000){2}{\rule{0.350pt}{0.400pt}}
\multiput(686.00,442.61)(0.462,0.447){3}{\rule{0.500pt}{0.108pt}}
\multiput(686.00,441.17)(1.962,3.000){2}{\rule{0.250pt}{0.400pt}}
\put(689,445.17){\rule{0.700pt}{0.400pt}}
\multiput(689.00,444.17)(1.547,2.000){2}{\rule{0.350pt}{0.400pt}}
\multiput(692.00,447.61)(0.462,0.447){3}{\rule{0.500pt}{0.108pt}}
\multiput(692.00,446.17)(1.962,3.000){2}{\rule{0.250pt}{0.400pt}}
\put(695,450.17){\rule{0.700pt}{0.400pt}}
\multiput(695.00,449.17)(1.547,2.000){2}{\rule{0.350pt}{0.400pt}}
\multiput(698.00,452.61)(0.462,0.447){3}{\rule{0.500pt}{0.108pt}}
\multiput(698.00,451.17)(1.962,3.000){2}{\rule{0.250pt}{0.400pt}}
\put(701,455.17){\rule{0.700pt}{0.400pt}}
\multiput(701.00,454.17)(1.547,2.000){2}{\rule{0.350pt}{0.400pt}}
\multiput(704.00,457.61)(0.462,0.447){3}{\rule{0.500pt}{0.108pt}}
\multiput(704.00,456.17)(1.962,3.000){2}{\rule{0.250pt}{0.400pt}}
\put(707,460.17){\rule{0.700pt}{0.400pt}}
\multiput(707.00,459.17)(1.547,2.000){2}{\rule{0.350pt}{0.400pt}}
\multiput(710.00,462.61)(0.462,0.447){3}{\rule{0.500pt}{0.108pt}}
\multiput(710.00,461.17)(1.962,3.000){2}{\rule{0.250pt}{0.400pt}}
\put(713,465.17){\rule{0.700pt}{0.400pt}}
\multiput(713.00,464.17)(1.547,2.000){2}{\rule{0.350pt}{0.400pt}}
\multiput(716.00,467.61)(0.462,0.447){3}{\rule{0.500pt}{0.108pt}}
\multiput(716.00,466.17)(1.962,3.000){2}{\rule{0.250pt}{0.400pt}}
\put(719,470.17){\rule{0.700pt}{0.400pt}}
\multiput(719.00,469.17)(1.547,2.000){2}{\rule{0.350pt}{0.400pt}}
\multiput(722.00,472.61)(0.462,0.447){3}{\rule{0.500pt}{0.108pt}}
\multiput(722.00,471.17)(1.962,3.000){2}{\rule{0.250pt}{0.400pt}}
\put(725,475.17){\rule{0.700pt}{0.400pt}}
\multiput(725.00,474.17)(1.547,2.000){2}{\rule{0.350pt}{0.400pt}}
\multiput(728.00,477.61)(0.462,0.447){3}{\rule{0.500pt}{0.108pt}}
\multiput(728.00,476.17)(1.962,3.000){2}{\rule{0.250pt}{0.400pt}}
\put(731,480.17){\rule{0.700pt}{0.400pt}}
\multiput(731.00,479.17)(1.547,2.000){2}{\rule{0.350pt}{0.400pt}}
\put(734.17,482){\rule{0.400pt}{0.700pt}}
\multiput(733.17,482.00)(2.000,1.547){2}{\rule{0.400pt}{0.350pt}}
\put(736,485.17){\rule{0.700pt}{0.400pt}}
\multiput(736.00,484.17)(1.547,2.000){2}{\rule{0.350pt}{0.400pt}}
\multiput(739.00,487.61)(0.462,0.447){3}{\rule{0.500pt}{0.108pt}}
\multiput(739.00,486.17)(1.962,3.000){2}{\rule{0.250pt}{0.400pt}}
\put(742,490.17){\rule{0.700pt}{0.400pt}}
\multiput(742.00,489.17)(1.547,2.000){2}{\rule{0.350pt}{0.400pt}}
\multiput(745.00,492.61)(0.462,0.447){3}{\rule{0.500pt}{0.108pt}}
\multiput(745.00,491.17)(1.962,3.000){2}{\rule{0.250pt}{0.400pt}}
\put(748,495.17){\rule{0.700pt}{0.400pt}}
\multiput(748.00,494.17)(1.547,2.000){2}{\rule{0.350pt}{0.400pt}}
\multiput(751.00,497.61)(0.462,0.447){3}{\rule{0.500pt}{0.108pt}}
\multiput(751.00,496.17)(1.962,3.000){2}{\rule{0.250pt}{0.400pt}}
\put(754,500.17){\rule{0.700pt}{0.400pt}}
\multiput(754.00,499.17)(1.547,2.000){2}{\rule{0.350pt}{0.400pt}}
\multiput(757.00,502.61)(0.462,0.447){3}{\rule{0.500pt}{0.108pt}}
\multiput(757.00,501.17)(1.962,3.000){2}{\rule{0.250pt}{0.400pt}}
\put(760,505.17){\rule{0.700pt}{0.400pt}}
\multiput(760.00,504.17)(1.547,2.000){2}{\rule{0.350pt}{0.400pt}}
\multiput(763.00,507.61)(0.462,0.447){3}{\rule{0.500pt}{0.108pt}}
\multiput(763.00,506.17)(1.962,3.000){2}{\rule{0.250pt}{0.400pt}}
\put(766,510.17){\rule{0.700pt}{0.400pt}}
\multiput(766.00,509.17)(1.547,2.000){2}{\rule{0.350pt}{0.400pt}}
\multiput(769.00,512.61)(0.462,0.447){3}{\rule{0.500pt}{0.108pt}}
\multiput(769.00,511.17)(1.962,3.000){2}{\rule{0.250pt}{0.400pt}}
\put(772,515.17){\rule{0.700pt}{0.400pt}}
\multiput(772.00,514.17)(1.547,2.000){2}{\rule{0.350pt}{0.400pt}}
\multiput(775.00,517.61)(0.462,0.447){3}{\rule{0.500pt}{0.108pt}}
\multiput(775.00,516.17)(1.962,3.000){2}{\rule{0.250pt}{0.400pt}}
\multiput(778.00,520.61)(0.462,0.447){3}{\rule{0.500pt}{0.108pt}}
\multiput(778.00,519.17)(1.962,3.000){2}{\rule{0.250pt}{0.400pt}}
\put(781,523.17){\rule{0.700pt}{0.400pt}}
\multiput(781.00,522.17)(1.547,2.000){2}{\rule{0.350pt}{0.400pt}}
\multiput(784.00,525.61)(0.462,0.447){3}{\rule{0.500pt}{0.108pt}}
\multiput(784.00,524.17)(1.962,3.000){2}{\rule{0.250pt}{0.400pt}}
\put(787,528.17){\rule{0.700pt}{0.400pt}}
\multiput(787.00,527.17)(1.547,2.000){2}{\rule{0.350pt}{0.400pt}}
\put(790.17,530){\rule{0.400pt}{0.700pt}}
\multiput(789.17,530.00)(2.000,1.547){2}{\rule{0.400pt}{0.350pt}}
\put(792,533.17){\rule{0.700pt}{0.400pt}}
\multiput(792.00,532.17)(1.547,2.000){2}{\rule{0.350pt}{0.400pt}}
\multiput(795.00,535.61)(0.462,0.447){3}{\rule{0.500pt}{0.108pt}}
\multiput(795.00,534.17)(1.962,3.000){2}{\rule{0.250pt}{0.400pt}}
\put(798,538.17){\rule{0.700pt}{0.400pt}}
\multiput(798.00,537.17)(1.547,2.000){2}{\rule{0.350pt}{0.400pt}}
\multiput(801.00,540.61)(0.462,0.447){3}{\rule{0.500pt}{0.108pt}}
\multiput(801.00,539.17)(1.962,3.000){2}{\rule{0.250pt}{0.400pt}}
\put(804,543.17){\rule{0.700pt}{0.400pt}}
\multiput(804.00,542.17)(1.547,2.000){2}{\rule{0.350pt}{0.400pt}}
\multiput(807.00,545.61)(0.462,0.447){3}{\rule{0.500pt}{0.108pt}}
\multiput(807.00,544.17)(1.962,3.000){2}{\rule{0.250pt}{0.400pt}}
\put(810,548.17){\rule{0.700pt}{0.400pt}}
\multiput(810.00,547.17)(1.547,2.000){2}{\rule{0.350pt}{0.400pt}}
\multiput(813.00,550.61)(0.462,0.447){3}{\rule{0.500pt}{0.108pt}}
\multiput(813.00,549.17)(1.962,3.000){2}{\rule{0.250pt}{0.400pt}}
\put(816,553.17){\rule{0.700pt}{0.400pt}}
\multiput(816.00,552.17)(1.547,2.000){2}{\rule{0.350pt}{0.400pt}}
\multiput(819.00,555.61)(0.462,0.447){3}{\rule{0.500pt}{0.108pt}}
\multiput(819.00,554.17)(1.962,3.000){2}{\rule{0.250pt}{0.400pt}}
\put(822,558.17){\rule{0.700pt}{0.400pt}}
\multiput(822.00,557.17)(1.547,2.000){2}{\rule{0.350pt}{0.400pt}}
\multiput(825.00,560.61)(0.462,0.447){3}{\rule{0.500pt}{0.108pt}}
\multiput(825.00,559.17)(1.962,3.000){2}{\rule{0.250pt}{0.400pt}}
\put(828,563.17){\rule{0.700pt}{0.400pt}}
\multiput(828.00,562.17)(1.547,2.000){2}{\rule{0.350pt}{0.400pt}}
\multiput(831.00,565.61)(0.462,0.447){3}{\rule{0.500pt}{0.108pt}}
\multiput(831.00,564.17)(1.962,3.000){2}{\rule{0.250pt}{0.400pt}}
\put(834,568.17){\rule{0.700pt}{0.400pt}}
\multiput(834.00,567.17)(1.547,2.000){2}{\rule{0.350pt}{0.400pt}}
\multiput(837.00,570.61)(0.462,0.447){3}{\rule{0.500pt}{0.108pt}}
\multiput(837.00,569.17)(1.962,3.000){2}{\rule{0.250pt}{0.400pt}}
\put(840,573.17){\rule{0.700pt}{0.400pt}}
\multiput(840.00,572.17)(1.547,2.000){2}{\rule{0.350pt}{0.400pt}}
\multiput(843.00,575.61)(0.462,0.447){3}{\rule{0.500pt}{0.108pt}}
\multiput(843.00,574.17)(1.962,3.000){2}{\rule{0.250pt}{0.400pt}}
\put(846,578.17){\rule{0.482pt}{0.400pt}}
\multiput(846.00,577.17)(1.000,2.000){2}{\rule{0.241pt}{0.400pt}}
\multiput(848.00,580.61)(0.462,0.447){3}{\rule{0.500pt}{0.108pt}}
\multiput(848.00,579.17)(1.962,3.000){2}{\rule{0.250pt}{0.400pt}}
\put(851,583.17){\rule{0.700pt}{0.400pt}}
\multiput(851.00,582.17)(1.547,2.000){2}{\rule{0.350pt}{0.400pt}}
\multiput(854.00,585.61)(0.462,0.447){3}{\rule{0.500pt}{0.108pt}}
\multiput(854.00,584.17)(1.962,3.000){2}{\rule{0.250pt}{0.400pt}}
\multiput(857.00,588.61)(0.462,0.447){3}{\rule{0.500pt}{0.108pt}}
\multiput(857.00,587.17)(1.962,3.000){2}{\rule{0.250pt}{0.400pt}}
\put(860,591.17){\rule{0.700pt}{0.400pt}}
\multiput(860.00,590.17)(1.547,2.000){2}{\rule{0.350pt}{0.400pt}}
\multiput(863.00,593.61)(0.462,0.447){3}{\rule{0.500pt}{0.108pt}}
\multiput(863.00,592.17)(1.962,3.000){2}{\rule{0.250pt}{0.400pt}}
\put(866,596.17){\rule{0.700pt}{0.400pt}}
\multiput(866.00,595.17)(1.547,2.000){2}{\rule{0.350pt}{0.400pt}}
\multiput(869.00,598.61)(0.462,0.447){3}{\rule{0.500pt}{0.108pt}}
\multiput(869.00,597.17)(1.962,3.000){2}{\rule{0.250pt}{0.400pt}}
\put(872,601.17){\rule{0.700pt}{0.400pt}}
\multiput(872.00,600.17)(1.547,2.000){2}{\rule{0.350pt}{0.400pt}}
\multiput(875.00,603.61)(0.462,0.447){3}{\rule{0.500pt}{0.108pt}}
\multiput(875.00,602.17)(1.962,3.000){2}{\rule{0.250pt}{0.400pt}}
\put(878,606.17){\rule{0.700pt}{0.400pt}}
\multiput(878.00,605.17)(1.547,2.000){2}{\rule{0.350pt}{0.400pt}}
\multiput(881.00,608.61)(0.462,0.447){3}{\rule{0.500pt}{0.108pt}}
\multiput(881.00,607.17)(1.962,3.000){2}{\rule{0.250pt}{0.400pt}}
\put(884,611.17){\rule{0.700pt}{0.400pt}}
\multiput(884.00,610.17)(1.547,2.000){2}{\rule{0.350pt}{0.400pt}}
\multiput(887.00,613.61)(0.462,0.447){3}{\rule{0.500pt}{0.108pt}}
\multiput(887.00,612.17)(1.962,3.000){2}{\rule{0.250pt}{0.400pt}}
\put(890,616.17){\rule{0.700pt}{0.400pt}}
\multiput(890.00,615.17)(1.547,2.000){2}{\rule{0.350pt}{0.400pt}}
\multiput(893.00,618.61)(0.462,0.447){3}{\rule{0.500pt}{0.108pt}}
\multiput(893.00,617.17)(1.962,3.000){2}{\rule{0.250pt}{0.400pt}}
\put(896,621.17){\rule{0.700pt}{0.400pt}}
\multiput(896.00,620.17)(1.547,2.000){2}{\rule{0.350pt}{0.400pt}}
\put(362.0,228.0){\rule[-0.200pt]{0.723pt}{0.400pt}}
\put(899.0,623.0){\usebox{\plotpoint}}
\multiput(412.00,207.61)(0.462,0.447){3}{\rule{0.500pt}{0.108pt}}
\multiput(412.00,206.17)(1.962,3.000){2}{\rule{0.250pt}{0.400pt}}
\put(415,210.17){\rule{0.700pt}{0.400pt}}
\multiput(415.00,209.17)(1.547,2.000){2}{\rule{0.350pt}{0.400pt}}
\multiput(418.00,212.61)(0.462,0.447){3}{\rule{0.500pt}{0.108pt}}
\multiput(418.00,211.17)(1.962,3.000){2}{\rule{0.250pt}{0.400pt}}
\multiput(421.00,215.61)(0.462,0.447){3}{\rule{0.500pt}{0.108pt}}
\multiput(421.00,214.17)(1.962,3.000){2}{\rule{0.250pt}{0.400pt}}
\put(424,218.17){\rule{0.700pt}{0.400pt}}
\multiput(424.00,217.17)(1.547,2.000){2}{\rule{0.350pt}{0.400pt}}
\multiput(427.00,220.61)(0.462,0.447){3}{\rule{0.500pt}{0.108pt}}
\multiput(427.00,219.17)(1.962,3.000){2}{\rule{0.250pt}{0.400pt}}
\put(430,223.17){\rule{0.700pt}{0.400pt}}
\multiput(430.00,222.17)(1.547,2.000){2}{\rule{0.350pt}{0.400pt}}
\multiput(433.00,225.61)(0.462,0.447){3}{\rule{0.500pt}{0.108pt}}
\multiput(433.00,224.17)(1.962,3.000){2}{\rule{0.250pt}{0.400pt}}
\put(436,228.17){\rule{0.700pt}{0.400pt}}
\multiput(436.00,227.17)(1.547,2.000){2}{\rule{0.350pt}{0.400pt}}
\multiput(439.00,230.61)(0.462,0.447){3}{\rule{0.500pt}{0.108pt}}
\multiput(439.00,229.17)(1.962,3.000){2}{\rule{0.250pt}{0.400pt}}
\put(442,233.17){\rule{0.700pt}{0.400pt}}
\multiput(442.00,232.17)(1.547,2.000){2}{\rule{0.350pt}{0.400pt}}
\multiput(445.00,235.61)(0.462,0.447){3}{\rule{0.500pt}{0.108pt}}
\multiput(445.00,234.17)(1.962,3.000){2}{\rule{0.250pt}{0.400pt}}
\put(448,238.17){\rule{0.700pt}{0.400pt}}
\multiput(448.00,237.17)(1.547,2.000){2}{\rule{0.350pt}{0.400pt}}
\multiput(451.00,240.61)(0.462,0.447){3}{\rule{0.500pt}{0.108pt}}
\multiput(451.00,239.17)(1.962,3.000){2}{\rule{0.250pt}{0.400pt}}
\put(454,243.17){\rule{0.482pt}{0.400pt}}
\multiput(454.00,242.17)(1.000,2.000){2}{\rule{0.241pt}{0.400pt}}
\multiput(456.00,245.61)(0.462,0.447){3}{\rule{0.500pt}{0.108pt}}
\multiput(456.00,244.17)(1.962,3.000){2}{\rule{0.250pt}{0.400pt}}
\put(459,248.17){\rule{0.700pt}{0.400pt}}
\multiput(459.00,247.17)(1.547,2.000){2}{\rule{0.350pt}{0.400pt}}
\multiput(462.00,250.61)(0.462,0.447){3}{\rule{0.500pt}{0.108pt}}
\multiput(462.00,249.17)(1.962,3.000){2}{\rule{0.250pt}{0.400pt}}
\put(465,253.17){\rule{0.700pt}{0.400pt}}
\multiput(465.00,252.17)(1.547,2.000){2}{\rule{0.350pt}{0.400pt}}
\multiput(468.00,255.61)(0.462,0.447){3}{\rule{0.500pt}{0.108pt}}
\multiput(468.00,254.17)(1.962,3.000){2}{\rule{0.250pt}{0.400pt}}
\put(471,258.17){\rule{0.700pt}{0.400pt}}
\multiput(471.00,257.17)(1.547,2.000){2}{\rule{0.350pt}{0.400pt}}
\multiput(474.00,260.61)(0.462,0.447){3}{\rule{0.500pt}{0.108pt}}
\multiput(474.00,259.17)(1.962,3.000){2}{\rule{0.250pt}{0.400pt}}
\put(477,263.17){\rule{0.700pt}{0.400pt}}
\multiput(477.00,262.17)(1.547,2.000){2}{\rule{0.350pt}{0.400pt}}
\multiput(480.00,265.61)(0.462,0.447){3}{\rule{0.500pt}{0.108pt}}
\multiput(480.00,264.17)(1.962,3.000){2}{\rule{0.250pt}{0.400pt}}
\multiput(483.00,268.61)(0.462,0.447){3}{\rule{0.500pt}{0.108pt}}
\multiput(483.00,267.17)(1.962,3.000){2}{\rule{0.250pt}{0.400pt}}
\put(486,271.17){\rule{0.700pt}{0.400pt}}
\multiput(486.00,270.17)(1.547,2.000){2}{\rule{0.350pt}{0.400pt}}
\multiput(489.00,273.61)(0.462,0.447){3}{\rule{0.500pt}{0.108pt}}
\multiput(489.00,272.17)(1.962,3.000){2}{\rule{0.250pt}{0.400pt}}
\put(492,276.17){\rule{0.700pt}{0.400pt}}
\multiput(492.00,275.17)(1.547,2.000){2}{\rule{0.350pt}{0.400pt}}
\multiput(495.00,278.61)(0.462,0.447){3}{\rule{0.500pt}{0.108pt}}
\multiput(495.00,277.17)(1.962,3.000){2}{\rule{0.250pt}{0.400pt}}
\put(498,281.17){\rule{0.700pt}{0.400pt}}
\multiput(498.00,280.17)(1.547,2.000){2}{\rule{0.350pt}{0.400pt}}
\multiput(501.00,283.61)(0.462,0.447){3}{\rule{0.500pt}{0.108pt}}
\multiput(501.00,282.17)(1.962,3.000){2}{\rule{0.250pt}{0.400pt}}
\put(504,286.17){\rule{0.700pt}{0.400pt}}
\multiput(504.00,285.17)(1.547,2.000){2}{\rule{0.350pt}{0.400pt}}
\multiput(507.00,288.61)(0.462,0.447){3}{\rule{0.500pt}{0.108pt}}
\multiput(507.00,287.17)(1.962,3.000){2}{\rule{0.250pt}{0.400pt}}
\put(510,291.17){\rule{0.482pt}{0.400pt}}
\multiput(510.00,290.17)(1.000,2.000){2}{\rule{0.241pt}{0.400pt}}
\multiput(512.00,293.61)(0.462,0.447){3}{\rule{0.500pt}{0.108pt}}
\multiput(512.00,292.17)(1.962,3.000){2}{\rule{0.250pt}{0.400pt}}
\put(515,296.17){\rule{0.700pt}{0.400pt}}
\multiput(515.00,295.17)(1.547,2.000){2}{\rule{0.350pt}{0.400pt}}
\multiput(518.00,298.61)(0.462,0.447){3}{\rule{0.500pt}{0.108pt}}
\multiput(518.00,297.17)(1.962,3.000){2}{\rule{0.250pt}{0.400pt}}
\put(521,301.17){\rule{0.700pt}{0.400pt}}
\multiput(521.00,300.17)(1.547,2.000){2}{\rule{0.350pt}{0.400pt}}
\multiput(524.00,303.61)(0.462,0.447){3}{\rule{0.500pt}{0.108pt}}
\multiput(524.00,302.17)(1.962,3.000){2}{\rule{0.250pt}{0.400pt}}
\put(527,306.17){\rule{0.700pt}{0.400pt}}
\multiput(527.00,305.17)(1.547,2.000){2}{\rule{0.350pt}{0.400pt}}
\multiput(530.00,308.61)(0.462,0.447){3}{\rule{0.500pt}{0.108pt}}
\multiput(530.00,307.17)(1.962,3.000){2}{\rule{0.250pt}{0.400pt}}
\put(533,311.17){\rule{0.700pt}{0.400pt}}
\multiput(533.00,310.17)(1.547,2.000){2}{\rule{0.350pt}{0.400pt}}
\multiput(536.00,313.61)(0.462,0.447){3}{\rule{0.500pt}{0.108pt}}
\multiput(536.00,312.17)(1.962,3.000){2}{\rule{0.250pt}{0.400pt}}
\put(539,316.17){\rule{0.700pt}{0.400pt}}
\multiput(539.00,315.17)(1.547,2.000){2}{\rule{0.350pt}{0.400pt}}
\multiput(542.00,318.61)(0.462,0.447){3}{\rule{0.500pt}{0.108pt}}
\multiput(542.00,317.17)(1.962,3.000){2}{\rule{0.250pt}{0.400pt}}
\multiput(545.00,321.61)(0.462,0.447){3}{\rule{0.500pt}{0.108pt}}
\multiput(545.00,320.17)(1.962,3.000){2}{\rule{0.250pt}{0.400pt}}
\put(548,324.17){\rule{0.700pt}{0.400pt}}
\multiput(548.00,323.17)(1.547,2.000){2}{\rule{0.350pt}{0.400pt}}
\multiput(551.00,326.61)(0.462,0.447){3}{\rule{0.500pt}{0.108pt}}
\multiput(551.00,325.17)(1.962,3.000){2}{\rule{0.250pt}{0.400pt}}
\put(554,329.17){\rule{0.700pt}{0.400pt}}
\multiput(554.00,328.17)(1.547,2.000){2}{\rule{0.350pt}{0.400pt}}
\multiput(557.00,331.61)(0.462,0.447){3}{\rule{0.500pt}{0.108pt}}
\multiput(557.00,330.17)(1.962,3.000){2}{\rule{0.250pt}{0.400pt}}
\put(560,334.17){\rule{0.700pt}{0.400pt}}
\multiput(560.00,333.17)(1.547,2.000){2}{\rule{0.350pt}{0.400pt}}
\multiput(563.00,336.61)(0.462,0.447){3}{\rule{0.500pt}{0.108pt}}
\multiput(563.00,335.17)(1.962,3.000){2}{\rule{0.250pt}{0.400pt}}
\put(566,339.17){\rule{0.482pt}{0.400pt}}
\multiput(566.00,338.17)(1.000,2.000){2}{\rule{0.241pt}{0.400pt}}
\multiput(568.00,341.61)(0.462,0.447){3}{\rule{0.500pt}{0.108pt}}
\multiput(568.00,340.17)(1.962,3.000){2}{\rule{0.250pt}{0.400pt}}
\put(571,344.17){\rule{0.700pt}{0.400pt}}
\multiput(571.00,343.17)(1.547,2.000){2}{\rule{0.350pt}{0.400pt}}
\multiput(574.00,346.61)(0.462,0.447){3}{\rule{0.500pt}{0.108pt}}
\multiput(574.00,345.17)(1.962,3.000){2}{\rule{0.250pt}{0.400pt}}
\put(577,349.17){\rule{0.700pt}{0.400pt}}
\multiput(577.00,348.17)(1.547,2.000){2}{\rule{0.350pt}{0.400pt}}
\multiput(580.00,351.61)(0.462,0.447){3}{\rule{0.500pt}{0.108pt}}
\multiput(580.00,350.17)(1.962,3.000){2}{\rule{0.250pt}{0.400pt}}
\put(583,354.17){\rule{0.700pt}{0.400pt}}
\multiput(583.00,353.17)(1.547,2.000){2}{\rule{0.350pt}{0.400pt}}
\multiput(586.00,356.61)(0.462,0.447){3}{\rule{0.500pt}{0.108pt}}
\multiput(586.00,355.17)(1.962,3.000){2}{\rule{0.250pt}{0.400pt}}
\put(589,359.17){\rule{0.700pt}{0.400pt}}
\multiput(589.00,358.17)(1.547,2.000){2}{\rule{0.350pt}{0.400pt}}
\multiput(592.00,361.61)(0.462,0.447){3}{\rule{0.500pt}{0.108pt}}
\multiput(592.00,360.17)(1.962,3.000){2}{\rule{0.250pt}{0.400pt}}
\put(595,364.17){\rule{0.700pt}{0.400pt}}
\multiput(595.00,363.17)(1.547,2.000){2}{\rule{0.350pt}{0.400pt}}
\multiput(598.00,366.61)(0.462,0.447){3}{\rule{0.500pt}{0.108pt}}
\multiput(598.00,365.17)(1.962,3.000){2}{\rule{0.250pt}{0.400pt}}
\put(601,369.17){\rule{0.700pt}{0.400pt}}
\multiput(601.00,368.17)(1.547,2.000){2}{\rule{0.350pt}{0.400pt}}
\multiput(604.00,371.61)(0.462,0.447){3}{\rule{0.500pt}{0.108pt}}
\multiput(604.00,370.17)(1.962,3.000){2}{\rule{0.250pt}{0.400pt}}
\multiput(607.00,374.61)(0.462,0.447){3}{\rule{0.500pt}{0.108pt}}
\multiput(607.00,373.17)(1.962,3.000){2}{\rule{0.250pt}{0.400pt}}
\put(610,377.17){\rule{0.700pt}{0.400pt}}
\multiput(610.00,376.17)(1.547,2.000){2}{\rule{0.350pt}{0.400pt}}
\multiput(613.00,379.61)(0.462,0.447){3}{\rule{0.500pt}{0.108pt}}
\multiput(613.00,378.17)(1.962,3.000){2}{\rule{0.250pt}{0.400pt}}
\put(616,382.17){\rule{0.700pt}{0.400pt}}
\multiput(616.00,381.17)(1.547,2.000){2}{\rule{0.350pt}{0.400pt}}
\multiput(619.00,384.61)(0.462,0.447){3}{\rule{0.500pt}{0.108pt}}
\multiput(619.00,383.17)(1.962,3.000){2}{\rule{0.250pt}{0.400pt}}
\put(622,387.17){\rule{0.482pt}{0.400pt}}
\multiput(622.00,386.17)(1.000,2.000){2}{\rule{0.241pt}{0.400pt}}
\multiput(624.00,389.61)(0.462,0.447){3}{\rule{0.500pt}{0.108pt}}
\multiput(624.00,388.17)(1.962,3.000){2}{\rule{0.250pt}{0.400pt}}
\put(627,392.17){\rule{0.700pt}{0.400pt}}
\multiput(627.00,391.17)(1.547,2.000){2}{\rule{0.350pt}{0.400pt}}
\multiput(630.00,394.61)(0.462,0.447){3}{\rule{0.500pt}{0.108pt}}
\multiput(630.00,393.17)(1.962,3.000){2}{\rule{0.250pt}{0.400pt}}
\put(633,397.17){\rule{0.700pt}{0.400pt}}
\multiput(633.00,396.17)(1.547,2.000){2}{\rule{0.350pt}{0.400pt}}
\multiput(636.00,399.61)(0.462,0.447){3}{\rule{0.500pt}{0.108pt}}
\multiput(636.00,398.17)(1.962,3.000){2}{\rule{0.250pt}{0.400pt}}
\put(639,402.17){\rule{0.700pt}{0.400pt}}
\multiput(639.00,401.17)(1.547,2.000){2}{\rule{0.350pt}{0.400pt}}
\multiput(642.00,404.61)(0.462,0.447){3}{\rule{0.500pt}{0.108pt}}
\multiput(642.00,403.17)(1.962,3.000){2}{\rule{0.250pt}{0.400pt}}
\put(645,407.17){\rule{0.700pt}{0.400pt}}
\multiput(645.00,406.17)(1.547,2.000){2}{\rule{0.350pt}{0.400pt}}
\multiput(648.00,409.61)(0.462,0.447){3}{\rule{0.500pt}{0.108pt}}
\multiput(648.00,408.17)(1.962,3.000){2}{\rule{0.250pt}{0.400pt}}
\put(651,412.17){\rule{0.700pt}{0.400pt}}
\multiput(651.00,411.17)(1.547,2.000){2}{\rule{0.350pt}{0.400pt}}
\multiput(654.00,414.61)(0.462,0.447){3}{\rule{0.500pt}{0.108pt}}
\multiput(654.00,413.17)(1.962,3.000){2}{\rule{0.250pt}{0.400pt}}
\put(657,417.17){\rule{0.700pt}{0.400pt}}
\multiput(657.00,416.17)(1.547,2.000){2}{\rule{0.350pt}{0.400pt}}
\multiput(660.00,419.61)(0.462,0.447){3}{\rule{0.500pt}{0.108pt}}
\multiput(660.00,418.17)(1.962,3.000){2}{\rule{0.250pt}{0.400pt}}
\put(663,422.17){\rule{0.700pt}{0.400pt}}
\multiput(663.00,421.17)(1.547,2.000){2}{\rule{0.350pt}{0.400pt}}
\multiput(666.00,424.61)(0.462,0.447){3}{\rule{0.500pt}{0.108pt}}
\multiput(666.00,423.17)(1.962,3.000){2}{\rule{0.250pt}{0.400pt}}
\multiput(669.00,427.61)(0.462,0.447){3}{\rule{0.500pt}{0.108pt}}
\multiput(669.00,426.17)(1.962,3.000){2}{\rule{0.250pt}{0.400pt}}
\put(672,430.17){\rule{0.700pt}{0.400pt}}
\multiput(672.00,429.17)(1.547,2.000){2}{\rule{0.350pt}{0.400pt}}
\multiput(675.00,432.61)(0.462,0.447){3}{\rule{0.500pt}{0.108pt}}
\multiput(675.00,431.17)(1.962,3.000){2}{\rule{0.250pt}{0.400pt}}
\put(678,435.17){\rule{0.482pt}{0.400pt}}
\multiput(678.00,434.17)(1.000,2.000){2}{\rule{0.241pt}{0.400pt}}
\multiput(680.00,437.61)(0.462,0.447){3}{\rule{0.500pt}{0.108pt}}
\multiput(680.00,436.17)(1.962,3.000){2}{\rule{0.250pt}{0.400pt}}
\put(683,440.17){\rule{0.700pt}{0.400pt}}
\multiput(683.00,439.17)(1.547,2.000){2}{\rule{0.350pt}{0.400pt}}
\multiput(686.00,442.61)(0.462,0.447){3}{\rule{0.500pt}{0.108pt}}
\multiput(686.00,441.17)(1.962,3.000){2}{\rule{0.250pt}{0.400pt}}
\put(689,445.17){\rule{0.700pt}{0.400pt}}
\multiput(689.00,444.17)(1.547,2.000){2}{\rule{0.350pt}{0.400pt}}
\multiput(692.00,447.61)(0.462,0.447){3}{\rule{0.500pt}{0.108pt}}
\multiput(692.00,446.17)(1.962,3.000){2}{\rule{0.250pt}{0.400pt}}
\put(695,450.17){\rule{0.700pt}{0.400pt}}
\multiput(695.00,449.17)(1.547,2.000){2}{\rule{0.350pt}{0.400pt}}
\multiput(698.00,452.61)(0.462,0.447){3}{\rule{0.500pt}{0.108pt}}
\multiput(698.00,451.17)(1.962,3.000){2}{\rule{0.250pt}{0.400pt}}
\put(701,455.17){\rule{0.700pt}{0.400pt}}
\multiput(701.00,454.17)(1.547,2.000){2}{\rule{0.350pt}{0.400pt}}
\multiput(704.00,457.61)(0.462,0.447){3}{\rule{0.500pt}{0.108pt}}
\multiput(704.00,456.17)(1.962,3.000){2}{\rule{0.250pt}{0.400pt}}
\put(707,460.17){\rule{0.700pt}{0.400pt}}
\multiput(707.00,459.17)(1.547,2.000){2}{\rule{0.350pt}{0.400pt}}
\multiput(710.00,462.61)(0.462,0.447){3}{\rule{0.500pt}{0.108pt}}
\multiput(710.00,461.17)(1.962,3.000){2}{\rule{0.250pt}{0.400pt}}
\put(713,465.17){\rule{0.700pt}{0.400pt}}
\multiput(713.00,464.17)(1.547,2.000){2}{\rule{0.350pt}{0.400pt}}
\multiput(716.00,467.61)(0.462,0.447){3}{\rule{0.500pt}{0.108pt}}
\multiput(716.00,466.17)(1.962,3.000){2}{\rule{0.250pt}{0.400pt}}
\put(719,470.17){\rule{0.700pt}{0.400pt}}
\multiput(719.00,469.17)(1.547,2.000){2}{\rule{0.350pt}{0.400pt}}
\multiput(722.00,472.61)(0.462,0.447){3}{\rule{0.500pt}{0.108pt}}
\multiput(722.00,471.17)(1.962,3.000){2}{\rule{0.250pt}{0.400pt}}
\put(725,475.17){\rule{0.700pt}{0.400pt}}
\multiput(725.00,474.17)(1.547,2.000){2}{\rule{0.350pt}{0.400pt}}
\multiput(728.00,477.61)(0.462,0.447){3}{\rule{0.500pt}{0.108pt}}
\multiput(728.00,476.17)(1.962,3.000){2}{\rule{0.250pt}{0.400pt}}
\multiput(731.00,480.61)(0.462,0.447){3}{\rule{0.500pt}{0.108pt}}
\multiput(731.00,479.17)(1.962,3.000){2}{\rule{0.250pt}{0.400pt}}
\put(734,483.17){\rule{0.482pt}{0.400pt}}
\multiput(734.00,482.17)(1.000,2.000){2}{\rule{0.241pt}{0.400pt}}
\multiput(736.00,485.61)(0.462,0.447){3}{\rule{0.500pt}{0.108pt}}
\multiput(736.00,484.17)(1.962,3.000){2}{\rule{0.250pt}{0.400pt}}
\put(739,488.17){\rule{0.700pt}{0.400pt}}
\multiput(739.00,487.17)(1.547,2.000){2}{\rule{0.350pt}{0.400pt}}
\multiput(742.00,490.61)(0.462,0.447){3}{\rule{0.500pt}{0.108pt}}
\multiput(742.00,489.17)(1.962,3.000){2}{\rule{0.250pt}{0.400pt}}
\put(745,493.17){\rule{0.700pt}{0.400pt}}
\multiput(745.00,492.17)(1.547,2.000){2}{\rule{0.350pt}{0.400pt}}
\multiput(748.00,495.61)(0.462,0.447){3}{\rule{0.500pt}{0.108pt}}
\multiput(748.00,494.17)(1.962,3.000){2}{\rule{0.250pt}{0.400pt}}
\put(751,498.17){\rule{0.700pt}{0.400pt}}
\multiput(751.00,497.17)(1.547,2.000){2}{\rule{0.350pt}{0.400pt}}
\multiput(754.00,500.61)(0.462,0.447){3}{\rule{0.500pt}{0.108pt}}
\multiput(754.00,499.17)(1.962,3.000){2}{\rule{0.250pt}{0.400pt}}
\put(757,503.17){\rule{0.700pt}{0.400pt}}
\multiput(757.00,502.17)(1.547,2.000){2}{\rule{0.350pt}{0.400pt}}
\multiput(760.00,505.61)(0.462,0.447){3}{\rule{0.500pt}{0.108pt}}
\multiput(760.00,504.17)(1.962,3.000){2}{\rule{0.250pt}{0.400pt}}
\put(763,508.17){\rule{0.700pt}{0.400pt}}
\multiput(763.00,507.17)(1.547,2.000){2}{\rule{0.350pt}{0.400pt}}
\multiput(766.00,510.61)(0.462,0.447){3}{\rule{0.500pt}{0.108pt}}
\multiput(766.00,509.17)(1.962,3.000){2}{\rule{0.250pt}{0.400pt}}
\put(769,513.17){\rule{0.700pt}{0.400pt}}
\multiput(769.00,512.17)(1.547,2.000){2}{\rule{0.350pt}{0.400pt}}
\multiput(772.00,515.61)(0.462,0.447){3}{\rule{0.500pt}{0.108pt}}
\multiput(772.00,514.17)(1.962,3.000){2}{\rule{0.250pt}{0.400pt}}
\put(775,518.17){\rule{0.700pt}{0.400pt}}
\multiput(775.00,517.17)(1.547,2.000){2}{\rule{0.350pt}{0.400pt}}
\multiput(778.00,520.61)(0.462,0.447){3}{\rule{0.500pt}{0.108pt}}
\multiput(778.00,519.17)(1.962,3.000){2}{\rule{0.250pt}{0.400pt}}
\put(781,523.17){\rule{0.700pt}{0.400pt}}
\multiput(781.00,522.17)(1.547,2.000){2}{\rule{0.350pt}{0.400pt}}
\multiput(784.00,525.61)(0.462,0.447){3}{\rule{0.500pt}{0.108pt}}
\multiput(784.00,524.17)(1.962,3.000){2}{\rule{0.250pt}{0.400pt}}
\put(787,528.17){\rule{0.700pt}{0.400pt}}
\multiput(787.00,527.17)(1.547,2.000){2}{\rule{0.350pt}{0.400pt}}
\put(790.17,530){\rule{0.400pt}{0.700pt}}
\multiput(789.17,530.00)(2.000,1.547){2}{\rule{0.400pt}{0.350pt}}
\multiput(792.00,533.61)(0.462,0.447){3}{\rule{0.500pt}{0.108pt}}
\multiput(792.00,532.17)(1.962,3.000){2}{\rule{0.250pt}{0.400pt}}
\put(795,536.17){\rule{0.700pt}{0.400pt}}
\multiput(795.00,535.17)(1.547,2.000){2}{\rule{0.350pt}{0.400pt}}
\multiput(798.00,538.61)(0.462,0.447){3}{\rule{0.500pt}{0.108pt}}
\multiput(798.00,537.17)(1.962,3.000){2}{\rule{0.250pt}{0.400pt}}
\put(801,541.17){\rule{0.700pt}{0.400pt}}
\multiput(801.00,540.17)(1.547,2.000){2}{\rule{0.350pt}{0.400pt}}
\multiput(804.00,543.61)(0.462,0.447){3}{\rule{0.500pt}{0.108pt}}
\multiput(804.00,542.17)(1.962,3.000){2}{\rule{0.250pt}{0.400pt}}
\put(807,546.17){\rule{0.700pt}{0.400pt}}
\multiput(807.00,545.17)(1.547,2.000){2}{\rule{0.350pt}{0.400pt}}
\multiput(810.00,548.61)(0.462,0.447){3}{\rule{0.500pt}{0.108pt}}
\multiput(810.00,547.17)(1.962,3.000){2}{\rule{0.250pt}{0.400pt}}
\put(813,551.17){\rule{0.700pt}{0.400pt}}
\multiput(813.00,550.17)(1.547,2.000){2}{\rule{0.350pt}{0.400pt}}
\multiput(816.00,553.61)(0.462,0.447){3}{\rule{0.500pt}{0.108pt}}
\multiput(816.00,552.17)(1.962,3.000){2}{\rule{0.250pt}{0.400pt}}
\put(819,556.17){\rule{0.700pt}{0.400pt}}
\multiput(819.00,555.17)(1.547,2.000){2}{\rule{0.350pt}{0.400pt}}
\multiput(822.00,558.61)(0.462,0.447){3}{\rule{0.500pt}{0.108pt}}
\multiput(822.00,557.17)(1.962,3.000){2}{\rule{0.250pt}{0.400pt}}
\put(825,561.17){\rule{0.700pt}{0.400pt}}
\multiput(825.00,560.17)(1.547,2.000){2}{\rule{0.350pt}{0.400pt}}
\multiput(828.00,563.61)(0.462,0.447){3}{\rule{0.500pt}{0.108pt}}
\multiput(828.00,562.17)(1.962,3.000){2}{\rule{0.250pt}{0.400pt}}
\put(831,566.17){\rule{0.700pt}{0.400pt}}
\multiput(831.00,565.17)(1.547,2.000){2}{\rule{0.350pt}{0.400pt}}
\multiput(834.00,568.61)(0.462,0.447){3}{\rule{0.500pt}{0.108pt}}
\multiput(834.00,567.17)(1.962,3.000){2}{\rule{0.250pt}{0.400pt}}
\put(837,571.17){\rule{0.700pt}{0.400pt}}
\multiput(837.00,570.17)(1.547,2.000){2}{\rule{0.350pt}{0.400pt}}
\multiput(840.00,573.61)(0.462,0.447){3}{\rule{0.500pt}{0.108pt}}
\multiput(840.00,572.17)(1.962,3.000){2}{\rule{0.250pt}{0.400pt}}
\put(843,576.17){\rule{0.700pt}{0.400pt}}
\multiput(843.00,575.17)(1.547,2.000){2}{\rule{0.350pt}{0.400pt}}
\put(846.17,578){\rule{0.400pt}{0.700pt}}
\multiput(845.17,578.00)(2.000,1.547){2}{\rule{0.400pt}{0.350pt}}
\put(848,581.17){\rule{0.700pt}{0.400pt}}
\multiput(848.00,580.17)(1.547,2.000){2}{\rule{0.350pt}{0.400pt}}
\multiput(851.00,583.61)(0.462,0.447){3}{\rule{0.500pt}{0.108pt}}
\multiput(851.00,582.17)(1.962,3.000){2}{\rule{0.250pt}{0.400pt}}
\multiput(854.00,586.61)(0.462,0.447){3}{\rule{0.500pt}{0.108pt}}
\multiput(854.00,585.17)(1.962,3.000){2}{\rule{0.250pt}{0.400pt}}
\put(857,589.17){\rule{0.700pt}{0.400pt}}
\multiput(857.00,588.17)(1.547,2.000){2}{\rule{0.350pt}{0.400pt}}
\multiput(860.00,591.61)(0.462,0.447){3}{\rule{0.500pt}{0.108pt}}
\multiput(860.00,590.17)(1.962,3.000){2}{\rule{0.250pt}{0.400pt}}
\put(863,594.17){\rule{0.700pt}{0.400pt}}
\multiput(863.00,593.17)(1.547,2.000){2}{\rule{0.350pt}{0.400pt}}
\multiput(866.00,596.61)(0.462,0.447){3}{\rule{0.500pt}{0.108pt}}
\multiput(866.00,595.17)(1.962,3.000){2}{\rule{0.250pt}{0.400pt}}
\put(869,599.17){\rule{0.700pt}{0.400pt}}
\multiput(869.00,598.17)(1.547,2.000){2}{\rule{0.350pt}{0.400pt}}
\multiput(872.00,601.61)(0.462,0.447){3}{\rule{0.500pt}{0.108pt}}
\multiput(872.00,600.17)(1.962,3.000){2}{\rule{0.250pt}{0.400pt}}
\put(875,604.17){\rule{0.700pt}{0.400pt}}
\multiput(875.00,603.17)(1.547,2.000){2}{\rule{0.350pt}{0.400pt}}
\multiput(878.00,606.61)(0.462,0.447){3}{\rule{0.500pt}{0.108pt}}
\multiput(878.00,605.17)(1.962,3.000){2}{\rule{0.250pt}{0.400pt}}
\put(881,609.17){\rule{0.700pt}{0.400pt}}
\multiput(881.00,608.17)(1.547,2.000){2}{\rule{0.350pt}{0.400pt}}
\multiput(884.00,611.61)(0.462,0.447){3}{\rule{0.500pt}{0.108pt}}
\multiput(884.00,610.17)(1.962,3.000){2}{\rule{0.250pt}{0.400pt}}
\put(887,614.17){\rule{0.700pt}{0.400pt}}
\multiput(887.00,613.17)(1.547,2.000){2}{\rule{0.350pt}{0.400pt}}
\multiput(890.00,616.61)(0.462,0.447){3}{\rule{0.500pt}{0.108pt}}
\multiput(890.00,615.17)(1.962,3.000){2}{\rule{0.250pt}{0.400pt}}
\put(893,619.17){\rule{0.700pt}{0.400pt}}
\multiput(893.00,618.17)(1.547,2.000){2}{\rule{0.350pt}{0.400pt}}
\multiput(896.00,621.61)(0.462,0.447){3}{\rule{0.500pt}{0.108pt}}
\multiput(896.00,620.17)(1.962,3.000){2}{\rule{0.250pt}{0.400pt}}
\put(131.0,207.0){\rule[-0.200pt]{67.693pt}{0.400pt}}
\put(899,624){\usebox{\plotpoint}}
\put(131.0,131.0){\rule[-0.200pt]{0.400pt}{132.013pt}}
\put(131.0,131.0){\rule[-0.200pt]{185.011pt}{0.400pt}}
\put(899.0,131.0){\rule[-0.200pt]{0.400pt}{132.013pt}}
\put(131.0,679.0){\rule[-0.200pt]{185.011pt}{0.400pt}}
\end{picture}

%% file: rv3.tex
\setlength{\unitlength}{0.240900pt}
\ifx\plotpoint\undefined\newsavebox{\plotpoint}\fi
\sbox{\plotpoint}{\rule[-0.200pt]{0.400pt}{0.400pt}}%
\begin{picture}(960,720)(0,0)
\sbox{\plotpoint}{\rule[-0.200pt]{0.400pt}{0.400pt}}%
\put(170.0,131.0){\rule[-0.200pt]{4.818pt}{0.400pt}}
\put(150,131){\makebox(0,0)[r]{ 0.2}}
\put(879.0,131.0){\rule[-0.200pt]{4.818pt}{0.400pt}}
\put(170.0,268.0){\rule[-0.200pt]{4.818pt}{0.400pt}}
\put(150,268){\makebox(0,0)[r]{ 0.201}}
\put(879.0,268.0){\rule[-0.200pt]{4.818pt}{0.400pt}}
\put(170.0,405.0){\rule[-0.200pt]{4.818pt}{0.400pt}}
\put(150,405){\makebox(0,0)[r]{ 0.202}}
\put(879.0,405.0){\rule[-0.200pt]{4.818pt}{0.400pt}}
\put(170.0,542.0){\rule[-0.200pt]{4.818pt}{0.400pt}}
\put(150,542){\makebox(0,0)[r]{ 0.203}}
\put(879.0,542.0){\rule[-0.200pt]{4.818pt}{0.400pt}}
\put(170.0,679.0){\rule[-0.200pt]{4.818pt}{0.400pt}}
\put(150,679){\makebox(0,0)[r]{ 0.204}}
\put(879.0,679.0){\rule[-0.200pt]{4.818pt}{0.400pt}}
\put(170.0,131.0){\rule[-0.200pt]{0.400pt}{4.818pt}}
\put(170,90){\makebox(0,0){ 0}}
\put(170.0,659.0){\rule[-0.200pt]{0.400pt}{4.818pt}}
\put(344.0,131.0){\rule[-0.200pt]{0.400pt}{4.818pt}}
\put(344,90){\makebox(0,0){ 0.0001}}
\put(344.0,659.0){\rule[-0.200pt]{0.400pt}{4.818pt}}
\put(517.0,131.0){\rule[-0.200pt]{0.400pt}{4.818pt}}
\put(517,90){\makebox(0,0){ 0.0002}}
\put(517.0,659.0){\rule[-0.200pt]{0.400pt}{4.818pt}}
\put(691.0,131.0){\rule[-0.200pt]{0.400pt}{4.818pt}}
\put(691,90){\makebox(0,0){ 0.0003}}
\put(691.0,659.0){\rule[-0.200pt]{0.400pt}{4.818pt}}
\put(864.0,131.0){\rule[-0.200pt]{0.400pt}{4.818pt}}
\put(864,90){\makebox(0,0){ 0.0004}}
\put(864.0,659.0){\rule[-0.200pt]{0.400pt}{4.818pt}}
\put(170.0,131.0){\rule[-0.200pt]{0.400pt}{132.013pt}}
\put(170.0,131.0){\rule[-0.200pt]{175.616pt}{0.400pt}}
\put(899.0,131.0){\rule[-0.200pt]{0.400pt}{132.013pt}}
\put(170.0,679.0){\rule[-0.200pt]{175.616pt}{0.400pt}}
\put(534,29){\makebox(0,0){$ 1/n ^2 $}}
\put(190.0,536.0){\rule[-0.200pt]{0.400pt}{29.631pt}}
\put(190.0,659.0){\rule[-0.200pt]{72.270pt}{0.400pt}}
\put(490.0,536.0){\rule[-0.200pt]{0.400pt}{29.631pt}}
\put(190.0,536.0){\rule[-0.200pt]{72.270pt}{0.400pt}}
\put(190.0,659.0){\rule[-0.200pt]{72.270pt}{0.400pt}}
\put(350,639){\makebox(0,0)[r]{$\mu _n (0)$}}
\multiput(895.68,552.92)(-0.880,-0.496){37}{\rule{0.800pt}{0.119pt}}
\multiput(897.34,553.17)(-33.340,-20.000){2}{\rule{0.400pt}{0.400pt}}
\multiput(860.75,532.92)(-0.856,-0.499){245}{\rule{0.784pt}{0.120pt}}
\multiput(862.37,533.17)(-210.373,-124.000){2}{\rule{0.392pt}{0.400pt}}
\multiput(648.75,408.92)(-0.854,-0.499){147}{\rule{0.783pt}{0.120pt}}
\multiput(650.38,409.17)(-126.376,-75.000){2}{\rule{0.391pt}{0.400pt}}
\multiput(520.71,333.92)(-0.866,-0.498){93}{\rule{0.792pt}{0.120pt}}
\multiput(522.36,334.17)(-81.357,-48.000){2}{\rule{0.396pt}{0.400pt}}
\multiput(437.80,285.92)(-0.841,-0.498){65}{\rule{0.771pt}{0.120pt}}
\multiput(439.40,286.17)(-55.401,-34.000){2}{\rule{0.385pt}{0.400pt}}
\multiput(380.82,251.92)(-0.837,-0.496){45}{\rule{0.767pt}{0.120pt}}
\multiput(382.41,252.17)(-38.409,-24.000){2}{\rule{0.383pt}{0.400pt}}
\multiput(340.56,227.92)(-0.919,-0.495){31}{\rule{0.829pt}{0.119pt}}
\multiput(342.28,228.17)(-29.279,-17.000){2}{\rule{0.415pt}{0.400pt}}
\multiput(309.77,210.92)(-0.853,-0.493){23}{\rule{0.777pt}{0.119pt}}
\multiput(311.39,211.17)(-20.387,-13.000){2}{\rule{0.388pt}{0.400pt}}
\multiput(287.87,197.92)(-0.826,-0.492){19}{\rule{0.755pt}{0.118pt}}
\multiput(289.43,198.17)(-16.434,-11.000){2}{\rule{0.377pt}{0.400pt}}
\multiput(269.68,186.93)(-0.890,-0.488){13}{\rule{0.800pt}{0.117pt}}
\multiput(271.34,187.17)(-12.340,-8.000){2}{\rule{0.400pt}{0.400pt}}
\multiput(255.26,178.93)(-1.033,-0.482){9}{\rule{0.900pt}{0.116pt}}
\multiput(257.13,179.17)(-10.132,-6.000){2}{\rule{0.450pt}{0.400pt}}
\multiput(244.09,172.93)(-0.762,-0.482){9}{\rule{0.700pt}{0.116pt}}
\multiput(245.55,173.17)(-7.547,-6.000){2}{\rule{0.350pt}{0.400pt}}
\multiput(234.26,166.94)(-1.066,-0.468){5}{\rule{0.900pt}{0.113pt}}
\multiput(236.13,167.17)(-6.132,-4.000){2}{\rule{0.450pt}{0.400pt}}
\multiput(227.09,162.94)(-0.774,-0.468){5}{\rule{0.700pt}{0.113pt}}
\multiput(228.55,163.17)(-4.547,-4.000){2}{\rule{0.350pt}{0.400pt}}
\multiput(220.26,158.95)(-1.132,-0.447){3}{\rule{0.900pt}{0.108pt}}
\multiput(222.13,159.17)(-4.132,-3.000){2}{\rule{0.450pt}{0.400pt}}
\put(213,155.17){\rule{1.100pt}{0.400pt}}
\multiput(215.72,156.17)(-2.717,-2.000){2}{\rule{0.550pt}{0.400pt}}
\multiput(210.37,153.95)(-0.685,-0.447){3}{\rule{0.633pt}{0.108pt}}
\multiput(211.69,154.17)(-2.685,-3.000){2}{\rule{0.317pt}{0.400pt}}
\put(206,150.17){\rule{0.700pt}{0.400pt}}
\multiput(207.55,151.17)(-1.547,-2.000){2}{\rule{0.350pt}{0.400pt}}
\put(203,148.67){\rule{0.723pt}{0.400pt}}
\multiput(204.50,149.17)(-1.500,-1.000){2}{\rule{0.361pt}{0.400pt}}
\put(200,147.17){\rule{0.700pt}{0.400pt}}
\multiput(201.55,148.17)(-1.547,-2.000){2}{\rule{0.350pt}{0.400pt}}
\put(198,145.67){\rule{0.482pt}{0.400pt}}
\multiput(199.00,146.17)(-1.000,-1.000){2}{\rule{0.241pt}{0.400pt}}
\put(864,534){\makebox(0,0){$+$}}
\put(652,410){\makebox(0,0){$+$}}
\put(524,335){\makebox(0,0){$+$}}
\put(441,287){\makebox(0,0){$+$}}
\put(384,253){\makebox(0,0){$+$}}
\put(344,229){\makebox(0,0){$+$}}
\put(313,212){\makebox(0,0){$+$}}
\put(291,199){\makebox(0,0){$+$}}
\put(273,188){\makebox(0,0){$+$}}
\put(259,180){\makebox(0,0){$+$}}
\put(247,174){\makebox(0,0){$+$}}
\put(238,168){\makebox(0,0){$+$}}
\put(230,164){\makebox(0,0){$+$}}
\put(224,160){\makebox(0,0){$+$}}
\put(218,157){\makebox(0,0){$+$}}
\put(213,155){\makebox(0,0){$+$}}
\put(209,152){\makebox(0,0){$+$}}
\put(206,150){\makebox(0,0){$+$}}
\put(203,149){\makebox(0,0){$+$}}
\put(200,147){\makebox(0,0){$+$}}
\put(198,146){\makebox(0,0){$+$}}
\put(420,639){\makebox(0,0){$+$}}
\put(370.0,639.0){\rule[-0.200pt]{24.090pt}{0.400pt}}
\put(350,598){\makebox(0,0)[r]{$\mu _n (1)$}}
\multiput(895.94,581.92)(-0.799,-0.496){41}{\rule{0.736pt}{0.120pt}}
\multiput(897.47,582.17)(-33.472,-22.000){2}{\rule{0.368pt}{0.400pt}}
\multiput(860.98,559.92)(-0.786,-0.499){267}{\rule{0.728pt}{0.120pt}}
\multiput(862.49,560.17)(-210.489,-135.000){2}{\rule{0.364pt}{0.400pt}}
\multiput(648.96,424.92)(-0.791,-0.499){159}{\rule{0.732pt}{0.120pt}}
\multiput(650.48,425.17)(-126.480,-81.000){2}{\rule{0.366pt}{0.400pt}}
\multiput(520.93,343.92)(-0.799,-0.498){101}{\rule{0.738pt}{0.120pt}}
\multiput(522.47,344.17)(-81.467,-52.000){2}{\rule{0.369pt}{0.400pt}}
\multiput(437.88,291.92)(-0.816,-0.498){67}{\rule{0.751pt}{0.120pt}}
\multiput(439.44,292.17)(-55.440,-35.000){2}{\rule{0.376pt}{0.400pt}}
\multiput(380.93,256.92)(-0.803,-0.497){47}{\rule{0.740pt}{0.120pt}}
\multiput(382.46,257.17)(-38.464,-25.000){2}{\rule{0.370pt}{0.400pt}}
\multiput(340.88,231.92)(-0.820,-0.495){35}{\rule{0.753pt}{0.119pt}}
\multiput(342.44,232.17)(-29.438,-19.000){2}{\rule{0.376pt}{0.400pt}}
\multiput(309.77,212.92)(-0.853,-0.493){23}{\rule{0.777pt}{0.119pt}}
\multiput(311.39,213.17)(-20.387,-13.000){2}{\rule{0.388pt}{0.400pt}}
\multiput(287.87,199.92)(-0.826,-0.492){19}{\rule{0.755pt}{0.118pt}}
\multiput(289.43,200.17)(-16.434,-11.000){2}{\rule{0.377pt}{0.400pt}}
\multiput(270.00,188.93)(-0.786,-0.489){15}{\rule{0.722pt}{0.118pt}}
\multiput(271.50,189.17)(-12.501,-9.000){2}{\rule{0.361pt}{0.400pt}}
\multiput(255.26,179.93)(-1.033,-0.482){9}{\rule{0.900pt}{0.116pt}}
\multiput(257.13,180.17)(-10.132,-6.000){2}{\rule{0.450pt}{0.400pt}}
\multiput(244.09,173.93)(-0.762,-0.482){9}{\rule{0.700pt}{0.116pt}}
\multiput(245.55,174.17)(-7.547,-6.000){2}{\rule{0.350pt}{0.400pt}}
\multiput(234.26,167.94)(-1.066,-0.468){5}{\rule{0.900pt}{0.113pt}}
\multiput(236.13,168.17)(-6.132,-4.000){2}{\rule{0.450pt}{0.400pt}}
\multiput(227.09,163.94)(-0.774,-0.468){5}{\rule{0.700pt}{0.113pt}}
\multiput(228.55,164.17)(-4.547,-4.000){2}{\rule{0.350pt}{0.400pt}}
\multiput(220.26,159.95)(-1.132,-0.447){3}{\rule{0.900pt}{0.108pt}}
\multiput(222.13,160.17)(-4.132,-3.000){2}{\rule{0.450pt}{0.400pt}}
\multiput(214.82,156.95)(-0.909,-0.447){3}{\rule{0.767pt}{0.108pt}}
\multiput(216.41,157.17)(-3.409,-3.000){2}{\rule{0.383pt}{0.400pt}}
\put(209,153.17){\rule{0.900pt}{0.400pt}}
\multiput(211.13,154.17)(-2.132,-2.000){2}{\rule{0.450pt}{0.400pt}}
\put(206,151.17){\rule{0.700pt}{0.400pt}}
\multiput(207.55,152.17)(-1.547,-2.000){2}{\rule{0.350pt}{0.400pt}}
\put(203,149.17){\rule{0.700pt}{0.400pt}}
\multiput(204.55,150.17)(-1.547,-2.000){2}{\rule{0.350pt}{0.400pt}}
\put(200,147.67){\rule{0.723pt}{0.400pt}}
\multiput(201.50,148.17)(-1.500,-1.000){2}{\rule{0.361pt}{0.400pt}}
\put(198,146.17){\rule{0.482pt}{0.400pt}}
\multiput(199.00,147.17)(-1.000,-2.000){2}{\rule{0.241pt}{0.400pt}}
\put(864,561){\makebox(0,0){$\times$}}
\put(652,426){\makebox(0,0){$\times$}}
\put(524,345){\makebox(0,0){$\times$}}
\put(441,293){\makebox(0,0){$\times$}}
\put(384,258){\makebox(0,0){$\times$}}
\put(344,233){\makebox(0,0){$\times$}}
\put(313,214){\makebox(0,0){$\times$}}
\put(291,201){\makebox(0,0){$\times$}}
\put(273,190){\makebox(0,0){$\times$}}
\put(259,181){\makebox(0,0){$\times$}}
\put(247,175){\makebox(0,0){$\times$}}
\put(238,169){\makebox(0,0){$\times$}}
\put(230,165){\makebox(0,0){$\times$}}
\put(224,161){\makebox(0,0){$\times$}}
\put(218,158){\makebox(0,0){$\times$}}
\put(213,155){\makebox(0,0){$\times$}}
\put(209,153){\makebox(0,0){$\times$}}
\put(206,151){\makebox(0,0){$\times$}}
\put(203,149){\makebox(0,0){$\times$}}
\put(200,148){\makebox(0,0){$\times$}}
\put(198,146){\makebox(0,0){$\times$}}
\put(420,598){\makebox(0,0){$\times$}}
\put(370.0,598.0){\rule[-0.200pt]{24.090pt}{0.400pt}}
\sbox{\plotpoint}{\rule[-0.400pt]{0.800pt}{0.800pt}}%
\sbox{\plotpoint}{\rule[-0.200pt]{0.400pt}{0.400pt}}%
\put(350,557){\makebox(0,0)[r]{$\mu _n (10)$}}
\sbox{\plotpoint}{\rule[-0.400pt]{0.800pt}{0.800pt}}%
\sbox{\plotpoint}{\rule[-0.200pt]{0.400pt}{0.400pt}}%
\multiput(896.43,646.92)(-0.649,-0.497){51}{\rule{0.619pt}{0.120pt}}
\multiput(897.72,647.17)(-33.716,-27.000){2}{\rule{0.309pt}{0.400pt}}
\multiput(861.40,619.92)(-0.658,-0.499){319}{\rule{0.627pt}{0.120pt}}
\multiput(862.70,620.17)(-210.699,-161.000){2}{\rule{0.313pt}{0.400pt}}
\multiput(649.32,458.92)(-0.681,-0.499){185}{\rule{0.645pt}{0.120pt}}
\multiput(650.66,459.17)(-126.662,-94.000){2}{\rule{0.322pt}{0.400pt}}
\multiput(521.25,364.92)(-0.704,-0.499){115}{\rule{0.663pt}{0.120pt}}
\multiput(522.62,365.17)(-81.625,-59.000){2}{\rule{0.331pt}{0.400pt}}
\multiput(438.22,305.92)(-0.713,-0.498){77}{\rule{0.670pt}{0.120pt}}
\multiput(439.61,306.17)(-55.609,-40.000){2}{\rule{0.335pt}{0.400pt}}
\multiput(381.21,265.92)(-0.716,-0.497){53}{\rule{0.671pt}{0.120pt}}
\multiput(382.61,266.17)(-38.606,-28.000){2}{\rule{0.336pt}{0.400pt}}
\multiput(341.01,237.92)(-0.778,-0.496){37}{\rule{0.720pt}{0.119pt}}
\multiput(342.51,238.17)(-29.506,-20.000){2}{\rule{0.360pt}{0.400pt}}
\multiput(310.15,217.92)(-0.737,-0.494){27}{\rule{0.687pt}{0.119pt}}
\multiput(311.57,218.17)(-20.575,-15.000){2}{\rule{0.343pt}{0.400pt}}
\multiput(287.87,202.92)(-0.826,-0.492){19}{\rule{0.755pt}{0.118pt}}
\multiput(289.43,203.17)(-16.434,-11.000){2}{\rule{0.377pt}{0.400pt}}
\multiput(270.00,191.93)(-0.786,-0.489){15}{\rule{0.722pt}{0.118pt}}
\multiput(271.50,192.17)(-12.501,-9.000){2}{\rule{0.361pt}{0.400pt}}
\multiput(255.74,182.93)(-0.874,-0.485){11}{\rule{0.786pt}{0.117pt}}
\multiput(257.37,183.17)(-10.369,-7.000){2}{\rule{0.393pt}{0.400pt}}
\multiput(244.09,175.93)(-0.762,-0.482){9}{\rule{0.700pt}{0.116pt}}
\multiput(245.55,176.17)(-7.547,-6.000){2}{\rule{0.350pt}{0.400pt}}
\multiput(234.93,169.93)(-0.821,-0.477){7}{\rule{0.740pt}{0.115pt}}
\multiput(236.46,170.17)(-6.464,-5.000){2}{\rule{0.370pt}{0.400pt}}
\multiput(227.09,164.94)(-0.774,-0.468){5}{\rule{0.700pt}{0.113pt}}
\multiput(228.55,165.17)(-4.547,-4.000){2}{\rule{0.350pt}{0.400pt}}
\multiput(220.26,160.95)(-1.132,-0.447){3}{\rule{0.900pt}{0.108pt}}
\multiput(222.13,161.17)(-4.132,-3.000){2}{\rule{0.450pt}{0.400pt}}
\multiput(214.82,157.95)(-0.909,-0.447){3}{\rule{0.767pt}{0.108pt}}
\multiput(216.41,158.17)(-3.409,-3.000){2}{\rule{0.383pt}{0.400pt}}
\multiput(210.37,154.95)(-0.685,-0.447){3}{\rule{0.633pt}{0.108pt}}
\multiput(211.69,155.17)(-2.685,-3.000){2}{\rule{0.317pt}{0.400pt}}
\put(206,151.17){\rule{0.700pt}{0.400pt}}
\multiput(207.55,152.17)(-1.547,-2.000){2}{\rule{0.350pt}{0.400pt}}
\put(203,149.67){\rule{0.723pt}{0.400pt}}
\multiput(204.50,150.17)(-1.500,-1.000){2}{\rule{0.361pt}{0.400pt}}
\put(200,148.17){\rule{0.700pt}{0.400pt}}
\multiput(201.55,149.17)(-1.547,-2.000){2}{\rule{0.350pt}{0.400pt}}
\put(198,146.67){\rule{0.482pt}{0.400pt}}
\multiput(199.00,147.17)(-1.000,-1.000){2}{\rule{0.241pt}{0.400pt}}
\put(864,621){\makebox(0,0){$\ast$}}
\put(652,460){\makebox(0,0){$\ast$}}
\put(524,366){\makebox(0,0){$\ast$}}
\put(441,307){\makebox(0,0){$\ast$}}
\put(384,267){\makebox(0,0){$\ast$}}
\put(344,239){\makebox(0,0){$\ast$}}
\put(313,219){\makebox(0,0){$\ast$}}
\put(291,204){\makebox(0,0){$\ast$}}
\put(273,193){\makebox(0,0){$\ast$}}
\put(259,184){\makebox(0,0){$\ast$}}
\put(247,177){\makebox(0,0){$\ast$}}
\put(238,171){\makebox(0,0){$\ast$}}
\put(230,166){\makebox(0,0){$\ast$}}
\put(224,162){\makebox(0,0){$\ast$}}
\put(218,159){\makebox(0,0){$\ast$}}
\put(213,156){\makebox(0,0){$\ast$}}
\put(209,153){\makebox(0,0){$\ast$}}
\put(206,151){\makebox(0,0){$\ast$}}
\put(203,150){\makebox(0,0){$\ast$}}
\put(200,148){\makebox(0,0){$\ast$}}
\put(198,147){\makebox(0,0){$\ast$}}
\put(420,557){\makebox(0,0){$\ast$}}
\put(370.0,557.0){\rule[-0.200pt]{24.090pt}{0.400pt}}
\put(170.0,131.0){\rule[-0.200pt]{0.400pt}{132.013pt}}
\put(170.0,131.0){\rule[-0.200pt]{175.616pt}{0.400pt}}
\put(899.0,131.0){\rule[-0.200pt]{0.400pt}{132.013pt}}
\put(170.0,679.0){\rule[-0.200pt]{175.616pt}{0.400pt}}
\end{picture}

%% file: rv2.tex
\setlength{\unitlength}{0.240900pt}
\ifx\plotpoint\undefined\newsavebox{\plotpoint}\fi
\sbox{\plotpoint}{\rule[-0.200pt]{0.400pt}{0.400pt}}%
\begin{picture}(1500,675)(0,0)
\sbox{\plotpoint}{\rule[-0.200pt]{0.400pt}{0.400pt}}%
\put(211.0,131.0){\rule[-0.200pt]{4.818pt}{0.400pt}}
\put(191,131){\makebox(0,0)[r]{ 0.2}}
\put(459.0,131.0){\rule[-0.200pt]{4.818pt}{0.400pt}}
\put(211.0,262.0){\rule[-0.200pt]{4.818pt}{0.400pt}}
\put(191,262){\makebox(0,0)[r]{ 0.201}}
\put(459.0,262.0){\rule[-0.200pt]{4.818pt}{0.400pt}}
\put(211.0,394.0){\rule[-0.200pt]{4.818pt}{0.400pt}}
\put(191,394){\makebox(0,0)[r]{ 0.202}}
\put(459.0,394.0){\rule[-0.200pt]{4.818pt}{0.400pt}}
\put(211.0,525.0){\rule[-0.200pt]{4.818pt}{0.400pt}}
\put(191,525){\makebox(0,0)[r]{ 0.203}}
\put(459.0,525.0){\rule[-0.200pt]{4.818pt}{0.400pt}}
\put(211.0,131.0){\rule[-0.200pt]{0.400pt}{4.818pt}}
\put(211,90){\makebox(0,0){ 0}}
\put(211.0,531.0){\rule[-0.200pt]{0.400pt}{4.818pt}}
\put(318.0,131.0){\rule[-0.200pt]{0.400pt}{4.818pt}}
\put(318,90){\makebox(0,0){ 0.0002}}
\put(318.0,531.0){\rule[-0.200pt]{0.400pt}{4.818pt}}
\put(425.0,131.0){\rule[-0.200pt]{0.400pt}{4.818pt}}
\put(425,90){\makebox(0,0){ 0.0004}}
\put(425.0,531.0){\rule[-0.200pt]{0.400pt}{4.818pt}}
\put(211.0,131.0){\rule[-0.200pt]{0.400pt}{101.178pt}}
\put(211.0,131.0){\rule[-0.200pt]{64.561pt}{0.400pt}}
\put(479.0,131.0){\rule[-0.200pt]{0.400pt}{101.178pt}}
\put(211.0,551.0){\rule[-0.200pt]{64.561pt}{0.400pt}}
\put(30,341){\makebox(0,0){$\lambda _n$}}
\put(345,29){\makebox(0,0){$ 1/n ^2 $}}
\put(345,613){\makebox(0,0){$a=1$}}
\put(425,517){\makebox(0,0){$+$}}
\put(360,399){\makebox(0,0){$+$}}
\put(320,327){\makebox(0,0){$+$}}
\put(295,280){\makebox(0,0){$+$}}
\put(277,248){\makebox(0,0){$+$}}
\put(265,225){\makebox(0,0){$+$}}
\put(255,209){\makebox(0,0){$+$}}
\put(248,196){\makebox(0,0){$+$}}
\put(243,186){\makebox(0,0){$+$}}
\put(238,178){\makebox(0,0){$+$}}
\put(235,172){\makebox(0,0){$+$}}
\put(232,167){\makebox(0,0){$+$}}
\put(230,163){\makebox(0,0){$+$}}
\put(228,159){\makebox(0,0){$+$}}
\put(226,156){\makebox(0,0){$+$}}
\put(224,154){\makebox(0,0){$+$}}
\put(211.0,131.0){\rule[-0.200pt]{0.400pt}{101.178pt}}
\put(211.0,131.0){\rule[-0.200pt]{64.561pt}{0.400pt}}
\put(479.0,131.0){\rule[-0.200pt]{0.400pt}{101.178pt}}
\put(211.0,551.0){\rule[-0.200pt]{64.561pt}{0.400pt}}
\put(617.0,210.0){\rule[-0.200pt]{4.818pt}{0.400pt}}
\put(597,210){\makebox(0,0)[r]{ 0.428}}
\put(901.0,210.0){\rule[-0.200pt]{4.818pt}{0.400pt}}
\put(617.0,369.0){\rule[-0.200pt]{4.818pt}{0.400pt}}
\put(597,369){\makebox(0,0)[r]{ 0.4284}}
\put(901.0,369.0){\rule[-0.200pt]{4.818pt}{0.400pt}}
\put(617.0,527.0){\rule[-0.200pt]{4.818pt}{0.400pt}}
\put(597,527){\makebox(0,0)[r]{ 0.4288}}
\put(901.0,527.0){\rule[-0.200pt]{4.818pt}{0.400pt}}
\put(617.0,131.0){\rule[-0.200pt]{0.400pt}{4.818pt}}
\put(617,90){\makebox(0,0){ 0}}
\put(617.0,531.0){\rule[-0.200pt]{0.400pt}{4.818pt}}
\put(739.0,131.0){\rule[-0.200pt]{0.400pt}{4.818pt}}
\put(739,90){\makebox(0,0){ 0.0002}}
\put(739.0,531.0){\rule[-0.200pt]{0.400pt}{4.818pt}}
\put(860.0,131.0){\rule[-0.200pt]{0.400pt}{4.818pt}}
\put(860,90){\makebox(0,0){ 0.0004}}
\put(860.0,531.0){\rule[-0.200pt]{0.400pt}{4.818pt}}
\put(617.0,131.0){\rule[-0.200pt]{0.400pt}{101.178pt}}
\put(617.0,131.0){\rule[-0.200pt]{73.234pt}{0.400pt}}
\put(921.0,131.0){\rule[-0.200pt]{0.400pt}{101.178pt}}
\put(617.0,551.0){\rule[-0.200pt]{73.234pt}{0.400pt}}
\put(769,29){\makebox(0,0){$ 1/n ^2 $}}
\put(769,613){\makebox(0,0){$a=5$}}
\put(860,525){\makebox(0,0){$+$}}
\put(786,396){\makebox(0,0){$+$}}
\put(741,320){\makebox(0,0){$+$}}
\put(712,273){\makebox(0,0){$+$}}
\put(692,241){\makebox(0,0){$+$}}
\put(678,218){\makebox(0,0){$+$}}
\put(667,202){\makebox(0,0){$+$}}
\put(659,190){\makebox(0,0){$+$}}
\put(653,181){\makebox(0,0){$+$}}
\put(648,174){\makebox(0,0){$+$}}
\put(644,168){\makebox(0,0){$+$}}
\put(641,163){\makebox(0,0){$+$}}
\put(638,159){\makebox(0,0){$+$}}
\put(636,156){\makebox(0,0){$+$}}
\put(634,153){\makebox(0,0){$+$}}
\put(632,151){\makebox(0,0){$+$}}
\put(617.0,131.0){\rule[-0.200pt]{0.400pt}{101.178pt}}
\put(617.0,131.0){\rule[-0.200pt]{73.234pt}{0.400pt}}
\put(921.0,131.0){\rule[-0.200pt]{0.400pt}{101.178pt}}
\put(617.0,551.0){\rule[-0.200pt]{73.234pt}{0.400pt}}
\put(1060.0,155.0){\rule[-0.200pt]{4.818pt}{0.400pt}}
\put(1040,155){\makebox(0,0)[r]{ 0.4638}}
\put(1344.0,155.0){\rule[-0.200pt]{4.818pt}{0.400pt}}
\put(1060.0,275.0){\rule[-0.200pt]{4.818pt}{0.400pt}}
\put(1040,275){\makebox(0,0)[r]{ 0.464}}
\put(1344.0,275.0){\rule[-0.200pt]{4.818pt}{0.400pt}}
\put(1060.0,395.0){\rule[-0.200pt]{4.818pt}{0.400pt}}
\put(1040,395){\makebox(0,0)[r]{ 0.4642}}
\put(1344.0,395.0){\rule[-0.200pt]{4.818pt}{0.400pt}}
\put(1060.0,515.0){\rule[-0.200pt]{4.818pt}{0.400pt}}
\put(1040,515){\makebox(0,0)[r]{ 0.4644}}
\put(1344.0,515.0){\rule[-0.200pt]{4.818pt}{0.400pt}}
\put(1060.0,131.0){\rule[-0.200pt]{0.400pt}{4.818pt}}
\put(1060,90){\makebox(0,0){ 0}}
\put(1060.0,531.0){\rule[-0.200pt]{0.400pt}{4.818pt}}
\put(1182.0,131.0){\rule[-0.200pt]{0.400pt}{4.818pt}}
\put(1182,90){\makebox(0,0){ 0.0002}}
\put(1182.0,531.0){\rule[-0.200pt]{0.400pt}{4.818pt}}
\put(1303.0,131.0){\rule[-0.200pt]{0.400pt}{4.818pt}}
\put(1303,90){\makebox(0,0){ 0.0004}}
\put(1303.0,531.0){\rule[-0.200pt]{0.400pt}{4.818pt}}
\put(1060.0,131.0){\rule[-0.200pt]{0.400pt}{101.178pt}}
\put(1060.0,131.0){\rule[-0.200pt]{73.234pt}{0.400pt}}
\put(1364.0,131.0){\rule[-0.200pt]{0.400pt}{101.178pt}}
\put(1060.0,551.0){\rule[-0.200pt]{73.234pt}{0.400pt}}
\put(1212,29){\makebox(0,0){$ 1/n ^2 $}}
\put(1212,613){\makebox(0,0){$a=10$}}
\put(1303,475){\makebox(0,0){$+$}}
\put(1229,362){\makebox(0,0){$+$}}
\put(1184,298){\makebox(0,0){$+$}}
\put(1155,258){\makebox(0,0){$+$}}
\put(1135,231){\makebox(0,0){$+$}}
\put(1121,213){\makebox(0,0){$+$}}
\put(1110,200){\makebox(0,0){$+$}}
\put(1102,190){\makebox(0,0){$+$}}
\put(1096,182){\makebox(0,0){$+$}}
\put(1091,177){\makebox(0,0){$+$}}
\put(1087,172){\makebox(0,0){$+$}}
\put(1084,168){\makebox(0,0){$+$}}
\put(1081,165){\makebox(0,0){$+$}}
\put(1079,162){\makebox(0,0){$+$}}
\put(1077,160){\makebox(0,0){$+$}}
\put(1075,158){\makebox(0,0){$+$}}
\put(1060.0,131.0){\rule[-0.200pt]{0.400pt}{101.178pt}}
\put(1060.0,131.0){\rule[-0.200pt]{73.234pt}{0.400pt}}
\put(1364.0,131.0){\rule[-0.200pt]{0.400pt}{101.178pt}}
\put(1060.0,551.0){\rule[-0.200pt]{73.234pt}{0.400pt}}
\end{picture}

%% file: desordrenonrel.tex
\setlength{\unitlength}{0.240900pt}
\ifx\plotpoint\undefined\newsavebox{\plotpoint}\fi
\sbox{\plotpoint}{\rule[-0.200pt]{0.400pt}{0.400pt}}%
\begin{picture}(1500,720)(0,0)
\sbox{\plotpoint}{\rule[-0.200pt]{0.400pt}{0.400pt}}%
\put(770.0,167.0){\rule[-0.200pt]{4.818pt}{0.400pt}}
\put(750,167){\makebox(0,0)[r]{-16}}
\put(1299.0,167.0){\rule[-0.200pt]{4.818pt}{0.400pt}}
\put(770.0,238.0){\rule[-0.200pt]{4.818pt}{0.400pt}}
\put(750,238){\makebox(0,0)[r]{-14}}
\put(1299.0,238.0){\rule[-0.200pt]{4.818pt}{0.400pt}}
\put(770.0,310.0){\rule[-0.200pt]{4.818pt}{0.400pt}}
\put(750,310){\makebox(0,0)[r]{-12}}
\put(1299.0,310.0){\rule[-0.200pt]{4.818pt}{0.400pt}}
\put(770.0,381.0){\rule[-0.200pt]{4.818pt}{0.400pt}}
\put(750,381){\makebox(0,0)[r]{-10}}
\put(1299.0,381.0){\rule[-0.200pt]{4.818pt}{0.400pt}}
\put(770.0,453.0){\rule[-0.200pt]{4.818pt}{0.400pt}}
\put(750,453){\makebox(0,0)[r]{-8}}
\put(1299.0,453.0){\rule[-0.200pt]{4.818pt}{0.400pt}}
\put(770.0,524.0){\rule[-0.200pt]{4.818pt}{0.400pt}}
\put(750,524){\makebox(0,0)[r]{-6}}
\put(1299.0,524.0){\rule[-0.200pt]{4.818pt}{0.400pt}}
\put(770.0,596.0){\rule[-0.200pt]{4.818pt}{0.400pt}}
\put(750,596){\makebox(0,0)[r]{-4}}
\put(1299.0,596.0){\rule[-0.200pt]{4.818pt}{0.400pt}}
\put(770.0,131.0){\rule[-0.200pt]{0.400pt}{4.818pt}}
\put(770,90){\makebox(0,0){-7}}
\put(770.0,576.0){\rule[-0.200pt]{0.400pt}{4.818pt}}
\put(870.0,131.0){\rule[-0.200pt]{0.400pt}{4.818pt}}
\put(870,90){\makebox(0,0){-6}}
\put(870.0,576.0){\rule[-0.200pt]{0.400pt}{4.818pt}}
\put(970.0,131.0){\rule[-0.200pt]{0.400pt}{4.818pt}}
\put(970,90){\makebox(0,0){-5}}
\put(970.0,576.0){\rule[-0.200pt]{0.400pt}{4.818pt}}
\put(1069.0,131.0){\rule[-0.200pt]{0.400pt}{4.818pt}}
\put(1069,90){\makebox(0,0){-4}}
\put(1069.0,576.0){\rule[-0.200pt]{0.400pt}{4.818pt}}
\put(1169.0,131.0){\rule[-0.200pt]{0.400pt}{4.818pt}}
\put(1169,90){\makebox(0,0){-3}}
\put(1169.0,576.0){\rule[-0.200pt]{0.400pt}{4.818pt}}
\put(1269.0,131.0){\rule[-0.200pt]{0.400pt}{4.818pt}}
\put(1269,90){\makebox(0,0){-2}}
\put(1269.0,576.0){\rule[-0.200pt]{0.400pt}{4.818pt}}
\put(770.0,131.0){\rule[-0.200pt]{0.400pt}{112.018pt}}
\put(770.0,131.0){\rule[-0.200pt]{132.254pt}{0.400pt}}
\put(1319.0,131.0){\rule[-0.200pt]{0.400pt}{112.018pt}}
\put(770.0,596.0){\rule[-0.200pt]{132.254pt}{0.400pt}}
\put(1044,29){\makebox(0,0){$ \log(\lambda -\lambda _c)$}}
\put(1044,658){\makebox(0,0){$\log \left( \dfrac{\left< X_n \right>}{2^n} \right) $ for $n=10$, $12$, $\dots$, $30$}}
\put(803,133){\rule{1pt}{1pt}}
\put(809,138){\rule{1pt}{1pt}}
\put(814,143){\rule{1pt}{1pt}}
\put(820,148){\rule{1pt}{1pt}}
\put(825,152){\rule{1pt}{1pt}}
\put(831,157){\rule{1pt}{1pt}}
\put(837,162){\rule{1pt}{1pt}}
\put(842,167){\rule{1pt}{1pt}}
\put(848,171){\rule{1pt}{1pt}}
\put(853,176){\rule{1pt}{1pt}}
\put(859,181){\rule{1pt}{1pt}}
\put(864,186){\rule{1pt}{1pt}}
\put(870,190){\rule{1pt}{1pt}}
\put(875,195){\rule{1pt}{1pt}}
\put(881,200){\rule{1pt}{1pt}}
\put(886,205){\rule{1pt}{1pt}}
\put(892,209){\rule{1pt}{1pt}}
\put(898,214){\rule{1pt}{1pt}}
\put(903,219){\rule{1pt}{1pt}}
\put(909,224){\rule{1pt}{1pt}}
\put(914,228){\rule{1pt}{1pt}}
\put(920,233){\rule{1pt}{1pt}}
\put(925,238){\rule{1pt}{1pt}}
\put(931,243){\rule{1pt}{1pt}}
\put(936,247){\rule{1pt}{1pt}}
\put(942,252){\rule{1pt}{1pt}}
\put(947,257){\rule{1pt}{1pt}}
\put(953,262){\rule{1pt}{1pt}}
\put(959,266){\rule{1pt}{1pt}}
\put(964,271){\rule{1pt}{1pt}}
\put(970,276){\rule{1pt}{1pt}}
\put(975,281){\rule{1pt}{1pt}}
\put(981,285){\rule{1pt}{1pt}}
\put(986,290){\rule{1pt}{1pt}}
\put(992,295){\rule{1pt}{1pt}}
\put(997,300){\rule{1pt}{1pt}}
\put(1003,304){\rule{1pt}{1pt}}
\put(1008,309){\rule{1pt}{1pt}}
\put(1014,314){\rule{1pt}{1pt}}
\put(1020,319){\rule{1pt}{1pt}}
\put(1025,323){\rule{1pt}{1pt}}
\put(1031,328){\rule{1pt}{1pt}}
\put(1036,333){\rule{1pt}{1pt}}
\put(1042,338){\rule{1pt}{1pt}}
\put(1047,342){\rule{1pt}{1pt}}
\put(1053,347){\rule{1pt}{1pt}}
\put(1058,352){\rule{1pt}{1pt}}
\put(1064,357){\rule{1pt}{1pt}}
\put(1069,361){\rule{1pt}{1pt}}
\put(1075,366){\rule{1pt}{1pt}}
\put(1081,371){\rule{1pt}{1pt}}
\put(1086,376){\rule{1pt}{1pt}}
\put(1092,380){\rule{1pt}{1pt}}
\put(1097,385){\rule{1pt}{1pt}}
\put(1103,390){\rule{1pt}{1pt}}
\put(1108,395){\rule{1pt}{1pt}}
\put(1114,399){\rule{1pt}{1pt}}
\put(1119,404){\rule{1pt}{1pt}}
\put(1125,409){\rule{1pt}{1pt}}
\put(1130,414){\rule{1pt}{1pt}}
\put(1136,418){\rule{1pt}{1pt}}
\put(1142,423){\rule{1pt}{1pt}}
\put(1147,428){\rule{1pt}{1pt}}
\put(1153,433){\rule{1pt}{1pt}}
\put(1158,437){\rule{1pt}{1pt}}
\put(1164,442){\rule{1pt}{1pt}}
\put(1169,447){\rule{1pt}{1pt}}
\put(1175,452){\rule{1pt}{1pt}}
\put(1180,456){\rule{1pt}{1pt}}
\put(1186,461){\rule{1pt}{1pt}}
\put(1191,466){\rule{1pt}{1pt}}
\put(1197,471){\rule{1pt}{1pt}}
\put(1203,475){\rule{1pt}{1pt}}
\put(1208,480){\rule{1pt}{1pt}}
\put(1214,485){\rule{1pt}{1pt}}
\put(1219,490){\rule{1pt}{1pt}}
\put(1225,494){\rule{1pt}{1pt}}
\put(1230,499){\rule{1pt}{1pt}}
\put(1236,504){\rule{1pt}{1pt}}
\put(1241,509){\rule{1pt}{1pt}}
\put(1247,513){\rule{1pt}{1pt}}
\put(1252,518){\rule{1pt}{1pt}}
\put(1258,523){\rule{1pt}{1pt}}
\put(1264,528){\rule{1pt}{1pt}}
\put(1269,532){\rule{1pt}{1pt}}
\put(1275,537){\rule{1pt}{1pt}}
\put(1280,542){\rule{1pt}{1pt}}
\put(1286,547){\rule{1pt}{1pt}}
\put(1291,551){\rule{1pt}{1pt}}
\put(1297,556){\rule{1pt}{1pt}}
\put(1302,561){\rule{1pt}{1pt}}
\put(1308,566){\rule{1pt}{1pt}}
\put(1313,570){\rule{1pt}{1pt}}
\put(1319,575){\rule{1pt}{1pt}}
\put(910,207){\usebox{\plotpoint}}
\multiput(910.58,207.00)(0.491,5.551){17}{\rule{0.118pt}{4.380pt}}
\multiput(909.17,207.00)(10.000,97.909){2}{\rule{0.400pt}{2.190pt}}
\multiput(920.58,314.00)(0.491,1.277){17}{\rule{0.118pt}{1.100pt}}
\multiput(919.17,314.00)(10.000,22.717){2}{\rule{0.400pt}{0.550pt}}
\multiput(930.58,339.00)(0.491,0.860){17}{\rule{0.118pt}{0.780pt}}
\multiput(929.17,339.00)(10.000,15.381){2}{\rule{0.400pt}{0.390pt}}
\multiput(940.58,356.00)(0.491,0.600){17}{\rule{0.118pt}{0.580pt}}
\multiput(939.17,356.00)(10.000,10.796){2}{\rule{0.400pt}{0.290pt}}
\multiput(950.00,368.58)(0.495,0.491){17}{\rule{0.500pt}{0.118pt}}
\multiput(950.00,367.17)(8.962,10.000){2}{\rule{0.250pt}{0.400pt}}
\multiput(960.00,378.59)(0.553,0.489){15}{\rule{0.544pt}{0.118pt}}
\multiput(960.00,377.17)(8.870,9.000){2}{\rule{0.272pt}{0.400pt}}
\multiput(970.00,387.59)(0.626,0.488){13}{\rule{0.600pt}{0.117pt}}
\multiput(970.00,386.17)(8.755,8.000){2}{\rule{0.300pt}{0.400pt}}
\multiput(980.00,395.59)(0.721,0.485){11}{\rule{0.671pt}{0.117pt}}
\multiput(980.00,394.17)(8.606,7.000){2}{\rule{0.336pt}{0.400pt}}
\multiput(990.00,402.59)(0.852,0.482){9}{\rule{0.767pt}{0.116pt}}
\multiput(990.00,401.17)(8.409,6.000){2}{\rule{0.383pt}{0.400pt}}
\multiput(1000.00,408.59)(0.721,0.485){11}{\rule{0.671pt}{0.117pt}}
\multiput(1000.00,407.17)(8.606,7.000){2}{\rule{0.336pt}{0.400pt}}
\multiput(1010.00,415.59)(0.852,0.482){9}{\rule{0.767pt}{0.116pt}}
\multiput(1010.00,414.17)(8.409,6.000){2}{\rule{0.383pt}{0.400pt}}
\multiput(1020.00,421.59)(0.852,0.482){9}{\rule{0.767pt}{0.116pt}}
\multiput(1020.00,420.17)(8.409,6.000){2}{\rule{0.383pt}{0.400pt}}
\multiput(1030.00,427.59)(1.044,0.477){7}{\rule{0.900pt}{0.115pt}}
\multiput(1030.00,426.17)(8.132,5.000){2}{\rule{0.450pt}{0.400pt}}
\multiput(1040.00,432.59)(0.762,0.482){9}{\rule{0.700pt}{0.116pt}}
\multiput(1040.00,431.17)(7.547,6.000){2}{\rule{0.350pt}{0.400pt}}
\multiput(1049.00,438.59)(1.044,0.477){7}{\rule{0.900pt}{0.115pt}}
\multiput(1049.00,437.17)(8.132,5.000){2}{\rule{0.450pt}{0.400pt}}
\multiput(1059.00,443.59)(0.852,0.482){9}{\rule{0.767pt}{0.116pt}}
\multiput(1059.00,442.17)(8.409,6.000){2}{\rule{0.383pt}{0.400pt}}
\multiput(1069.00,449.59)(1.044,0.477){7}{\rule{0.900pt}{0.115pt}}
\multiput(1069.00,448.17)(8.132,5.000){2}{\rule{0.450pt}{0.400pt}}
\multiput(1079.00,454.59)(0.852,0.482){9}{\rule{0.767pt}{0.116pt}}
\multiput(1079.00,453.17)(8.409,6.000){2}{\rule{0.383pt}{0.400pt}}
\multiput(1089.00,460.59)(1.044,0.477){7}{\rule{0.900pt}{0.115pt}}
\multiput(1089.00,459.17)(8.132,5.000){2}{\rule{0.450pt}{0.400pt}}
\multiput(1099.00,465.59)(0.852,0.482){9}{\rule{0.767pt}{0.116pt}}
\multiput(1099.00,464.17)(8.409,6.000){2}{\rule{0.383pt}{0.400pt}}
\multiput(1109.00,471.59)(1.044,0.477){7}{\rule{0.900pt}{0.115pt}}
\multiput(1109.00,470.17)(8.132,5.000){2}{\rule{0.450pt}{0.400pt}}
\multiput(1119.00,476.59)(0.852,0.482){9}{\rule{0.767pt}{0.116pt}}
\multiput(1119.00,475.17)(8.409,6.000){2}{\rule{0.383pt}{0.400pt}}
\multiput(1129.00,482.59)(0.852,0.482){9}{\rule{0.767pt}{0.116pt}}
\multiput(1129.00,481.17)(8.409,6.000){2}{\rule{0.383pt}{0.400pt}}
\multiput(1139.00,488.59)(0.852,0.482){9}{\rule{0.767pt}{0.116pt}}
\multiput(1139.00,487.17)(8.409,6.000){2}{\rule{0.383pt}{0.400pt}}
\multiput(1149.00,494.59)(0.852,0.482){9}{\rule{0.767pt}{0.116pt}}
\multiput(1149.00,493.17)(8.409,6.000){2}{\rule{0.383pt}{0.400pt}}
\multiput(1159.00,500.59)(0.852,0.482){9}{\rule{0.767pt}{0.116pt}}
\multiput(1159.00,499.17)(8.409,6.000){2}{\rule{0.383pt}{0.400pt}}
\multiput(1169.00,506.59)(0.721,0.485){11}{\rule{0.671pt}{0.117pt}}
\multiput(1169.00,505.17)(8.606,7.000){2}{\rule{0.336pt}{0.400pt}}
\multiput(1179.00,513.59)(0.852,0.482){9}{\rule{0.767pt}{0.116pt}}
\multiput(1179.00,512.17)(8.409,6.000){2}{\rule{0.383pt}{0.400pt}}
\multiput(1189.00,519.59)(0.721,0.485){11}{\rule{0.671pt}{0.117pt}}
\multiput(1189.00,518.17)(8.606,7.000){2}{\rule{0.336pt}{0.400pt}}
\multiput(1199.00,526.59)(0.721,0.485){11}{\rule{0.671pt}{0.117pt}}
\multiput(1199.00,525.17)(8.606,7.000){2}{\rule{0.336pt}{0.400pt}}
\multiput(1209.00,533.59)(0.721,0.485){11}{\rule{0.671pt}{0.117pt}}
\multiput(1209.00,532.17)(8.606,7.000){2}{\rule{0.336pt}{0.400pt}}
\multiput(1219.00,540.59)(0.721,0.485){11}{\rule{0.671pt}{0.117pt}}
\multiput(1219.00,539.17)(8.606,7.000){2}{\rule{0.336pt}{0.400pt}}
\multiput(1229.00,547.59)(0.721,0.485){11}{\rule{0.671pt}{0.117pt}}
\multiput(1229.00,546.17)(8.606,7.000){2}{\rule{0.336pt}{0.400pt}}
\multiput(1239.00,554.59)(0.626,0.488){13}{\rule{0.600pt}{0.117pt}}
\multiput(1239.00,553.17)(8.755,8.000){2}{\rule{0.300pt}{0.400pt}}
\multiput(1249.00,562.59)(0.721,0.485){11}{\rule{0.671pt}{0.117pt}}
\multiput(1249.00,561.17)(8.606,7.000){2}{\rule{0.336pt}{0.400pt}}
\multiput(1259.00,569.59)(0.721,0.485){11}{\rule{0.671pt}{0.117pt}}
\multiput(1259.00,568.17)(8.606,7.000){2}{\rule{0.336pt}{0.400pt}}
\multiput(1269.00,576.59)(0.626,0.488){13}{\rule{0.600pt}{0.117pt}}
\multiput(1269.00,575.17)(8.755,8.000){2}{\rule{0.300pt}{0.400pt}}
\multiput(1279.00,584.59)(0.721,0.485){11}{\rule{0.671pt}{0.117pt}}
\multiput(1279.00,583.17)(8.606,7.000){2}{\rule{0.336pt}{0.400pt}}
\sbox{\plotpoint}{\rule[-0.400pt]{0.800pt}{0.800pt}}%
\sbox{\plotpoint}{\rule[-0.200pt]{0.400pt}{0.400pt}}%
\put(850,236){\usebox{\plotpoint}}
\multiput(850.58,236.00)(0.491,2.059){17}{\rule{0.118pt}{1.700pt}}
\multiput(849.17,236.00)(10.000,36.472){2}{\rule{0.400pt}{0.850pt}}
\multiput(860.58,276.00)(0.491,1.017){17}{\rule{0.118pt}{0.900pt}}
\multiput(859.17,276.00)(10.000,18.132){2}{\rule{0.400pt}{0.450pt}}
\multiput(870.58,296.00)(0.491,0.704){17}{\rule{0.118pt}{0.660pt}}
\multiput(869.17,296.00)(10.000,12.630){2}{\rule{0.400pt}{0.330pt}}
\multiput(880.58,310.00)(0.491,0.547){17}{\rule{0.118pt}{0.540pt}}
\multiput(879.17,310.00)(10.000,9.879){2}{\rule{0.400pt}{0.270pt}}
\multiput(890.00,321.59)(0.553,0.489){15}{\rule{0.544pt}{0.118pt}}
\multiput(890.00,320.17)(8.870,9.000){2}{\rule{0.272pt}{0.400pt}}
\multiput(900.00,330.59)(0.553,0.489){15}{\rule{0.544pt}{0.118pt}}
\multiput(900.00,329.17)(8.870,9.000){2}{\rule{0.272pt}{0.400pt}}
\multiput(910.00,339.59)(0.721,0.485){11}{\rule{0.671pt}{0.117pt}}
\multiput(910.00,338.17)(8.606,7.000){2}{\rule{0.336pt}{0.400pt}}
\multiput(920.00,346.59)(0.721,0.485){11}{\rule{0.671pt}{0.117pt}}
\multiput(920.00,345.17)(8.606,7.000){2}{\rule{0.336pt}{0.400pt}}
\multiput(930.00,353.59)(0.721,0.485){11}{\rule{0.671pt}{0.117pt}}
\multiput(930.00,352.17)(8.606,7.000){2}{\rule{0.336pt}{0.400pt}}
\multiput(940.00,360.59)(0.852,0.482){9}{\rule{0.767pt}{0.116pt}}
\multiput(940.00,359.17)(8.409,6.000){2}{\rule{0.383pt}{0.400pt}}
\multiput(950.00,366.59)(0.852,0.482){9}{\rule{0.767pt}{0.116pt}}
\multiput(950.00,365.17)(8.409,6.000){2}{\rule{0.383pt}{0.400pt}}
\multiput(960.00,372.59)(1.044,0.477){7}{\rule{0.900pt}{0.115pt}}
\multiput(960.00,371.17)(8.132,5.000){2}{\rule{0.450pt}{0.400pt}}
\multiput(970.00,377.59)(0.852,0.482){9}{\rule{0.767pt}{0.116pt}}
\multiput(970.00,376.17)(8.409,6.000){2}{\rule{0.383pt}{0.400pt}}
\multiput(980.00,383.59)(1.044,0.477){7}{\rule{0.900pt}{0.115pt}}
\multiput(980.00,382.17)(8.132,5.000){2}{\rule{0.450pt}{0.400pt}}
\multiput(990.00,388.59)(0.852,0.482){9}{\rule{0.767pt}{0.116pt}}
\multiput(990.00,387.17)(8.409,6.000){2}{\rule{0.383pt}{0.400pt}}
\multiput(1000.00,394.59)(1.044,0.477){7}{\rule{0.900pt}{0.115pt}}
\multiput(1000.00,393.17)(8.132,5.000){2}{\rule{0.450pt}{0.400pt}}
\multiput(1010.00,399.59)(0.852,0.482){9}{\rule{0.767pt}{0.116pt}}
\multiput(1010.00,398.17)(8.409,6.000){2}{\rule{0.383pt}{0.400pt}}
\multiput(1020.00,405.59)(1.044,0.477){7}{\rule{0.900pt}{0.115pt}}
\multiput(1020.00,404.17)(8.132,5.000){2}{\rule{0.450pt}{0.400pt}}
\multiput(1030.00,410.59)(0.852,0.482){9}{\rule{0.767pt}{0.116pt}}
\multiput(1030.00,409.17)(8.409,6.000){2}{\rule{0.383pt}{0.400pt}}
\multiput(1040.00,416.59)(0.933,0.477){7}{\rule{0.820pt}{0.115pt}}
\multiput(1040.00,415.17)(7.298,5.000){2}{\rule{0.410pt}{0.400pt}}
\multiput(1049.00,421.59)(0.852,0.482){9}{\rule{0.767pt}{0.116pt}}
\multiput(1049.00,420.17)(8.409,6.000){2}{\rule{0.383pt}{0.400pt}}
\multiput(1059.00,427.59)(0.852,0.482){9}{\rule{0.767pt}{0.116pt}}
\multiput(1059.00,426.17)(8.409,6.000){2}{\rule{0.383pt}{0.400pt}}
\multiput(1069.00,433.59)(1.044,0.477){7}{\rule{0.900pt}{0.115pt}}
\multiput(1069.00,432.17)(8.132,5.000){2}{\rule{0.450pt}{0.400pt}}
\multiput(1079.00,438.59)(0.852,0.482){9}{\rule{0.767pt}{0.116pt}}
\multiput(1079.00,437.17)(8.409,6.000){2}{\rule{0.383pt}{0.400pt}}
\multiput(1089.00,444.59)(0.721,0.485){11}{\rule{0.671pt}{0.117pt}}
\multiput(1089.00,443.17)(8.606,7.000){2}{\rule{0.336pt}{0.400pt}}
\multiput(1099.00,451.59)(0.852,0.482){9}{\rule{0.767pt}{0.116pt}}
\multiput(1099.00,450.17)(8.409,6.000){2}{\rule{0.383pt}{0.400pt}}
\multiput(1109.00,457.59)(0.721,0.485){11}{\rule{0.671pt}{0.117pt}}
\multiput(1109.00,456.17)(8.606,7.000){2}{\rule{0.336pt}{0.400pt}}
\multiput(1119.00,464.59)(0.852,0.482){9}{\rule{0.767pt}{0.116pt}}
\multiput(1119.00,463.17)(8.409,6.000){2}{\rule{0.383pt}{0.400pt}}
\multiput(1129.00,470.59)(0.721,0.485){11}{\rule{0.671pt}{0.117pt}}
\multiput(1129.00,469.17)(8.606,7.000){2}{\rule{0.336pt}{0.400pt}}
\multiput(1139.00,477.59)(0.721,0.485){11}{\rule{0.671pt}{0.117pt}}
\multiput(1139.00,476.17)(8.606,7.000){2}{\rule{0.336pt}{0.400pt}}
\multiput(1149.00,484.59)(0.626,0.488){13}{\rule{0.600pt}{0.117pt}}
\multiput(1149.00,483.17)(8.755,8.000){2}{\rule{0.300pt}{0.400pt}}
\multiput(1159.00,492.59)(0.721,0.485){11}{\rule{0.671pt}{0.117pt}}
\multiput(1159.00,491.17)(8.606,7.000){2}{\rule{0.336pt}{0.400pt}}
\multiput(1169.00,499.59)(0.721,0.485){11}{\rule{0.671pt}{0.117pt}}
\multiput(1169.00,498.17)(8.606,7.000){2}{\rule{0.336pt}{0.400pt}}
\multiput(1179.00,506.59)(0.626,0.488){13}{\rule{0.600pt}{0.117pt}}
\multiput(1179.00,505.17)(8.755,8.000){2}{\rule{0.300pt}{0.400pt}}
\multiput(1189.00,514.59)(0.626,0.488){13}{\rule{0.600pt}{0.117pt}}
\multiput(1189.00,513.17)(8.755,8.000){2}{\rule{0.300pt}{0.400pt}}
\multiput(1199.00,522.59)(0.721,0.485){11}{\rule{0.671pt}{0.117pt}}
\multiput(1199.00,521.17)(8.606,7.000){2}{\rule{0.336pt}{0.400pt}}
\multiput(1209.00,529.59)(0.626,0.488){13}{\rule{0.600pt}{0.117pt}}
\multiput(1209.00,528.17)(8.755,8.000){2}{\rule{0.300pt}{0.400pt}}
\multiput(1219.00,537.59)(0.626,0.488){13}{\rule{0.600pt}{0.117pt}}
\multiput(1219.00,536.17)(8.755,8.000){2}{\rule{0.300pt}{0.400pt}}
\multiput(1229.00,545.59)(0.721,0.485){11}{\rule{0.671pt}{0.117pt}}
\multiput(1229.00,544.17)(8.606,7.000){2}{\rule{0.336pt}{0.400pt}}
\multiput(1239.00,552.59)(0.626,0.488){13}{\rule{0.600pt}{0.117pt}}
\multiput(1239.00,551.17)(8.755,8.000){2}{\rule{0.300pt}{0.400pt}}
\multiput(1249.00,560.59)(0.626,0.488){13}{\rule{0.600pt}{0.117pt}}
\multiput(1249.00,559.17)(8.755,8.000){2}{\rule{0.300pt}{0.400pt}}
\multiput(1259.00,568.59)(0.721,0.485){11}{\rule{0.671pt}{0.117pt}}
\multiput(1259.00,567.17)(8.606,7.000){2}{\rule{0.336pt}{0.400pt}}
\multiput(1269.00,575.59)(0.626,0.488){13}{\rule{0.600pt}{0.117pt}}
\multiput(1269.00,574.17)(8.755,8.000){2}{\rule{0.300pt}{0.400pt}}
\multiput(1279.00,583.59)(0.626,0.488){13}{\rule{0.600pt}{0.117pt}}
\multiput(1279.00,582.17)(8.755,8.000){2}{\rule{0.300pt}{0.400pt}}
\sbox{\plotpoint}{\rule[-0.500pt]{1.000pt}{1.000pt}}%
\sbox{\plotpoint}{\rule[-0.200pt]{0.400pt}{0.400pt}}%
\put(800,236){\usebox{\plotpoint}}
\multiput(800.58,236.00)(0.491,0.808){17}{\rule{0.118pt}{0.740pt}}
\multiput(799.17,236.00)(10.000,14.464){2}{\rule{0.400pt}{0.370pt}}
\multiput(810.58,252.00)(0.491,0.600){17}{\rule{0.118pt}{0.580pt}}
\multiput(809.17,252.00)(10.000,10.796){2}{\rule{0.400pt}{0.290pt}}
\multiput(820.00,264.58)(0.495,0.491){17}{\rule{0.500pt}{0.118pt}}
\multiput(820.00,263.17)(8.962,10.000){2}{\rule{0.250pt}{0.400pt}}
\multiput(830.00,274.59)(0.553,0.489){15}{\rule{0.544pt}{0.118pt}}
\multiput(830.00,273.17)(8.870,9.000){2}{\rule{0.272pt}{0.400pt}}
\multiput(840.00,283.59)(0.626,0.488){13}{\rule{0.600pt}{0.117pt}}
\multiput(840.00,282.17)(8.755,8.000){2}{\rule{0.300pt}{0.400pt}}
\multiput(850.00,291.59)(0.721,0.485){11}{\rule{0.671pt}{0.117pt}}
\multiput(850.00,290.17)(8.606,7.000){2}{\rule{0.336pt}{0.400pt}}
\multiput(860.00,298.59)(0.852,0.482){9}{\rule{0.767pt}{0.116pt}}
\multiput(860.00,297.17)(8.409,6.000){2}{\rule{0.383pt}{0.400pt}}
\multiput(870.00,304.59)(0.721,0.485){11}{\rule{0.671pt}{0.117pt}}
\multiput(870.00,303.17)(8.606,7.000){2}{\rule{0.336pt}{0.400pt}}
\multiput(880.00,311.59)(0.852,0.482){9}{\rule{0.767pt}{0.116pt}}
\multiput(880.00,310.17)(8.409,6.000){2}{\rule{0.383pt}{0.400pt}}
\multiput(890.00,317.59)(1.044,0.477){7}{\rule{0.900pt}{0.115pt}}
\multiput(890.00,316.17)(8.132,5.000){2}{\rule{0.450pt}{0.400pt}}
\multiput(900.00,322.59)(0.852,0.482){9}{\rule{0.767pt}{0.116pt}}
\multiput(900.00,321.17)(8.409,6.000){2}{\rule{0.383pt}{0.400pt}}
\multiput(910.00,328.59)(0.852,0.482){9}{\rule{0.767pt}{0.116pt}}
\multiput(910.00,327.17)(8.409,6.000){2}{\rule{0.383pt}{0.400pt}}
\multiput(920.00,334.59)(1.044,0.477){7}{\rule{0.900pt}{0.115pt}}
\multiput(920.00,333.17)(8.132,5.000){2}{\rule{0.450pt}{0.400pt}}
\multiput(930.00,339.59)(1.044,0.477){7}{\rule{0.900pt}{0.115pt}}
\multiput(930.00,338.17)(8.132,5.000){2}{\rule{0.450pt}{0.400pt}}
\multiput(940.00,344.59)(0.852,0.482){9}{\rule{0.767pt}{0.116pt}}
\multiput(940.00,343.17)(8.409,6.000){2}{\rule{0.383pt}{0.400pt}}
\multiput(950.00,350.59)(1.044,0.477){7}{\rule{0.900pt}{0.115pt}}
\multiput(950.00,349.17)(8.132,5.000){2}{\rule{0.450pt}{0.400pt}}
\multiput(960.00,355.59)(1.044,0.477){7}{\rule{0.900pt}{0.115pt}}
\multiput(960.00,354.17)(8.132,5.000){2}{\rule{0.450pt}{0.400pt}}
\multiput(970.00,360.59)(0.852,0.482){9}{\rule{0.767pt}{0.116pt}}
\multiput(970.00,359.17)(8.409,6.000){2}{\rule{0.383pt}{0.400pt}}
\multiput(980.00,366.59)(1.044,0.477){7}{\rule{0.900pt}{0.115pt}}
\multiput(980.00,365.17)(8.132,5.000){2}{\rule{0.450pt}{0.400pt}}
\multiput(990.00,371.59)(0.852,0.482){9}{\rule{0.767pt}{0.116pt}}
\multiput(990.00,370.17)(8.409,6.000){2}{\rule{0.383pt}{0.400pt}}
\multiput(1000.00,377.59)(0.852,0.482){9}{\rule{0.767pt}{0.116pt}}
\multiput(1000.00,376.17)(8.409,6.000){2}{\rule{0.383pt}{0.400pt}}
\multiput(1010.00,383.59)(0.852,0.482){9}{\rule{0.767pt}{0.116pt}}
\multiput(1010.00,382.17)(8.409,6.000){2}{\rule{0.383pt}{0.400pt}}
\multiput(1020.00,389.59)(0.852,0.482){9}{\rule{0.767pt}{0.116pt}}
\multiput(1020.00,388.17)(8.409,6.000){2}{\rule{0.383pt}{0.400pt}}
\multiput(1030.00,395.59)(0.852,0.482){9}{\rule{0.767pt}{0.116pt}}
\multiput(1030.00,394.17)(8.409,6.000){2}{\rule{0.383pt}{0.400pt}}
\multiput(1040.00,401.59)(0.762,0.482){9}{\rule{0.700pt}{0.116pt}}
\multiput(1040.00,400.17)(7.547,6.000){2}{\rule{0.350pt}{0.400pt}}
\multiput(1049.00,407.59)(0.721,0.485){11}{\rule{0.671pt}{0.117pt}}
\multiput(1049.00,406.17)(8.606,7.000){2}{\rule{0.336pt}{0.400pt}}
\multiput(1059.00,414.59)(0.721,0.485){11}{\rule{0.671pt}{0.117pt}}
\multiput(1059.00,413.17)(8.606,7.000){2}{\rule{0.336pt}{0.400pt}}
\multiput(1069.00,421.59)(0.721,0.485){11}{\rule{0.671pt}{0.117pt}}
\multiput(1069.00,420.17)(8.606,7.000){2}{\rule{0.336pt}{0.400pt}}
\multiput(1079.00,428.59)(0.721,0.485){11}{\rule{0.671pt}{0.117pt}}
\multiput(1079.00,427.17)(8.606,7.000){2}{\rule{0.336pt}{0.400pt}}
\multiput(1089.00,435.59)(0.721,0.485){11}{\rule{0.671pt}{0.117pt}}
\multiput(1089.00,434.17)(8.606,7.000){2}{\rule{0.336pt}{0.400pt}}
\multiput(1099.00,442.59)(0.626,0.488){13}{\rule{0.600pt}{0.117pt}}
\multiput(1099.00,441.17)(8.755,8.000){2}{\rule{0.300pt}{0.400pt}}
\multiput(1109.00,450.59)(0.626,0.488){13}{\rule{0.600pt}{0.117pt}}
\multiput(1109.00,449.17)(8.755,8.000){2}{\rule{0.300pt}{0.400pt}}
\multiput(1119.00,458.59)(0.721,0.485){11}{\rule{0.671pt}{0.117pt}}
\multiput(1119.00,457.17)(8.606,7.000){2}{\rule{0.336pt}{0.400pt}}
\multiput(1129.00,465.59)(0.626,0.488){13}{\rule{0.600pt}{0.117pt}}
\multiput(1129.00,464.17)(8.755,8.000){2}{\rule{0.300pt}{0.400pt}}
\multiput(1139.00,473.59)(0.626,0.488){13}{\rule{0.600pt}{0.117pt}}
\multiput(1139.00,472.17)(8.755,8.000){2}{\rule{0.300pt}{0.400pt}}
\multiput(1149.00,481.59)(0.626,0.488){13}{\rule{0.600pt}{0.117pt}}
\multiput(1149.00,480.17)(8.755,8.000){2}{\rule{0.300pt}{0.400pt}}
\multiput(1159.00,489.59)(0.626,0.488){13}{\rule{0.600pt}{0.117pt}}
\multiput(1159.00,488.17)(8.755,8.000){2}{\rule{0.300pt}{0.400pt}}
\multiput(1169.00,497.59)(0.626,0.488){13}{\rule{0.600pt}{0.117pt}}
\multiput(1169.00,496.17)(8.755,8.000){2}{\rule{0.300pt}{0.400pt}}
\multiput(1179.00,505.59)(0.626,0.488){13}{\rule{0.600pt}{0.117pt}}
\multiput(1179.00,504.17)(8.755,8.000){2}{\rule{0.300pt}{0.400pt}}
\multiput(1189.00,513.59)(0.626,0.488){13}{\rule{0.600pt}{0.117pt}}
\multiput(1189.00,512.17)(8.755,8.000){2}{\rule{0.300pt}{0.400pt}}
\multiput(1199.00,521.59)(0.721,0.485){11}{\rule{0.671pt}{0.117pt}}
\multiput(1199.00,520.17)(8.606,7.000){2}{\rule{0.336pt}{0.400pt}}
\multiput(1209.00,528.59)(0.626,0.488){13}{\rule{0.600pt}{0.117pt}}
\multiput(1209.00,527.17)(8.755,8.000){2}{\rule{0.300pt}{0.400pt}}
\multiput(1219.00,536.59)(0.626,0.488){13}{\rule{0.600pt}{0.117pt}}
\multiput(1219.00,535.17)(8.755,8.000){2}{\rule{0.300pt}{0.400pt}}
\multiput(1229.00,544.59)(0.626,0.488){13}{\rule{0.600pt}{0.117pt}}
\multiput(1229.00,543.17)(8.755,8.000){2}{\rule{0.300pt}{0.400pt}}
\multiput(1239.00,552.59)(0.626,0.488){13}{\rule{0.600pt}{0.117pt}}
\multiput(1239.00,551.17)(8.755,8.000){2}{\rule{0.300pt}{0.400pt}}
\multiput(1249.00,560.59)(0.721,0.485){11}{\rule{0.671pt}{0.117pt}}
\multiput(1249.00,559.17)(8.606,7.000){2}{\rule{0.336pt}{0.400pt}}
\sbox{\plotpoint}{\rule[-0.600pt]{1.200pt}{1.200pt}}%
\sbox{\plotpoint}{\rule[-0.200pt]{0.400pt}{0.400pt}}%
\put(800,250){\usebox{\plotpoint}}
\multiput(800.00,250.59)(0.852,0.482){9}{\rule{0.767pt}{0.116pt}}
\multiput(800.00,249.17)(8.409,6.000){2}{\rule{0.383pt}{0.400pt}}
\multiput(810.00,256.59)(0.852,0.482){9}{\rule{0.767pt}{0.116pt}}
\multiput(810.00,255.17)(8.409,6.000){2}{\rule{0.383pt}{0.400pt}}
\multiput(820.00,262.59)(0.852,0.482){9}{\rule{0.767pt}{0.116pt}}
\multiput(820.00,261.17)(8.409,6.000){2}{\rule{0.383pt}{0.400pt}}
\multiput(830.00,268.59)(0.852,0.482){9}{\rule{0.767pt}{0.116pt}}
\multiput(830.00,267.17)(8.409,6.000){2}{\rule{0.383pt}{0.400pt}}
\multiput(840.00,274.59)(1.044,0.477){7}{\rule{0.900pt}{0.115pt}}
\multiput(840.00,273.17)(8.132,5.000){2}{\rule{0.450pt}{0.400pt}}
\multiput(850.00,279.59)(1.044,0.477){7}{\rule{0.900pt}{0.115pt}}
\multiput(850.00,278.17)(8.132,5.000){2}{\rule{0.450pt}{0.400pt}}
\multiput(860.00,284.59)(0.852,0.482){9}{\rule{0.767pt}{0.116pt}}
\multiput(860.00,283.17)(8.409,6.000){2}{\rule{0.383pt}{0.400pt}}
\multiput(870.00,290.59)(1.044,0.477){7}{\rule{0.900pt}{0.115pt}}
\multiput(870.00,289.17)(8.132,5.000){2}{\rule{0.450pt}{0.400pt}}
\multiput(880.00,295.59)(1.044,0.477){7}{\rule{0.900pt}{0.115pt}}
\multiput(880.00,294.17)(8.132,5.000){2}{\rule{0.450pt}{0.400pt}}
\multiput(890.00,300.59)(0.852,0.482){9}{\rule{0.767pt}{0.116pt}}
\multiput(890.00,299.17)(8.409,6.000){2}{\rule{0.383pt}{0.400pt}}
\multiput(900.00,306.59)(1.044,0.477){7}{\rule{0.900pt}{0.115pt}}
\multiput(900.00,305.17)(8.132,5.000){2}{\rule{0.450pt}{0.400pt}}
\multiput(910.00,311.59)(1.044,0.477){7}{\rule{0.900pt}{0.115pt}}
\multiput(910.00,310.17)(8.132,5.000){2}{\rule{0.450pt}{0.400pt}}
\multiput(920.00,316.59)(0.852,0.482){9}{\rule{0.767pt}{0.116pt}}
\multiput(920.00,315.17)(8.409,6.000){2}{\rule{0.383pt}{0.400pt}}
\multiput(930.00,322.59)(1.044,0.477){7}{\rule{0.900pt}{0.115pt}}
\multiput(930.00,321.17)(8.132,5.000){2}{\rule{0.450pt}{0.400pt}}
\multiput(940.00,327.59)(0.852,0.482){9}{\rule{0.767pt}{0.116pt}}
\multiput(940.00,326.17)(8.409,6.000){2}{\rule{0.383pt}{0.400pt}}
\multiput(950.00,333.59)(0.852,0.482){9}{\rule{0.767pt}{0.116pt}}
\multiput(950.00,332.17)(8.409,6.000){2}{\rule{0.383pt}{0.400pt}}
\multiput(960.00,339.59)(0.852,0.482){9}{\rule{0.767pt}{0.116pt}}
\multiput(960.00,338.17)(8.409,6.000){2}{\rule{0.383pt}{0.400pt}}
\multiput(970.00,345.59)(0.852,0.482){9}{\rule{0.767pt}{0.116pt}}
\multiput(970.00,344.17)(8.409,6.000){2}{\rule{0.383pt}{0.400pt}}
\multiput(980.00,351.59)(0.721,0.485){11}{\rule{0.671pt}{0.117pt}}
\multiput(980.00,350.17)(8.606,7.000){2}{\rule{0.336pt}{0.400pt}}
\multiput(990.00,358.59)(0.852,0.482){9}{\rule{0.767pt}{0.116pt}}
\multiput(990.00,357.17)(8.409,6.000){2}{\rule{0.383pt}{0.400pt}}
\multiput(1000.00,364.59)(0.721,0.485){11}{\rule{0.671pt}{0.117pt}}
\multiput(1000.00,363.17)(8.606,7.000){2}{\rule{0.336pt}{0.400pt}}
\multiput(1010.00,371.59)(0.721,0.485){11}{\rule{0.671pt}{0.117pt}}
\multiput(1010.00,370.17)(8.606,7.000){2}{\rule{0.336pt}{0.400pt}}
\multiput(1020.00,378.59)(0.721,0.485){11}{\rule{0.671pt}{0.117pt}}
\multiput(1020.00,377.17)(8.606,7.000){2}{\rule{0.336pt}{0.400pt}}
\multiput(1030.00,385.59)(0.626,0.488){13}{\rule{0.600pt}{0.117pt}}
\multiput(1030.00,384.17)(8.755,8.000){2}{\rule{0.300pt}{0.400pt}}
\multiput(1040.00,393.59)(0.645,0.485){11}{\rule{0.614pt}{0.117pt}}
\multiput(1040.00,392.17)(7.725,7.000){2}{\rule{0.307pt}{0.400pt}}
\multiput(1049.00,400.59)(0.626,0.488){13}{\rule{0.600pt}{0.117pt}}
\multiput(1049.00,399.17)(8.755,8.000){2}{\rule{0.300pt}{0.400pt}}
\multiput(1059.00,408.59)(0.626,0.488){13}{\rule{0.600pt}{0.117pt}}
\multiput(1059.00,407.17)(8.755,8.000){2}{\rule{0.300pt}{0.400pt}}
\multiput(1069.00,416.59)(0.626,0.488){13}{\rule{0.600pt}{0.117pt}}
\multiput(1069.00,415.17)(8.755,8.000){2}{\rule{0.300pt}{0.400pt}}
\multiput(1079.00,424.59)(0.626,0.488){13}{\rule{0.600pt}{0.117pt}}
\multiput(1079.00,423.17)(8.755,8.000){2}{\rule{0.300pt}{0.400pt}}
\multiput(1089.00,432.59)(0.626,0.488){13}{\rule{0.600pt}{0.117pt}}
\multiput(1089.00,431.17)(8.755,8.000){2}{\rule{0.300pt}{0.400pt}}
\multiput(1099.00,440.59)(0.626,0.488){13}{\rule{0.600pt}{0.117pt}}
\multiput(1099.00,439.17)(8.755,8.000){2}{\rule{0.300pt}{0.400pt}}
\multiput(1109.00,448.59)(0.626,0.488){13}{\rule{0.600pt}{0.117pt}}
\multiput(1109.00,447.17)(8.755,8.000){2}{\rule{0.300pt}{0.400pt}}
\multiput(1119.00,456.59)(0.626,0.488){13}{\rule{0.600pt}{0.117pt}}
\multiput(1119.00,455.17)(8.755,8.000){2}{\rule{0.300pt}{0.400pt}}
\multiput(1129.00,464.59)(0.626,0.488){13}{\rule{0.600pt}{0.117pt}}
\multiput(1129.00,463.17)(8.755,8.000){2}{\rule{0.300pt}{0.400pt}}
\multiput(1139.00,472.59)(0.626,0.488){13}{\rule{0.600pt}{0.117pt}}
\multiput(1139.00,471.17)(8.755,8.000){2}{\rule{0.300pt}{0.400pt}}
\multiput(1149.00,480.59)(0.626,0.488){13}{\rule{0.600pt}{0.117pt}}
\multiput(1149.00,479.17)(8.755,8.000){2}{\rule{0.300pt}{0.400pt}}
\multiput(1159.00,488.59)(0.626,0.488){13}{\rule{0.600pt}{0.117pt}}
\multiput(1159.00,487.17)(8.755,8.000){2}{\rule{0.300pt}{0.400pt}}
\multiput(1169.00,496.59)(0.626,0.488){13}{\rule{0.600pt}{0.117pt}}
\multiput(1169.00,495.17)(8.755,8.000){2}{\rule{0.300pt}{0.400pt}}
\multiput(1179.00,504.59)(0.626,0.488){13}{\rule{0.600pt}{0.117pt}}
\multiput(1179.00,503.17)(8.755,8.000){2}{\rule{0.300pt}{0.400pt}}
\multiput(1189.00,512.59)(0.626,0.488){13}{\rule{0.600pt}{0.117pt}}
\multiput(1189.00,511.17)(8.755,8.000){2}{\rule{0.300pt}{0.400pt}}
\sbox{\plotpoint}{\rule[-0.500pt]{1.000pt}{1.000pt}}%
\sbox{\plotpoint}{\rule[-0.200pt]{0.400pt}{0.400pt}}%
\put(800,236){\usebox{\plotpoint}}
\multiput(800.00,236.59)(1.044,0.477){7}{\rule{0.900pt}{0.115pt}}
\multiput(800.00,235.17)(8.132,5.000){2}{\rule{0.450pt}{0.400pt}}
\multiput(810.00,241.59)(1.044,0.477){7}{\rule{0.900pt}{0.115pt}}
\multiput(810.00,240.17)(8.132,5.000){2}{\rule{0.450pt}{0.400pt}}
\multiput(820.00,246.59)(1.044,0.477){7}{\rule{0.900pt}{0.115pt}}
\multiput(820.00,245.17)(8.132,5.000){2}{\rule{0.450pt}{0.400pt}}
\multiput(830.00,251.59)(1.044,0.477){7}{\rule{0.900pt}{0.115pt}}
\multiput(830.00,250.17)(8.132,5.000){2}{\rule{0.450pt}{0.400pt}}
\multiput(840.00,256.59)(0.852,0.482){9}{\rule{0.767pt}{0.116pt}}
\multiput(840.00,255.17)(8.409,6.000){2}{\rule{0.383pt}{0.400pt}}
\multiput(850.00,262.59)(1.044,0.477){7}{\rule{0.900pt}{0.115pt}}
\multiput(850.00,261.17)(8.132,5.000){2}{\rule{0.450pt}{0.400pt}}
\multiput(860.00,267.59)(1.044,0.477){7}{\rule{0.900pt}{0.115pt}}
\multiput(860.00,266.17)(8.132,5.000){2}{\rule{0.450pt}{0.400pt}}
\multiput(870.00,272.59)(0.852,0.482){9}{\rule{0.767pt}{0.116pt}}
\multiput(870.00,271.17)(8.409,6.000){2}{\rule{0.383pt}{0.400pt}}
\multiput(880.00,278.59)(0.852,0.482){9}{\rule{0.767pt}{0.116pt}}
\multiput(880.00,277.17)(8.409,6.000){2}{\rule{0.383pt}{0.400pt}}
\multiput(890.00,284.59)(1.044,0.477){7}{\rule{0.900pt}{0.115pt}}
\multiput(890.00,283.17)(8.132,5.000){2}{\rule{0.450pt}{0.400pt}}
\multiput(900.00,289.59)(0.852,0.482){9}{\rule{0.767pt}{0.116pt}}
\multiput(900.00,288.17)(8.409,6.000){2}{\rule{0.383pt}{0.400pt}}
\multiput(910.00,295.59)(0.721,0.485){11}{\rule{0.671pt}{0.117pt}}
\multiput(910.00,294.17)(8.606,7.000){2}{\rule{0.336pt}{0.400pt}}
\multiput(920.00,302.59)(0.852,0.482){9}{\rule{0.767pt}{0.116pt}}
\multiput(920.00,301.17)(8.409,6.000){2}{\rule{0.383pt}{0.400pt}}
\multiput(930.00,308.59)(0.721,0.485){11}{\rule{0.671pt}{0.117pt}}
\multiput(930.00,307.17)(8.606,7.000){2}{\rule{0.336pt}{0.400pt}}
\multiput(940.00,315.59)(0.852,0.482){9}{\rule{0.767pt}{0.116pt}}
\multiput(940.00,314.17)(8.409,6.000){2}{\rule{0.383pt}{0.400pt}}
\multiput(950.00,321.59)(0.721,0.485){11}{\rule{0.671pt}{0.117pt}}
\multiput(950.00,320.17)(8.606,7.000){2}{\rule{0.336pt}{0.400pt}}
\multiput(960.00,328.59)(0.626,0.488){13}{\rule{0.600pt}{0.117pt}}
\multiput(960.00,327.17)(8.755,8.000){2}{\rule{0.300pt}{0.400pt}}
\multiput(970.00,336.59)(0.721,0.485){11}{\rule{0.671pt}{0.117pt}}
\multiput(970.00,335.17)(8.606,7.000){2}{\rule{0.336pt}{0.400pt}}
\multiput(980.00,343.59)(0.626,0.488){13}{\rule{0.600pt}{0.117pt}}
\multiput(980.00,342.17)(8.755,8.000){2}{\rule{0.300pt}{0.400pt}}
\multiput(990.00,351.59)(0.721,0.485){11}{\rule{0.671pt}{0.117pt}}
\multiput(990.00,350.17)(8.606,7.000){2}{\rule{0.336pt}{0.400pt}}
\multiput(1000.00,358.59)(0.626,0.488){13}{\rule{0.600pt}{0.117pt}}
\multiput(1000.00,357.17)(8.755,8.000){2}{\rule{0.300pt}{0.400pt}}
\multiput(1010.00,366.59)(0.626,0.488){13}{\rule{0.600pt}{0.117pt}}
\multiput(1010.00,365.17)(8.755,8.000){2}{\rule{0.300pt}{0.400pt}}
\multiput(1020.00,374.59)(0.626,0.488){13}{\rule{0.600pt}{0.117pt}}
\multiput(1020.00,373.17)(8.755,8.000){2}{\rule{0.300pt}{0.400pt}}
\multiput(1030.00,382.59)(0.626,0.488){13}{\rule{0.600pt}{0.117pt}}
\multiput(1030.00,381.17)(8.755,8.000){2}{\rule{0.300pt}{0.400pt}}
\multiput(1040.00,390.59)(0.560,0.488){13}{\rule{0.550pt}{0.117pt}}
\multiput(1040.00,389.17)(7.858,8.000){2}{\rule{0.275pt}{0.400pt}}
\multiput(1049.00,398.59)(0.626,0.488){13}{\rule{0.600pt}{0.117pt}}
\multiput(1049.00,397.17)(8.755,8.000){2}{\rule{0.300pt}{0.400pt}}
\multiput(1059.00,406.59)(0.626,0.488){13}{\rule{0.600pt}{0.117pt}}
\multiput(1059.00,405.17)(8.755,8.000){2}{\rule{0.300pt}{0.400pt}}
\multiput(1069.00,414.59)(0.553,0.489){15}{\rule{0.544pt}{0.118pt}}
\multiput(1069.00,413.17)(8.870,9.000){2}{\rule{0.272pt}{0.400pt}}
\multiput(1079.00,423.59)(0.626,0.488){13}{\rule{0.600pt}{0.117pt}}
\multiput(1079.00,422.17)(8.755,8.000){2}{\rule{0.300pt}{0.400pt}}
\multiput(1089.00,431.59)(0.626,0.488){13}{\rule{0.600pt}{0.117pt}}
\multiput(1089.00,430.17)(8.755,8.000){2}{\rule{0.300pt}{0.400pt}}
\multiput(1099.00,439.59)(0.626,0.488){13}{\rule{0.600pt}{0.117pt}}
\multiput(1099.00,438.17)(8.755,8.000){2}{\rule{0.300pt}{0.400pt}}
\multiput(1109.00,447.59)(0.626,0.488){13}{\rule{0.600pt}{0.117pt}}
\multiput(1109.00,446.17)(8.755,8.000){2}{\rule{0.300pt}{0.400pt}}
\multiput(1119.00,455.59)(0.553,0.489){15}{\rule{0.544pt}{0.118pt}}
\multiput(1119.00,454.17)(8.870,9.000){2}{\rule{0.272pt}{0.400pt}}
\multiput(1129.00,464.59)(0.721,0.485){11}{\rule{0.671pt}{0.117pt}}
\multiput(1129.00,463.17)(8.606,7.000){2}{\rule{0.336pt}{0.400pt}}
\put(800,218){\usebox{\plotpoint}}
\multiput(800.00,218.59)(1.044,0.477){7}{\rule{0.900pt}{0.115pt}}
\multiput(800.00,217.17)(8.132,5.000){2}{\rule{0.450pt}{0.400pt}}
\multiput(810.00,223.59)(0.852,0.482){9}{\rule{0.767pt}{0.116pt}}
\multiput(810.00,222.17)(8.409,6.000){2}{\rule{0.383pt}{0.400pt}}
\multiput(820.00,229.59)(1.044,0.477){7}{\rule{0.900pt}{0.115pt}}
\multiput(820.00,228.17)(8.132,5.000){2}{\rule{0.450pt}{0.400pt}}
\multiput(830.00,234.59)(0.852,0.482){9}{\rule{0.767pt}{0.116pt}}
\multiput(830.00,233.17)(8.409,6.000){2}{\rule{0.383pt}{0.400pt}}
\multiput(840.00,240.59)(0.852,0.482){9}{\rule{0.767pt}{0.116pt}}
\multiput(840.00,239.17)(8.409,6.000){2}{\rule{0.383pt}{0.400pt}}
\multiput(850.00,246.59)(0.852,0.482){9}{\rule{0.767pt}{0.116pt}}
\multiput(850.00,245.17)(8.409,6.000){2}{\rule{0.383pt}{0.400pt}}
\multiput(860.00,252.59)(0.852,0.482){9}{\rule{0.767pt}{0.116pt}}
\multiput(860.00,251.17)(8.409,6.000){2}{\rule{0.383pt}{0.400pt}}
\multiput(870.00,258.59)(0.721,0.485){11}{\rule{0.671pt}{0.117pt}}
\multiput(870.00,257.17)(8.606,7.000){2}{\rule{0.336pt}{0.400pt}}
\multiput(880.00,265.59)(0.721,0.485){11}{\rule{0.671pt}{0.117pt}}
\multiput(880.00,264.17)(8.606,7.000){2}{\rule{0.336pt}{0.400pt}}
\multiput(890.00,272.59)(0.721,0.485){11}{\rule{0.671pt}{0.117pt}}
\multiput(890.00,271.17)(8.606,7.000){2}{\rule{0.336pt}{0.400pt}}
\multiput(900.00,279.59)(0.721,0.485){11}{\rule{0.671pt}{0.117pt}}
\multiput(900.00,278.17)(8.606,7.000){2}{\rule{0.336pt}{0.400pt}}
\multiput(910.00,286.59)(0.721,0.485){11}{\rule{0.671pt}{0.117pt}}
\multiput(910.00,285.17)(8.606,7.000){2}{\rule{0.336pt}{0.400pt}}
\multiput(920.00,293.59)(0.626,0.488){13}{\rule{0.600pt}{0.117pt}}
\multiput(920.00,292.17)(8.755,8.000){2}{\rule{0.300pt}{0.400pt}}
\multiput(930.00,301.59)(0.721,0.485){11}{\rule{0.671pt}{0.117pt}}
\multiput(930.00,300.17)(8.606,7.000){2}{\rule{0.336pt}{0.400pt}}
\multiput(940.00,308.59)(0.626,0.488){13}{\rule{0.600pt}{0.117pt}}
\multiput(940.00,307.17)(8.755,8.000){2}{\rule{0.300pt}{0.400pt}}
\multiput(950.00,316.59)(0.626,0.488){13}{\rule{0.600pt}{0.117pt}}
\multiput(950.00,315.17)(8.755,8.000){2}{\rule{0.300pt}{0.400pt}}
\multiput(960.00,324.59)(0.626,0.488){13}{\rule{0.600pt}{0.117pt}}
\multiput(960.00,323.17)(8.755,8.000){2}{\rule{0.300pt}{0.400pt}}
\multiput(970.00,332.59)(0.626,0.488){13}{\rule{0.600pt}{0.117pt}}
\multiput(970.00,331.17)(8.755,8.000){2}{\rule{0.300pt}{0.400pt}}
\multiput(980.00,340.59)(0.626,0.488){13}{\rule{0.600pt}{0.117pt}}
\multiput(980.00,339.17)(8.755,8.000){2}{\rule{0.300pt}{0.400pt}}
\multiput(990.00,348.59)(0.626,0.488){13}{\rule{0.600pt}{0.117pt}}
\multiput(990.00,347.17)(8.755,8.000){2}{\rule{0.300pt}{0.400pt}}
\multiput(1000.00,356.59)(0.553,0.489){15}{\rule{0.544pt}{0.118pt}}
\multiput(1000.00,355.17)(8.870,9.000){2}{\rule{0.272pt}{0.400pt}}
\multiput(1010.00,365.59)(0.626,0.488){13}{\rule{0.600pt}{0.117pt}}
\multiput(1010.00,364.17)(8.755,8.000){2}{\rule{0.300pt}{0.400pt}}
\multiput(1020.00,373.59)(0.626,0.488){13}{\rule{0.600pt}{0.117pt}}
\multiput(1020.00,372.17)(8.755,8.000){2}{\rule{0.300pt}{0.400pt}}
\multiput(1030.00,381.59)(0.626,0.488){13}{\rule{0.600pt}{0.117pt}}
\multiput(1030.00,380.17)(8.755,8.000){2}{\rule{0.300pt}{0.400pt}}
\multiput(1040.00,389.59)(0.495,0.489){15}{\rule{0.500pt}{0.118pt}}
\multiput(1040.00,388.17)(7.962,9.000){2}{\rule{0.250pt}{0.400pt}}
\multiput(1049.00,398.59)(0.626,0.488){13}{\rule{0.600pt}{0.117pt}}
\multiput(1049.00,397.17)(8.755,8.000){2}{\rule{0.300pt}{0.400pt}}
\multiput(1059.00,406.59)(0.626,0.488){13}{\rule{0.600pt}{0.117pt}}
\multiput(1059.00,405.17)(8.755,8.000){2}{\rule{0.300pt}{0.400pt}}
\multiput(1069.00,414.59)(0.626,0.488){13}{\rule{0.600pt}{0.117pt}}
\multiput(1069.00,413.17)(8.755,8.000){2}{\rule{0.300pt}{0.400pt}}
\put(800,203){\usebox{\plotpoint}}
\multiput(800.00,203.59)(0.721,0.485){11}{\rule{0.671pt}{0.117pt}}
\multiput(800.00,202.17)(8.606,7.000){2}{\rule{0.336pt}{0.400pt}}
\multiput(810.00,210.59)(0.852,0.482){9}{\rule{0.767pt}{0.116pt}}
\multiput(810.00,209.17)(8.409,6.000){2}{\rule{0.383pt}{0.400pt}}
\multiput(820.00,216.59)(0.721,0.485){11}{\rule{0.671pt}{0.117pt}}
\multiput(820.00,215.17)(8.606,7.000){2}{\rule{0.336pt}{0.400pt}}
\multiput(830.00,223.59)(0.852,0.482){9}{\rule{0.767pt}{0.116pt}}
\multiput(830.00,222.17)(8.409,6.000){2}{\rule{0.383pt}{0.400pt}}
\multiput(840.00,229.59)(0.626,0.488){13}{\rule{0.600pt}{0.117pt}}
\multiput(840.00,228.17)(8.755,8.000){2}{\rule{0.300pt}{0.400pt}}
\multiput(850.00,237.59)(0.721,0.485){11}{\rule{0.671pt}{0.117pt}}
\multiput(850.00,236.17)(8.606,7.000){2}{\rule{0.336pt}{0.400pt}}
\multiput(860.00,244.59)(0.721,0.485){11}{\rule{0.671pt}{0.117pt}}
\multiput(860.00,243.17)(8.606,7.000){2}{\rule{0.336pt}{0.400pt}}
\multiput(870.00,251.59)(0.626,0.488){13}{\rule{0.600pt}{0.117pt}}
\multiput(870.00,250.17)(8.755,8.000){2}{\rule{0.300pt}{0.400pt}}
\multiput(880.00,259.59)(0.626,0.488){13}{\rule{0.600pt}{0.117pt}}
\multiput(880.00,258.17)(8.755,8.000){2}{\rule{0.300pt}{0.400pt}}
\multiput(890.00,267.59)(0.721,0.485){11}{\rule{0.671pt}{0.117pt}}
\multiput(890.00,266.17)(8.606,7.000){2}{\rule{0.336pt}{0.400pt}}
\multiput(900.00,274.59)(0.626,0.488){13}{\rule{0.600pt}{0.117pt}}
\multiput(900.00,273.17)(8.755,8.000){2}{\rule{0.300pt}{0.400pt}}
\multiput(910.00,282.59)(0.626,0.488){13}{\rule{0.600pt}{0.117pt}}
\multiput(910.00,281.17)(8.755,8.000){2}{\rule{0.300pt}{0.400pt}}
\multiput(920.00,290.59)(0.626,0.488){13}{\rule{0.600pt}{0.117pt}}
\multiput(920.00,289.17)(8.755,8.000){2}{\rule{0.300pt}{0.400pt}}
\multiput(930.00,298.59)(0.553,0.489){15}{\rule{0.544pt}{0.118pt}}
\multiput(930.00,297.17)(8.870,9.000){2}{\rule{0.272pt}{0.400pt}}
\multiput(940.00,307.59)(0.626,0.488){13}{\rule{0.600pt}{0.117pt}}
\multiput(940.00,306.17)(8.755,8.000){2}{\rule{0.300pt}{0.400pt}}
\multiput(950.00,315.59)(0.626,0.488){13}{\rule{0.600pt}{0.117pt}}
\multiput(950.00,314.17)(8.755,8.000){2}{\rule{0.300pt}{0.400pt}}
\multiput(960.00,323.59)(0.626,0.488){13}{\rule{0.600pt}{0.117pt}}
\multiput(960.00,322.17)(8.755,8.000){2}{\rule{0.300pt}{0.400pt}}
\multiput(970.00,331.59)(0.626,0.488){13}{\rule{0.600pt}{0.117pt}}
\multiput(970.00,330.17)(8.755,8.000){2}{\rule{0.300pt}{0.400pt}}
\multiput(980.00,339.59)(0.553,0.489){15}{\rule{0.544pt}{0.118pt}}
\multiput(980.00,338.17)(8.870,9.000){2}{\rule{0.272pt}{0.400pt}}
\multiput(990.00,348.59)(0.626,0.488){13}{\rule{0.600pt}{0.117pt}}
\multiput(990.00,347.17)(8.755,8.000){2}{\rule{0.300pt}{0.400pt}}
\multiput(1000.00,356.59)(0.626,0.488){13}{\rule{0.600pt}{0.117pt}}
\multiput(1000.00,355.17)(8.755,8.000){2}{\rule{0.300pt}{0.400pt}}
\multiput(1010.00,364.59)(0.626,0.488){13}{\rule{0.600pt}{0.117pt}}
\multiput(1010.00,363.17)(8.755,8.000){2}{\rule{0.300pt}{0.400pt}}
\sbox{\plotpoint}{\rule[-0.400pt]{0.800pt}{0.800pt}}%
\sbox{\plotpoint}{\rule[-0.200pt]{0.400pt}{0.400pt}}%
\put(800,195){\usebox{\plotpoint}}
\multiput(800.00,195.59)(0.626,0.488){13}{\rule{0.600pt}{0.117pt}}
\multiput(800.00,194.17)(8.755,8.000){2}{\rule{0.300pt}{0.400pt}}
\multiput(810.00,203.59)(0.721,0.485){11}{\rule{0.671pt}{0.117pt}}
\multiput(810.00,202.17)(8.606,7.000){2}{\rule{0.336pt}{0.400pt}}
\multiput(820.00,210.59)(0.626,0.488){13}{\rule{0.600pt}{0.117pt}}
\multiput(820.00,209.17)(8.755,8.000){2}{\rule{0.300pt}{0.400pt}}
\multiput(830.00,218.59)(0.721,0.485){11}{\rule{0.671pt}{0.117pt}}
\multiput(830.00,217.17)(8.606,7.000){2}{\rule{0.336pt}{0.400pt}}
\multiput(840.00,225.59)(0.626,0.488){13}{\rule{0.600pt}{0.117pt}}
\multiput(840.00,224.17)(8.755,8.000){2}{\rule{0.300pt}{0.400pt}}
\multiput(850.00,233.59)(0.626,0.488){13}{\rule{0.600pt}{0.117pt}}
\multiput(850.00,232.17)(8.755,8.000){2}{\rule{0.300pt}{0.400pt}}
\multiput(860.00,241.59)(0.626,0.488){13}{\rule{0.600pt}{0.117pt}}
\multiput(860.00,240.17)(8.755,8.000){2}{\rule{0.300pt}{0.400pt}}
\multiput(870.00,249.59)(0.626,0.488){13}{\rule{0.600pt}{0.117pt}}
\multiput(870.00,248.17)(8.755,8.000){2}{\rule{0.300pt}{0.400pt}}
\multiput(880.00,257.59)(0.626,0.488){13}{\rule{0.600pt}{0.117pt}}
\multiput(880.00,256.17)(8.755,8.000){2}{\rule{0.300pt}{0.400pt}}
\multiput(890.00,265.59)(0.626,0.488){13}{\rule{0.600pt}{0.117pt}}
\multiput(890.00,264.17)(8.755,8.000){2}{\rule{0.300pt}{0.400pt}}
\multiput(900.00,273.59)(0.626,0.488){13}{\rule{0.600pt}{0.117pt}}
\multiput(900.00,272.17)(8.755,8.000){2}{\rule{0.300pt}{0.400pt}}
\multiput(910.00,281.59)(0.553,0.489){15}{\rule{0.544pt}{0.118pt}}
\multiput(910.00,280.17)(8.870,9.000){2}{\rule{0.272pt}{0.400pt}}
\multiput(920.00,290.59)(0.626,0.488){13}{\rule{0.600pt}{0.117pt}}
\multiput(920.00,289.17)(8.755,8.000){2}{\rule{0.300pt}{0.400pt}}
\multiput(930.00,298.59)(0.626,0.488){13}{\rule{0.600pt}{0.117pt}}
\multiput(930.00,297.17)(8.755,8.000){2}{\rule{0.300pt}{0.400pt}}
\multiput(940.00,306.59)(0.626,0.488){13}{\rule{0.600pt}{0.117pt}}
\multiput(940.00,305.17)(8.755,8.000){2}{\rule{0.300pt}{0.400pt}}
\multiput(950.00,314.59)(0.626,0.488){13}{\rule{0.600pt}{0.117pt}}
\multiput(950.00,313.17)(8.755,8.000){2}{\rule{0.300pt}{0.400pt}}
\sbox{\plotpoint}{\rule[-0.500pt]{1.000pt}{1.000pt}}%
\sbox{\plotpoint}{\rule[-0.200pt]{0.400pt}{0.400pt}}%
\put(800,193){\usebox{\plotpoint}}
\multiput(800.00,193.59)(0.721,0.485){11}{\rule{0.671pt}{0.117pt}}
\multiput(800.00,192.17)(8.606,7.000){2}{\rule{0.336pt}{0.400pt}}
\multiput(810.00,200.59)(0.626,0.488){13}{\rule{0.600pt}{0.117pt}}
\multiput(810.00,199.17)(8.755,8.000){2}{\rule{0.300pt}{0.400pt}}
\multiput(820.00,208.59)(0.626,0.488){13}{\rule{0.600pt}{0.117pt}}
\multiput(820.00,207.17)(8.755,8.000){2}{\rule{0.300pt}{0.400pt}}
\multiput(830.00,216.59)(0.626,0.488){13}{\rule{0.600pt}{0.117pt}}
\multiput(830.00,215.17)(8.755,8.000){2}{\rule{0.300pt}{0.400pt}}
\multiput(840.00,224.59)(0.626,0.488){13}{\rule{0.600pt}{0.117pt}}
\multiput(840.00,223.17)(8.755,8.000){2}{\rule{0.300pt}{0.400pt}}
\multiput(850.00,232.59)(0.626,0.488){13}{\rule{0.600pt}{0.117pt}}
\multiput(850.00,231.17)(8.755,8.000){2}{\rule{0.300pt}{0.400pt}}
\multiput(860.00,240.59)(0.626,0.488){13}{\rule{0.600pt}{0.117pt}}
\multiput(860.00,239.17)(8.755,8.000){2}{\rule{0.300pt}{0.400pt}}
\multiput(870.00,248.59)(0.553,0.489){15}{\rule{0.544pt}{0.118pt}}
\multiput(870.00,247.17)(8.870,9.000){2}{\rule{0.272pt}{0.400pt}}
\multiput(880.00,257.59)(0.626,0.488){13}{\rule{0.600pt}{0.117pt}}
\multiput(880.00,256.17)(8.755,8.000){2}{\rule{0.300pt}{0.400pt}}
\multiput(890.00,265.59)(0.626,0.488){13}{\rule{0.600pt}{0.117pt}}
\multiput(890.00,264.17)(8.755,8.000){2}{\rule{0.300pt}{0.400pt}}
\sbox{\plotpoint}{\rule[-0.600pt]{1.200pt}{1.200pt}}%
\sbox{\plotpoint}{\rule[-0.200pt]{0.400pt}{0.400pt}}%
\put(800,192){\usebox{\plotpoint}}
\multiput(800.00,192.59)(0.626,0.488){13}{\rule{0.600pt}{0.117pt}}
\multiput(800.00,191.17)(8.755,8.000){2}{\rule{0.300pt}{0.400pt}}
\multiput(810.00,200.59)(0.626,0.488){13}{\rule{0.600pt}{0.117pt}}
\multiput(810.00,199.17)(8.755,8.000){2}{\rule{0.300pt}{0.400pt}}
\multiput(820.00,208.59)(0.626,0.488){13}{\rule{0.600pt}{0.117pt}}
\multiput(820.00,207.17)(8.755,8.000){2}{\rule{0.300pt}{0.400pt}}
\multiput(830.00,216.59)(0.626,0.488){13}{\rule{0.600pt}{0.117pt}}
\multiput(830.00,215.17)(8.755,8.000){2}{\rule{0.300pt}{0.400pt}}
\put(770.0,131.0){\rule[-0.200pt]{0.400pt}{112.018pt}}
\put(770.0,131.0){\rule[-0.200pt]{132.254pt}{0.400pt}}
\put(1319.0,131.0){\rule[-0.200pt]{0.400pt}{112.018pt}}
\put(770.0,596.0){\rule[-0.200pt]{132.254pt}{0.400pt}}
\put(190.0,183.0){\rule[-0.200pt]{4.818pt}{0.400pt}}
\put(170,183){\makebox(0,0)[r]{ 0.3776}}
\put(639.0,183.0){\rule[-0.200pt]{4.818pt}{0.400pt}}
\put(190.0,286.0){\rule[-0.200pt]{4.818pt}{0.400pt}}
\put(170,286){\makebox(0,0)[r]{ 0.378}}
\put(639.0,286.0){\rule[-0.200pt]{4.818pt}{0.400pt}}
\put(190.0,389.0){\rule[-0.200pt]{4.818pt}{0.400pt}}
\put(170,389){\makebox(0,0)[r]{ 0.3784}}
\put(639.0,389.0){\rule[-0.200pt]{4.818pt}{0.400pt}}
\put(190.0,493.0){\rule[-0.200pt]{4.818pt}{0.400pt}}
\put(170,493){\makebox(0,0)[r]{ 0.3788}}
\put(639.0,493.0){\rule[-0.200pt]{4.818pt}{0.400pt}}
\put(190.0,596.0){\rule[-0.200pt]{4.818pt}{0.400pt}}
\put(170,596){\makebox(0,0)[r]{ 0.3792}}
\put(639.0,596.0){\rule[-0.200pt]{4.818pt}{0.400pt}}
\put(190.0,131.0){\rule[-0.200pt]{0.400pt}{4.818pt}}
\put(190,90){\makebox(0,0){ 0}}
\put(190.0,576.0){\rule[-0.200pt]{0.400pt}{4.818pt}}
\put(346.0,131.0){\rule[-0.200pt]{0.400pt}{4.818pt}}
\put(346,90){\makebox(0,0){ 0.001}}
\put(346.0,576.0){\rule[-0.200pt]{0.400pt}{4.818pt}}
\put(503.0,131.0){\rule[-0.200pt]{0.400pt}{4.818pt}}
\put(503,90){\makebox(0,0){ 0.002}}
\put(503.0,576.0){\rule[-0.200pt]{0.400pt}{4.818pt}}
\put(659.0,131.0){\rule[-0.200pt]{0.400pt}{4.818pt}}
\put(659,90){\makebox(0,0){ 0.003}}
\put(659.0,576.0){\rule[-0.200pt]{0.400pt}{4.818pt}}
\put(190.0,131.0){\rule[-0.200pt]{0.400pt}{112.018pt}}
\put(190.0,131.0){\rule[-0.200pt]{112.982pt}{0.400pt}}
\put(659.0,131.0){\rule[-0.200pt]{0.400pt}{112.018pt}}
\put(190.0,596.0){\rule[-0.200pt]{112.982pt}{0.400pt}}
\put(424,29){\makebox(0,0){$ 1/n^2$}}
\put(424,658){\makebox(0,0){$\lambda _n$ for $n=20$, $22$, $\dots$, $70$}}
\put(581,562){\makebox(0,0){$+$}}
\put(513,386){\makebox(0,0){$+$}}
\put(461,289){\makebox(0,0){$+$}}
\put(421,235){\makebox(0,0){$+$}}
\put(389,204){\makebox(0,0){$+$}}
\put(364,188){\makebox(0,0){$+$}}
\put(343,178){\makebox(0,0){$+$}}
\put(325,173){\makebox(0,0){$+$}}
\put(311,170){\makebox(0,0){$+$}}
\put(298,168){\makebox(0,0){$+$}}
\put(288,167){\makebox(0,0){$+$}}
\put(279,167){\makebox(0,0){$+$}}
\put(271,167){\makebox(0,0){$+$}}
\put(264,166){\makebox(0,0){$+$}}
\put(258,166){\makebox(0,0){$+$}}
\put(253,166){\makebox(0,0){$+$}}
\put(248,166){\makebox(0,0){$+$}}
\put(244,166){\makebox(0,0){$+$}}
\put(240,166){\makebox(0,0){$+$}}
\put(236,166){\makebox(0,0){$+$}}
\put(233,166){\makebox(0,0){$+$}}
\put(231,166){\makebox(0,0){$+$}}
\put(228,166){\makebox(0,0){$+$}}
\put(226,166){\makebox(0,0){$+$}}
\put(224,166){\makebox(0,0){$+$}}
\put(190.0,131.0){\rule[-0.200pt]{0.400pt}{112.018pt}}
\put(190.0,131.0){\rule[-0.200pt]{112.982pt}{0.400pt}}
\put(659.0,131.0){\rule[-0.200pt]{0.400pt}{112.018pt}}
\put(190.0,596.0){\rule[-0.200pt]{112.982pt}{0.400pt}}
\end{picture}

%% file: bb.bbl
\begin{thebibliography}{}


\bibitem{AW}  Aizenman M., Wehr J., \textit{Rounding effects of quenched randomness on 1st-order phase transition}, Commun. Math. Phys. \textbf{130} (1990), 489-528.




\bibitem{Al} Alexander K.S., \textit{The effect of disorder on polymer depinning transitions}, Commun. Math. Phys. \textbf{279} (2008), 117-146.

\bibitem{AlS} Alexander K.S., Sidoravicius V., \textit{Pinning of polymers and interfaces by random potentials}, Annals Appl. Proba. \textbf{16} (2006), 636-669.

\bibitem{AZ}  Alexander K.S.,  Zygouras N., \textit{Quenched and annealed critical points in polymer pinning models}, Commun Math. Phys. \textbf{291} (2009), 659-689.

\bibitem{Ares} Ares S., Sanchez A., \textit{Modelling disorder: the cases of wetting and DNA denaturation}, Eur. Phys. J. B \textbf{56} (2007), 253-258.


\bibitem{Az1}  Azbel M.Y., \textit{Long-range interaction and heterogeneity yield a different kind of critical phenomenon}, Phys. Rev. E \textbf{68} (2003), 050901.

\bibitem{Az2} Azbel M.Y., \textit{Giant non-universal critical index and fluctuations in DNA phase transition}, Physica A - Stat. Mechanics Appl. \textbf{321} (2003), 571-576.

\bibitem{Berger} Berger Q., Toninelli F.L., \textit{Hierarchical pinning model in correlated random environment}, Annales de l'Institut Henri Poincar\'e - Prob. Stat. \textbf{49} (2013), 781-816.


\bibitem{MB1} Bhattacharjee S.M., Mukherji S., \textit{Directed polymers with random interaction - Marginal relevance and novel criticality}, Phys. Rev. Lett. \textbf{70} (1993), 49-52.

\bibitem{MB2} Bhattacharjee S.M., Mukherji S., \textit{Directed polymers with random interaction - An exactly solvable case}, Phys. Rev. E, \textbf{48} (1993), 3483-3496.

\bibitem{B1} Bolthausen E., Caravenna E., de Tiliere B., \textit{The quenched critical point of a diluted disordered polymer model}, Stochastic Process. Appl. \textbf{119} (2009), 1479-1504.

\bibitem{@@} Bolthausen E., Funaki T., Otobe T., \textit{Concentration under scaling limits for weakly pinned Gaussian random walks}, Proba Theory Relat. Fields \textbf{143} (2009), 441-480.

\bibitem{CG} Caravenna F., Giacomin G., \textit{On constrained annealed bounds for pinning and wetting models}, Elec. Commun. Proba. \textbf{10} (2005), 179-189.

\bibitem{Eckmann} Collet P., Eckmann J.P., Glaser V., Martin A., \textit{Study of the iterations of a mapping associated to a spin-glass model}, Commun. Math. Phys. \textbf{94}   (1984), 353-370.

\bibitem{CH} Cule D., Hwa T., \textit{Denaturation of heterogeneous DNA}, Phys. Rev. Lett.  \textbf{79} (1997), 2375-2378.

\bibitem{CY} Coluzzi B., Yeramian E., \textit{Numerical evidence for relevance of disorder in a Poland-Scheraga DNA denaturation model with self-avoidance: scaling behavior of average quantities}, Eur. Phys. J. B \textbf{56} (2007), 349-365.


\bibitem{Rel1} Derrida B., Giacomin G., Lacoin H., Toninelli F.L., \textit{Fractional moment bounds and disorder relevance for pinning models}, Commun. Math. Phys., \textbf{287} (2009), 867-887.

\bibitem{DHV} Derrida B., Hakim V., Vannimenus J., \textit{Effect of disorder on 2-dimensional wetting}, J. Stat. Phys. \textbf{66} (1992), 1189-1213.


\bibitem{Fisher} Fisher M.E., \textit{Walks, walls, wetting and melting}, J. Stat. Phys. \textbf{34} (1984), 684-689.

\bibitem{Flajolet} Flajolet P., Sedgewick R., \textit{Analytic Combinatorics}, Cambridge University Press, 2009.

\bibitem{FORG1} Forgacs G., Luck J.M., Nieuwenhuizen T.M., Orland H., \textit{Wetting of a disordered substrate -Exact critical behavior in 2 dimension}, Phys. Rev. Lett. \textbf{57} (1986), 2184-2187.

\bibitem{FORG2} Forgacs G., Luck J.M., Nieuwenhuizen T.M., Orland H., \textit{Exact critical behavior of two dimensional wetting problems with quenched disorder}, J. Stat. Phys. \textbf{51} (1988), 29-56.

\bibitem{GN} Gangardt D. M., Nechaev S. K., \textit{Wetting transition on a one-dimensional disorder}, J. Stat. Phys.  \textbf{130} (2008), 483-502.

\bibitem{GMO} Garel T., Monthus C., Orland H., \textit{A simple model for DNA denaturation}, EuroPhys. Lett.  \textbf{55}  (2001), 132-138.

\bibitem{MG3} Garel T., Monthus C., \textit{Two-dimensional wetting with binary disorder: a numerical study of the loop statistics}, Eur. Phys. J. B \textbf{46} (2005), 117-125.

\bibitem{G1} Giacomin G., \textit{Random Polymer Models}, Imperial College Press, World Scientific, London, 2007.

\bibitem{Rel2} Giacomin G.,  Lacoin H.,  Toninelli F.L., \textit{Disorder relevance at marginality and critical point shift}, Annales de l'Institut Henri Poincar\'e - Prob. Stat. \textbf{47} (2011), 148-175.

\bibitem{GT1} Giacomin G., Toninelli F.L., \textit{On the irrelevant disorder regime of pinning models}, Annals Proba. \textbf{37} (2009), 1841-1875.

\bibitem{Rel3} Giacomin G. ,  Lacoin H.,  Toninelli F.L., \textit{Marginal relevance of disorder for pinning models}, Commun. Pure and Applied Math. \textbf{63} (2010), 233-265.

\bibitem{L2} Giacomin G., Lacoin H., Toninelli F.L., \textit{Hierarchical pinning models, quadratic maps and quenched disorder}, Proba. Theory Relat. Fields \textbf{147} (2010), 185-216.

\bibitem{GT2}  Giacomin G.,  Toninelli F.L., \textit{Smoothing of depinning transitions for directed polymers with quenched disorder}, Phys. Rev. Lett. \textbf{96} (2006), 070602.


\bibitem{GT3} Giacomin G.,  Toninelli F.L., \textit{Smoothing effect of quenched disorder on polymer depinning transitions}, Commun. Math. Phys. \textbf{266} (2006),1-16.


\bibitem{harris}  Harris A.B., \textit{Effect of random defects on critical behavior of Ising model}, J. Phys. C - Solid State Phys. \textbf{7} (1974), 1671-1692.

\bibitem{Igloi} Igloi F., Monthus C., \textit{Strong disorder RG approach of random systems}, Phys. Reports-Rev. Section of Phys. Lett. \textbf{412} (2005), 277-431.


\bibitem{KMP} Kafri Y., Mukamel D., Peliti L., \textit{Why is the DNA denaturation transition first order?} Phys. Rev. Lett. \textbf{85} (2000), 4988-4991.

\bibitem{KM} Kafri Y., Mukamel D., \textit{Griffiths singularities in unbinding of strongly disordered polymers}, Phys. Rev. Lett. \textbf{91} (2003), 038103.


\bibitem{KT} Kosterlitz J.M., Thouless D.J., \textit{Early work on defect driven phase transitions}, 40 Years of Berezinskii-Kosterlitz-Thouless Theory, Ed. Jose Jorge V., World Scientific Publishing Co. Pte. Ltd. (2013), 1-67. 



\bibitem{KL} Kunz H., Livi R., \textit{DNA denaturation and wetting in the presence of disorder}, Eur. Phys. Lett. \textbf{99} (2012), 30001. 

\bibitem{L4} Lacoin H., \textit{Hierarchical pinning model with site disorder: disorder is marginally relevant}, Proba Theory Relat. Fields \textbf{148} (2010), 159-175.

\bibitem{L1}  Lacoin H., \textit{The martingale approach to disorder irrelevance for pinning models}, Elec. Commun. Prob. \textbf{15} (2010), 418-427.

\bibitem{L3} Lacoin H., Moreno G., \textit{Directed polymers on hierarchical lattices with site disorder}, Stochastic Process. Appl.  \textbf{120} (2010) , 467-493.

\bibitem{LT} Lacoin H., Toninelli F.L., \textit{A smoothing inequality for hierarchical pinning models}, Progress Proba. \textbf{62} (2009), 271-278. 


\bibitem{Lubensky-Nelson} Lubensky D.K., Nelson D.R., \textit{Single molecule statistics and the polynucleotide unzipping transition}, Phys. Rev. E \textbf{65} (2002), 031917. 


\bibitem{Monthus} Monthus C., \textit{Random walks and polymers in the presence of quenched disorder}, Lett. Math. Phys.  \textbf{78} (2006), 207-233.

\bibitem{MG1} Monthus C., Garel T., \textit{Distribution of pseudo-critical temperatures and lack of self-averaging in disordered Poland-Scheraga models with different loop
exponents}, Eur. Phys. J. B \textbf{48} (2005), 393-403.

\bibitem{MG4} Monthus C., Garel T., \textit{Multifractal statistics of the local order parameter at random critical points: Application to wetting transitions with disorder}, Phys. Rev. E \textbf{76} (2007), 021114.

\bibitem{MG2} Monthus C., Garel T., \textit{Critical points of quadratic renormalizations of random variables and phase transitions of disordered polymer models on diamond lattices}, Phys. Rev. E \textbf{77} (2008), 021132.


\bibitem{PS} Poland D., Scheraga H.A., \textit{Occurrence of a phase transition in nucleic acid models}, J. Chem. Phys. \textbf{45} (1966), 1464-1469.

\bibitem{Guttman}Richard C., Guttmann A. J., \textit{Poland-Scheraga models and the DNA denaturation transition}, J. Stat. Phys. \textbf{115} (2004), 943-965.

\bibitem{TC} Tang  L.H., Chat\' e H., \textit{Rare-event induced binding transition of heteropolymers}, Phys. Rev. Lett.  \textbf{86} (2001), 830-833.

\bibitem{T1} Toninelli F.L., \textit{A replica-coupling approach to disordered pinning models}, Commun. Math. Phys. \textbf{280} (2008), 389-401.

\bibitem{T2} Toninelli F.L., \textit{Disordered pinning models and copolymers: beyond annealed bounds}, Annals Appl. Prob. \textbf{18} (2008), 1569-1587.










%
%
%
%
%




%


%
%
%



%

%




%
%
%
%
%
%
%
%
%
%
%
%
%
%
%
%
%
%

\end{thebibliography}
